\documentclass[12pt]{article}
\usepackage[colorlinks=true, urlcolor=citecolor, linkcolor=citecolor, citecolor=citecolor]{hyperref}
\usepackage{colortbl}
\usepackage{graphicx}
\usepackage{amsmath,bm}
\usepackage{amsfonts}
\usepackage{mathrsfs}
\usepackage{natbib}
\usepackage{url}
\usepackage{enumitem}
\usepackage{booktabs}
\usepackage{makecell}
\usepackage{tabularx}
\usepackage{multirow}
\usepackage{arydshln}
\usepackage{color}
\usepackage{algorithm}
\usepackage{algpseudocode}
\usepackage{textcomp}
\usepackage{mathtools,comment}
\usepackage{pdflscape}
\usepackage{titling}

\graphicspath{{./figure/}}
   
\begin{document}
\def\spacingset#1{\renewcommand{\baselinestretch}%
{#1}\small\normalsize} \spacingset{1}

\newcommand{\e}{\operatorname{E}}
\newcommand{\prodn}{\prod_{i=1}^n}
\newcommand{\sumn}{\sum_{i=1}^n}
\newcommand{\dott}{$\,\boldsymbol{^\cdot}$}
\newcommand{\dottt}{$\,^\star$}
\newcommand{\dotttt}{$\,^{\star\star}$}
\newcommand{\dottttt}{$\,^{\star\star\star}$}
\newcommand{\Ab}{\boldsymbol{A}}
\newcommand{\Wb}{\boldsymbol{W}}
\newcommand{\sx}{\boldsymbol{\Sigma}_x}
\newcommand{\sz}{\boldsymbol{\Sigma}_z}
\newcommand{\sxy}{\boldsymbol{\Sigma}_{xy}}
\newcommand{\sy}{\boldsymbol{\Sigma}_{y}}
\newcommand{\Sx}{\boldsymbol{S}_x}
\newcommand{\Sxy}{\boldsymbol{S}_{xy}}
\newcommand{\Sy}{\boldsymbol{S}_{y}}
\newcommand{\oneb}{\boldsymbol{1}}
\newcommand{\Xb}{\boldsymbol{X}}
\newcommand{\Zb}{\boldsymbol{Z}}
\newcommand{\eb}{\boldsymbol{\varepsilon}}
\newcommand{\bb}{\boldsymbol{\beta}}
\newcommand{\cov}{\operatorname{Cov}}
\newcommand{\Ib}{\operatorname{I}}
\renewcommand{\vec}{\operatorname{Vec}}
\newcommand{\muxb}{\boldsymbol{\mu}_x}
\newcommand{\mub}{\boldsymbol{\mu}}
\newcommand{\sgb}{\boldsymbol{\Sigma}}
\newcommand{\Sb}{\boldsymbol{S}}
\newcommand{\Qb}{\boldsymbol{Q}}
\newcommand{\Phib}{\boldsymbol{\Phi}}
\newcommand{\Rb}{\boldsymbol{R}}
\newcommand{\rhob}{\boldsymbol{\rho}}
\newcommand{\alphab}{\boldsymbol{\alpha}}
\newcommand{\bOmega}{\boldsymbol{\Omega}}
\newcommand{\gammab}{\boldsymbol{\gamma}}
\def\gop{{\buildrel P \over \longrightarrow}}
\def\god{{\buildrel {d} \over \longrightarrow}}
\def\goas{{\buildrel {\rm a.s.} \over \longrightarrow}}
\newcommand{\Deltab}{\boldsymbol{\Delta}}
\renewcommand{\e}{\operatorname{E}}
\newcommand{\var}{\operatorname{Var}}
\def\bLambda{\boldsymbol{\Lambda}}
\def\bDelta{\boldsymbol{\Delta}}
\def\bTheta{\boldsymbol{\Theta}}
\def\btheta{\boldsymbol{\theta}}
\def\bSigma{\boldsymbol{\Sigma}}
\def\bPsi{\boldsymbol{\Psi}}
\def\bbeta{\boldsymbol{\beta}}
\def\blambda{\boldsymbol{\lambda}}
\def\bfeta{\boldsymbol{\eta}}
\def\balpha{\boldsymbol{\alpha}}
\def\bphi{\boldsymbol{\phi}}
\def\bpsi{\boldsymbol{\psi}}
\def\bxi{\boldsymbol{\xi}}
\def\bgamma{\boldsymbol{\gamma}}
\def\beps{\boldsymbol{\varepsilon}}
\def\bmu{\boldsymbol{\mu}}
\def\bx{\boldsymbol{x}}
\def\bfw{\boldsymbol{w}}
\def\by{\boldsymbol{y}}
\def\bz{\boldsymbol{z}}
\def\bv{\boldsymbol{v}}
\def\bb{\boldsymbol{b}}
\def\bd{\boldsymbol{d}}
\def\bD{\boldsymbol{D}}
\def\bC{\boldsymbol{C}}
\def\bg{\boldsymbol{g}}
\def\bJ{\boldsymbol{J}}
\def\bY{\boldsymbol{Y}}
\def\bS{\boldsymbol{S}}
\def\bs{\boldsymbol{s}}
\def\bU{\boldsymbol{U}}
\def\bH{\boldsymbol{H}}
\def\bF{\boldsymbol{F}}
\def\bI{\boldsymbol{I}}
\def\bG{\boldsymbol{G}}
\def\bX{\boldsymbol{X}}
\def\bW{\boldsymbol{W}}
\def\bZ{\boldsymbol{Z}}
\def\bN{\boldsymbol{N}}
\def\bP{\boldsymbol{P}}
\def\bQ{\boldsymbol{Q}}
\def\bV{\boldsymbol{V}}
\def\bL{\boldsymbol{L}}
\def\bA{\boldsymbol{A}}
\def\ba{\boldsymbol{a}}
\def\bB{\boldsymbol{B}}
\def\bfe{\boldsymbol{e}}
\def\bO{\boldsymbol{O}}
\def\bT{\boldsymbol{T}}
\def\bt{\boldsymbol{t}}
\def\b0{\boldsymbol{0}}
\def\bR{\boldsymbol{R}}
\def\br{\boldsymbol{r}}
\def \sgn{\mbox{sign}}
\def\bc{\boldsymbol{c}}
\def\blt{\begin{flushleft}}
\def\elt{\end{flushleft}}
\def\bse{\begin{eqnarray*}}
\def\ese{\end{eqnarray*}}
\def\be{\begin{eqnarray}}
\def\ee{\end{eqnarray}}
\def \sgn{\mbox{sign}}
\newcommand{\indep}{\perp \!\!\! \perp}

\newtheorem{theorem}{Theorem}
\newtheorem{corollary}{Corollary}

  \title{\bf Mixed-Response State-Space Model for Analyzing Multi-Dimensional Digital Phenotypes}
  \author{Tianchen Xu\\
  Department of Biostatistics\\
  Mailman School of Public Health, Columbia University, NY 10032, USA
  \and
  Yuan Chen\\
  Department of Epidemiology and Biostatistics\\
  Memorial Sloan Kettering Cancer Center, NY 10065, USA
  \and 
  Donglin Zeng\\
  Department of Biostatistics\\
  The University of North Carolina at Chapel Hill, NC 27599, USA
 \and 
  Yuanjia Wang\\
  Department of Biostatistics\\
  Mailman School of Public Health, Columbia University, NY 10032, USA
  }
\date{}
\maketitle

\thispagestyle{empty}

\enlargethispage*{20pt}
\begin{abstract}
Digital technologies (e.g., mobile phones) can be used to obtain objective, frequent, and real-world digital phenotypes from individuals. However, modeling these data poses substantial challenges since observational data are subject to confounding and various sources of variabilities. For example, signals on patients' underlying health status and treatment effects are mixed with variation due to the living environment and measurement noises. The digital phenotype data thus shows extensive variabilities between- and within-patient as well as across different health domains (e.g., motor, cognitive, and speaking). Motivated by a mobile health study of Parkinson's disease (PD), we develop a mixed-response state-space (MRSS) model to jointly capture multi-dimensional, multi-modal digital phenotypes and their measurement processes by a finite number of latent state time series. These latent states reflect the dynamic health status and personalized time-varying treatment effects and can be used to adjust for informative measurements. 
For computation, we use the Kalman filter for Gaussian phenotypes and importance sampling with Laplace approximation for non-Gaussian phenotypes. We conduct comprehensive simulation studies and demonstrate the advantage of MRSS in modeling a mobile health study that remotely collects real-time digital phenotypes  from PD patients.
\end{abstract}

\noindent%

\newpage
\spacingset{1.9}
 \setcounter{page}{1}
\section{Introduction}
Chronic neurological diseases such as Parkinson's disease (PD), Alzheimer's disease, and dystonia afflict millions worldwide and account for tremendous morbidity and mortality. Unfortunately, the current state of neurological care is inadequate. Reports from the World Health Organization (WHO) show that large swaths of the world lack access to proper neurological care \citep{world2004atlas}, and  ``most patients admitted with a neurological condition rarely see a neurologist''  \citep{dorsey2018teleneurology}. In the United States, more than 40\% of individuals with PD over 65 did not see a neurologist over four years \citep{willis2011neurologist}. The main barriers restricting access include distance, patient disability, neurologists' scarcity, and uneven distribution. 

With the global proliferation of smartphones and wearable sensors, a potential approach to narrow this gap is through modern teleneurology, which aims to provide real-time neurological monitoring and care remotely \citep{dorsey2018teleneurology}. Objective, frequent (or even continuous), high-granularity, and real-world assessments of patients can be collected from mobile devices, which are measurements of a patient's digital phenotype, i.e., ``the expression of an individual's disease state through the lens of interactions with the digital environment'' \citep{jain2015digital}. In our motivating study, the Mobile Parkinson's Observatory for Worldwide Evidence-based Research (mPower),  researchers remotely collected frequent digital information about the daily changes in symptom severity (including movement, cognition, and voice features) from PD patients and healthy controls through smartphones. During the study, a proportion of PD patients repeatedly took long-term dopaminomimetic therapy (i.e., Levodopa) to manage their motor symptoms. These digital phenotypes can potentially be used to monitor patients' health in their daily life to facilitate telehealth care, estimate personalized time-varying  treatment effects, and serve as a basis to inform clinical trial designs to create robust and generalizable digital endpoints \citep{dorsey2018teleneurology}. 

Despite their promises, digital phenotypes are subject to various sources of variability  and are affected by  a patient's living environment (e.g., passive versus active collection,  schedule of assessments). Thus, the ability of sensor-derived mobile measures to monitor disease progression is unknown \citep{baumeister2019digital}. Recently, efforts have been devoted to standardizing study protocols to control for sources of variation, improve signal processing methods for feature extraction, and validate digital assessments enabled by the sensors within smartphones  against traditional rating scales or diagnostic assessments of PD patients. Several studies found that digital markers can distinguish PD patients from controls at the population level, and the digital assessments are  correlated with clinical assessments \citep{neto2017analysis, sieberts2021crowdsourcing}.

Although preliminary results from studies of remote sensors are promising, statistical modeling of digital phenotype data and its applications are still in their infancy \citep{dorsey2018teleneurology}. Several significant challenges must be addressed before digital markers can provide helpful information for patient care and clinical trials. The first consideration is that  appropriate signal processing methods should be used for feature extraction to remove systematic biases and errors \citep{wroge2018parkinson}. The second consideration is that mobile devices do not directly measure a patient's true underlying health, and substantial variabilities between and within patients and across different health domains (e.g., motor, cognitive, voice) are present. Even with the best preprocessing techniques applied, the extracted information (e.g., walking and voice features) is subject to intrinsic unobserved variabilities best captured through a latent model \citep{ghosh2016deep, alaa2019attentive}. Third, the processed data are heterogeneous and from multiple modalities (e.g., movement, cognition, and voice domain), so their distributions do not conform to a single type (e.g., some distributions are  non-Gaussian). Lastly, these digital markers are often collected from observational studies and hence subject to selection bias  due to patients' self-choice to interact with digital devices (e.g.,  informative measurement process). In our motivating study mPower, patients were instructed to take measurements at specific times, i.e., right before and after taking Levodopa medication.  Such a self-initiated measurement process is often informative of a patient's underlying health status \citep{marsden1976off, bhidayasiri2008motor}. For example, patients may be more likely to take medication and record  measurements when  their symptoms worsen. Thus, selection bias may be induced if the measurement process is not accounted for in the modeling process. 


Methods based on functional data analysis and latent growth mixture models have been proposed to model digital phenotype data. \citet{lee2021phenotypes} considered a functional principal component analysis  to identify common structures of  behavioral measures from wearable devices in a natural environment. 
Machine learning methods such as recurrent neural network \citep{liang2019survey}, reinforcement learning \citep{wang2021optimizing},  and hidden Markov models \citep{hulme2020adaptive} have also been used to analyze intensive longitudinal digital phenotype data. {In the latent state-space model framework, Mixed-Effects State-Space Models (MESSM) were considered to model longitudinal data  \citep{jones1993longitudinal, icaza1999state,liu2011mixed}. 
The MESSM allowed both fixed and  random effects to model system matrices  to account for  between-subject variations but parameter estimation could be challenging.  Different distributions for continuous responses have been considered by \cite{zhou2014estimating}, \citet{liu2011mixed},  
and \citet{velascomixed}.}
{Exponential family and non-Gaussian state space models have also been studied   \citep{grunwald1993prediction, vidoni1999exponential, durbin1997monte,shephard1997likelihood,durbin2000time,klein2003state}.} 
However, there is no existing work on jointly modeling mixed type digital phenotypes (both continuous and discrete measures) in the latent state-space model framework. In addition, current methods do not handle  the practical challenges of digital health studies. 

{We make several novel contributions in this work. First, we jointly model heterogeneous types of digital phenotypes in a general state-space model framework. Second, we allow the measurement times of digital phenotypes to be informative of patients' health status, which is pervasive in mobile health studies. Third, we incorporate the medication process, which also relates to the patient's  underlying health,   into modeling. Lastly,  methods proposed for learning treatment policies from mobile health data usually 
assume Markov decision processes (MDP), which may be unrealistic  due to long-term dependencies and time-varying trends of health status. We propose a method that incorporates  latent processes with long-term effects. We present further details of the modeling framework in the next section.}

\begin{figure}[!b]
    \centering
        \includegraphics[width=0.7\textwidth]{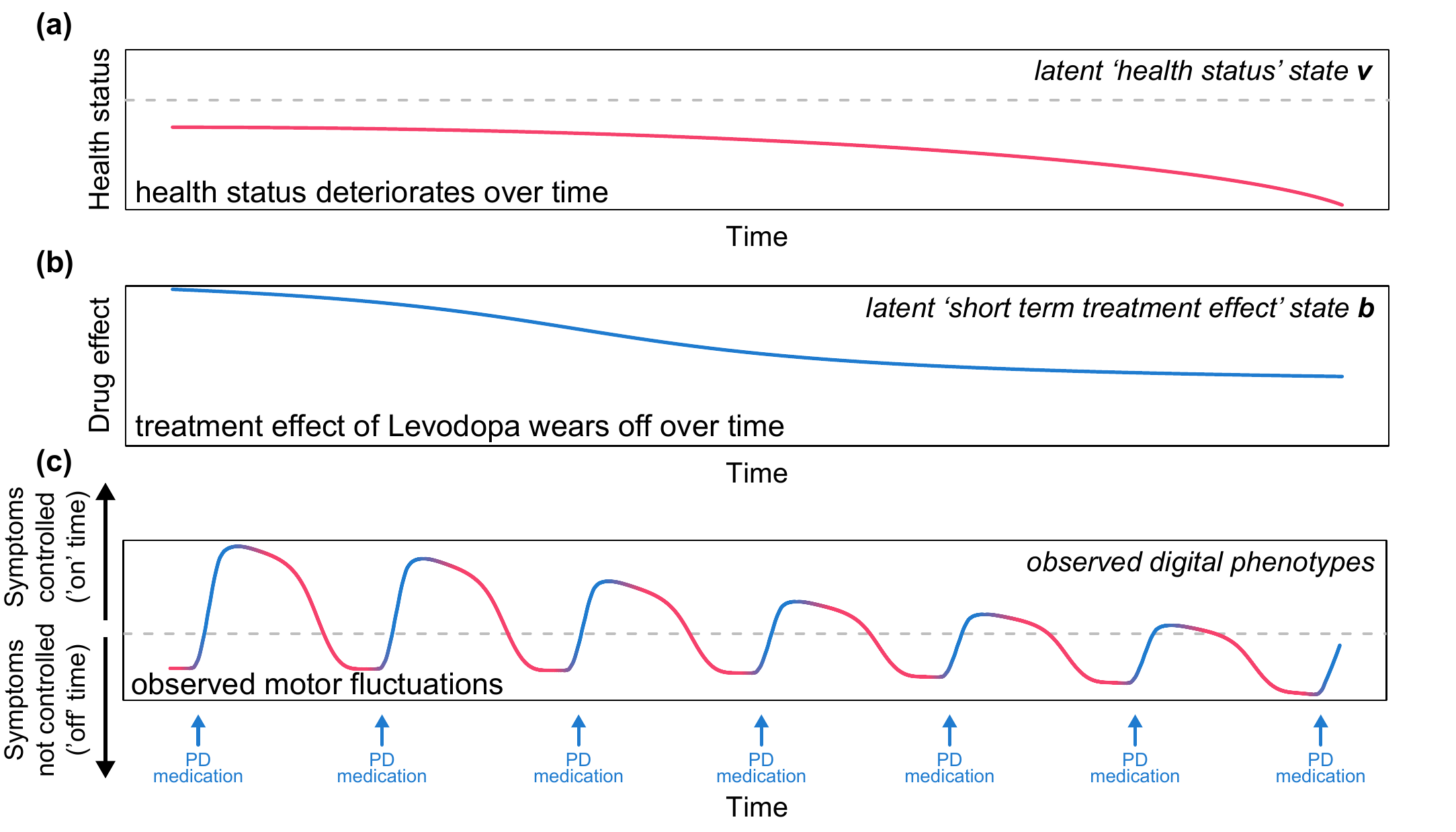}
    \caption{Demonstration of unobserved and observed states:  (a) Patient's health status (state variables $\bv$) deteriorates over time. (b) The treatment effect of Levodopa (state variables $\bb$) wears off over time. (c) Observed motor symptom fluctuations due to the short-term treatment effect.}\label{fig:state}
\end{figure}

\subsection{Mixed-Response State-Space Model (MRSS) for mPower}\label{sec:model}
To motivate our proposed modeling strategy, we visualize the latent health status of a PD patient  over time in Figure \ref{fig:state}(a). {The patient's health gradually deteriorates with time as represented by the health state vector $\bv$. It is well-known that Levodopa is a symptomatic treatment and its short-term effect wears off quickly ($\sim 50$ minutes half time) as shown by the state vector $\bb$ in Figure~\ref{fig:state}(b) \citep{bhidayasiri2008motor}. 
Both the health state and the treatment effect are not directly observed from digital phenotype data.
Figure~\ref{fig:state}(c) shows the observed health status of a PD patient under repeated treatments of Levodopa (overlapping state vectors $\bv$ and $\bb$). The symptoms are quickly controlled after each dose of Levodopa (blue segments of the curve) but the treatment effect wears off in the short-term (back to red segments).} The  magnitude of benefit following a dose of treatment becomes progressively weaker and the duration becomes shorter. These motor fluctuations are coined as the `on-off phenomenon', a signature of Levodopa treatment observed decades ago in PD patients \citep{marsden1976off}. This phenomenon suggests that multiple latent states are required to model PD patients' phenotypes under Levodopa treatment.

In this work, we assume a patient's underlying health status is governed by multiple unobserved latent processes, and digital phenotypes are reflective but noisy measurements of the underlying health. 
We propose a mixed-response state-space (MRSS) model that unites heterogeneous phenotypes with various data distributions  (a binary process) through shared latent processes. In addition, this joint modeling approach accounts for the informative measurements to mitigate selection bias. A patient's observed digital phenotypes  and their corresponding measurement process depend on the patient's underlying health state that is  modeled by a series of unobserved state vectors.
For Gaussian outcomes, the Kalman filter and Kalman smoother can be used to fit the parameters efficiently. Non-Gaussian outcomes pose substantial computational challenges. We use importance sampling with an informative proposal distribution to improve computational efficiency. We examine the theoretical properties of the proposed method and establish asymptotic distribution to facilitate inference. We conduct extensive simulations to compare MRSS with alternative time series methods and demonstrate its advantage in modeling mPower study data.

\section{Method}
   
Suppose we measure $m_Y$ phenotypes from $m_A$ modalities in $n$ subjects. Let $Y_{itj}$ denote the  $j$th  digital phenotype observed from the $i$th patient  at time $t$ and let $\bY_{it} = (Y_{it1},\cdots,Y_{itm_Y})^T$ be the vector including all $m_Y$ phenotypes, where $t=1,...,m_i$ so the number of the measurement times for each subject may be different. Let $\bA_{it}=(A_{it1},\cdots,A_{itm_A})^T$ be the binary  measurement time indicators for each modality, with $A_{itk}=1$ indicating  measuring phenotypes immediately after treatment and $A_{itk}=0$ indicating  measuring phenotypes without treatment. The usual statistical models for joint modeling use latent models to connect outcomes and measurement process \citep{van2004fitting}. With intensively measured longitudinal digital phenotype data and heterogeneous types of outcomes, however, it is difficult to simultaneously incorporate the time-dependence structure and correlation among heterogeneous outcomes, and the computation in nonlinear models poses a substantial challenge. Time series models such as Granger causality analysis or autoregressive (AR) models are often used to directly model observed processes for a single subject, which are not appropriate for our applications \citep{seth2015granger}. 

\begin{figure}[!t]
    \centering
        \includegraphics[width=0.8\textwidth]{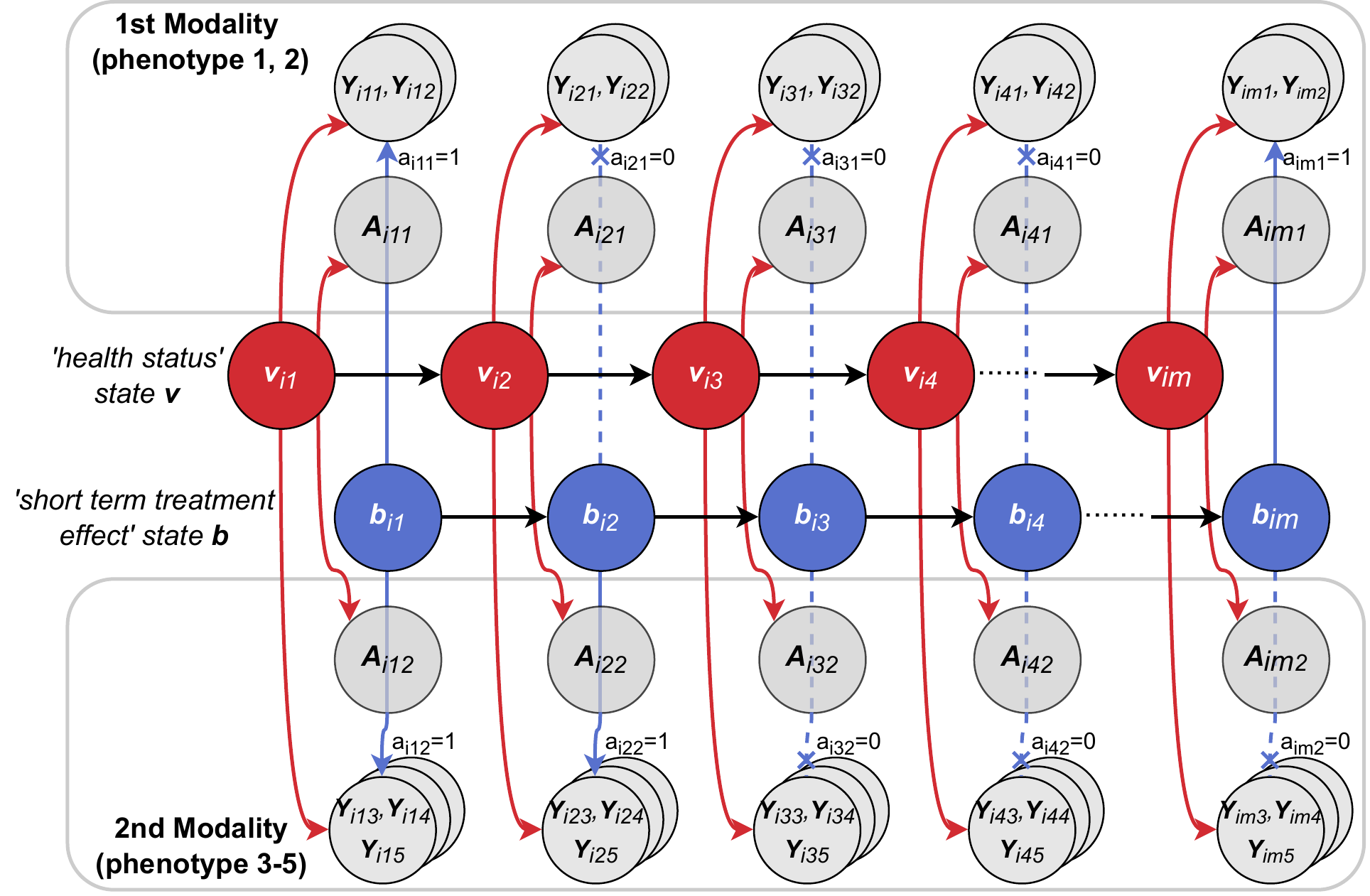}
    \caption{Illustration diagram of the proposed MRSS  that jointly models digital phenotypes $\bY$ and their  measurement processes $\bA$ (indicating measurement before treatment or immediately after treatment), latent underlying health states, $\bv$ and treatment effects $\bb$. }\label{fig:model}
\end{figure}

On the other hand, state-space models can incorporate  multiple observed digital phenotypes by connecting the observed and latent processes. Linear state-space models have been developed under Gaussian distribution. Here, we generalize them to accommodate different types of  phenotypes and use an exponential family to model their distributions \citep{durbin2012time}. We let $\bY_{it}^{(\ba_{it})}$ denote the potential outcome of the digital phenotype under treatment $\ba_{it}$. Motivated by our application of the short half-time of Levodopa medication, we assume no delayed treatment effect.  We further assume the potential outcomes  $ \bY_{it}^{(\ba_{it})}\in\mathbb{R}^{m_A+m_Y}$ are conditionally independent of the measurement process $\bA_{it},$ given lower-dimensional latent processes $(\bb_{it}, \bv_{it})^T \in \mathbb{R}^{m_b+m_v}$ with $m_b+m_v<m_A+m_Y$, and  
    \begin{equation}\label{eqn:obs_org}
    \binom{\bA_{it}}{\bY_{it}^{(\ba_{it})}}\Big|\bb_{it},\bv_{it}\sim p
    \left( \binom{\bA_{it}}{\bY_{it}^{(\ba_{it})}}\middle|\blambda_{it} \binom{\bb_{it}}{\bv_{it}}+\bbeta\bX_{it}  \right),
    \end{equation}
     \begin{gather*}
        \blambda_{it} = \begin{pmatrix}
            \textbf{0} & \blambda_{Av}\\
            \blambda_{Yb}\ba_{it} & \blambda_{Yv}
        \end{pmatrix}, {\quad i=1,\cdots,n;\; t=1,\cdots,m_i,}
    \end{gather*}
where $\bb_{it}$ is the latent treatment effect vector; $\bv_{it}$ is the latent health status vector;  $\bX_{it}\in\mathbb{R}^{q}$ are the potentially time-varying covariates (e.g., comorbidities); 
$\ba_{it}=\operatorname{diag}(\underbrace{a_{it1}, \cdots, a_{it1}}_{n_{(1)}}, \cdots,$ $\underbrace{a_{itm_A},\cdots, a_{itm_A}}_{n_{(m_A)}})$ is a diagonal matrix ($n_{(j)}$ is the number of phenotypes in $\bY_{it}^{(\ba_{it})}$ that belong to the $j$th modality) and $a_{itj}$ indicates treatment  of $j$th modality of subject $i$ at time $t$. Loading matrices $\blambda_{Av}\in \mathbb{R}^{m_A\times m_v}$, $\blambda_{Yb}\in\mathbb{R}^{m_Y\times m_b}$, and $\blambda_{Yv}\in\mathbb{R}^{m_Y\times m_v}$  map a higher dimensional phenotype space to a lower dimensional latent space, which effectively performs dimension reduction and facilitates computation. {The conditional distribution $p(\cdot|\bv,\bb, \bX)$ belongs to the exponential family with an appropriate link function chosen depending on each phenotype type. For each subject, time indexes $t=1,2,\cdots, m_i$ indicate $m_i$ evenly distributed measurements such as daily measurements in a year.  If any measurements are missing (e.g., no records on weekends), the corresponding response $\bY_{it}$ will be missing.  The approach to handle missing values will be introduced at the end of Section~\ref{sec:likelihood}.} Note that our approach can handle different numbers of measurements for each phenotype. However, we discuss the same measurement schedule across phenotypes hereafter for ease of notation. 

The schematic of MRSS modeling framework is shown in Figure~\ref{fig:model}. In the proposed model, the measurement process $\bA_{it}$ is associated with the patient's health status vector $\bv_{it}$ and covariates $\bX_{it}$, which accounts for patients' differential  behaviors when recording their phenotypes. This component is essential in modeling the informative measurement process to ensure valid estimation of other model parameters from observational data. Meanwhile, in the loading matrix $\blambda_{it}$, when $\ba_{it}=\boldsymbol{0}$ (measurement taken before a treatment), digital phenotype measure $\bY_{it}^{(\ba_{it})}$ only depends on health state $\bv_{it}$ and $\bX_{it}$; when  $\ba_{it}= \boldsymbol{1}$ (measurement taken right after treatment),  $\bY_{it}^{(\ba_{it})}$ takes into account the additional short-term treatment effect~$\bb_{it}$. 

The dependence over time for the latent processes  is modeled by the transition equation  in the latent space as
    \begin{align}\label{eqn:tran_org}
        \binom{\bb_{i,t+1}}{\bv_{i,t+1}} = \bT \binom{\bb_{it}}{\bv_{it}} + \binom{\beps_i^b}{\beps_i^v},\qquad i=1,\cdots,n;\; t=1,\cdots,m_i
    \end{align}
where $\bT\in\mathbb{R}^{(m_b+m_v)\times (m_b+m_v)}$ is a diagonal transition matrix, and $(\beps_i^b, \beps_i^v)^T
\sim MVN_{m_b+m_v}(\bc, \bQ)$ capture the random errors. Note that due to the autoregressive structure, the first $m_b$ diagonal components of $\bT$ reflect the attenuation of treatment effects over time often observed in practice. We mainly present the AR(1) model in \eqref{eqn:tran_org} for ease of presentation. The method can be extended to higher-order models  without technical difficulties when necessary.

\subsection{Likelihood Computation}\label{sec:likelihood}
Let ${\bY_{it}}$ denote the observed digital phenotypes. We assume that the measurement process at time $t$ is conditionally independent of the potential outcomes at $t$ given the lower-dimensional latent processes and the observed covariates up to  $t$. We also assume the standard stable unit treatment value assumption (SUTVA), i.e., $\bY^{(\ba_{it})}_{it}=\bY_{it}$ if $\bA_{it}=\ba_{it}$. For the $i$th subject, we re-express the model  \eqref{eqn:obs_org} and \eqref{eqn:tran_org} in Section~\ref{sec:model} in a more general form:
\begin{equation}\label{eqn:nmodel}
  \begin{alignedat}{2}
      \bZ_{it}|\balpha_{it} &= p(\bZ_{it}|\blambda_{it}\balpha_{it}+\bd_{it}),  \qquad & & i=1,\cdots,n;\; t=1, \ldots, m_i\\
      \balpha_{i, t+1} &=\bT \balpha_{it}+\bc+ \bfeta_{it}, \qquad && \bfeta_{it}  \sim \mathrm{N}_{m_b+m_v}\left(\boldsymbol{0}, \bQ\right),  \\
      &&& \balpha_{i1} \sim \operatorname{N}_{m_b+m_v}(\ba_{i1}, \bP_{i1})
  \end{alignedat}  
\end{equation}
where $\bZ_{it}=\displaystyle\binom{\bA_{it}}{\bY_{it}}$ is a $(m_A+m_Y) \times 1$ vector representing the observed processes and $\displaystyle\balpha_{it}=\binom{\bb_{it}}{\bv_{it}}$ is an unobserved $(m_b+m_v) \times 1$ vector denoting the latent processes. The distribution of initial state $\balpha_{i1}$ is assumed to be known in this section (see Supplementary A.1 for more discussion on diffuse initialization when $\balpha_{i1}$ is unknown). We refer to the first equation in \eqref{eqn:nmodel} as the observation equation and the second as the state equation, 
$\blambda_{it}$ is a $(m_A+m_Y)$ by $(m_b+m_v)$ loading matrix that maps lower dimensional latent state $\balpha_{it}$ to high dimensional phenotypes $\bZ_{it}$, and
vector $\bd_{it}=\bbeta \bX_{it}$. 

Let $\bpsi$ denote the vector of all unknown parameters, which includes $\blambda$, $\bbeta$, $\bT$, $\bc$, $\bQ$. We  estimate $\bpsi$ by maximum likelihood. The likelihood of the $i$th subject, $L_i(\bpsi)$, is
\begin{align*}
  L_i(\bpsi)=\int p(\balpha_i^*, \bZ_{i(m)})\ d\balpha_i^*.
\end{align*}
where $\bZ_{i(m)}=\left(\begin{matrix}\bZ_{i1}^T &\cdots &\bZ^T_{im} \end{matrix}\right)^T$, $\balpha_i^*=\left(\begin{matrix}\balpha_{i1}^T &\cdots &\balpha^T_{i,m+1}\end{matrix}\right)^T$ and $p(\balpha_i^*, \bZ_{i(m)})$ is the joint density. 
Since the distribution $p(\bZ_{it}|\balpha_{it})$ can be non-Gaussian in MRSS, the likelihood $L_i(\bpsi)$ cannot be easily obtained by existing methods such as the Kalman filter. 
{To tackle this challenge, we will use importance sampling for estimation \citep{durbin1997monte, shephard1997likelihood}.}  Generic importance sampling is extremely slow, 
but a good choice of importance proposal distribution will substantially speed up computation. An approximate Gaussian state-space model with the same conditional posterior mode is an example of good proposal distributions. Thus, we  adopt the Laplace approximation of the posterior density $p(\balpha_i^*|\bZ_{i(m)})$, and  the updates are estimated via Kalman filtering and smoothing from the approximating Gaussian model.  Specifically, we divide and multiply a Gaussian density $g(\balpha_i^*|\bZ_{i(m)})$ and rewrite the likelihood as the following (details are provided in Supplementary \ref{wbsec:approx}):
\begin{align*}
  L_i(\bpsi)&=\int \frac{p(\balpha_i^*, \bZ_{i(m)})g(\balpha_i^* |\bZ_{i(m)})}{g(\balpha _i^*|\bZ_{i(m)})}\ d\balpha_i^*\\
 & =g(\bZ_{i(m)})\int \frac{p(\balpha_i^*, \bZ_{i(m)})g(\balpha_i^* |\bZ_{i(m)})}{g(\balpha_i^*,\bZ_{i(m)})}\ d\balpha_i^*
  =g(\bZ_{i(m)})\e_{\balpha_i^* |\bZ_{i(m)}}\left\{\frac{p(\balpha_i^*, \bZ_{i(m)})}{g(\balpha_i^*,\bZ_{i(m)})}\right\},
\end{align*}
where $\e_{\balpha_i^* |\bZ_{i(m)}}( \cdot )$ is the expectation with respect to $g(\balpha_i^*|\bZ_{i(m)})$; $g(\bZ_{i(m)})$ is the likelihood of the approximating Gaussian model which can be easily obtained by the Kalman filter (see Supplementary A.1); and $g(\balpha_i^*,\bZ_{i(m)})$ is the corresponding joint density. 
Therefore, the log-likelihood of the $i$th subject can be decomposed into two parts:
\begin{align*}
  \log L_i(\bpsi)&=\log g(\bZ_{i(m)})+\log \e_{\balpha_i^*|\bZ_{i(m)})}\left\{\omega(\balpha_i^*, \bZ_{i(m)})\right\},
\end{align*} 
where  $\displaystyle\omega(\balpha_i^*, \bZ_{i(m)})=\frac{p(\balpha_i^*, \bZ_{i(m)})}{g(\balpha_i^*,\bZ_{i(m)})}$ is an adjustment factor (importance sampling weights).
Using importance sampling, we obtain the estimated log-likelihood $\hat L_i(\bpsi)$ as
\begin{align}\label{eqn:ind}
  \log \hat L_i(\bpsi) = \log g(\bZ_{i(m)})+\log \left\{\bar \omega(\balpha_i^*, \bZ_{i(m)})\right\},
\end{align}
where $\bar \omega(\balpha_i^*, \bZ_{i(m)}) = 1/N\sum_{j=1}^N \omega(\balpha_i^{*(j)}, \bZ_{i(m)})$ with $\balpha_i^{*(1)},\cdots,\balpha_i^{*(N)}$ being $N$ samples generated from $g(\balpha_i^*|\bZ_{i(m)})$. The log-likelihood of all $n$ subjects is the summation of each individual:
\begin{align}\label{eqn:all}
    \log \hat L(\bpsi)=\sum_{i=1}^n \log \hat L_i(\bpsi).
\end{align}

When a subject misses measurements during time points $t+1, \cdots, t+\tau$, the transition equation between $\balpha_{it}$ to $\balpha_{i,t+\tau+1}$ becomes: $\balpha_{i,t+\tau+1}=\bT^{\tau+1}\balpha_{it}+\bc^\star+\bfeta^\star_{t,t+\tau}$, where $\bc^\star=\sum_{k=0}^\tau \bT^k \bc$, $\bfeta^*_{t,t+\tau}\sim N_{m_b+m_v}\left(\boldsymbol{0},\sum_{k=0}^{\tau} \bT^{k}\bQ {\bT^{k}}^T \right)$ and the rest of the  procedure remains the same \citep[see][Section 4.10 for more details]{durbin2012time}. 

\subsection{Parameter Estimation and Computational Algorithm}\label{sec:estimation}
After obtaining the likelihood under MRSS, we adopt a cyclic block coordinate descent-type (CBCD-type) method, which performs iterative updates for a few coordinates (a block) simultaneously throughout the procedure to maximize the likelihood function \citep{beck2013convergence}. 
Specifically, regardless of the true distribution of the response,  we first pretend that $p(\bZ_{it}|\balpha_{it})$ follows a multivariate normal distribution with mean $\blambda_{it}\balpha_{it}+\bd_{it}$ and variance $\bH$ (unknown) to obtain reasonable initial parameter estimates.
Reliable initial values can accelerate the iterative process and help with convergence. This is particularly important in optimization problems with a dimension of parameters up to hundreds.
Section 7.3 in \citet{durbin2012time} elaborated on the optimization of a Gaussian state space model. 
After obtaining initial parameter values, we  update $\blambda$, $\bbeta$, $\bT$, $\bc$, and $\bQ$ in turn in a predefined sequence by maximizing the log-likelihood function until convergence. In practice, the following procedure can boost the convergence speed and provide a more stable result: 1) First update $\blambda$, $\bbeta$ simultaneously; 2) The estimation of $\bT$ is sensitive to other parameters and thus we re-estimate $\bT$ whenever another block of parameters are updated; 3) Variance component $\bQ$ has little influence on estimating other parameters, and thus we update $\bQ$ after $c$ rounds of estimating other parameters  (we use $c=3$). We summarize the optimization procedure in Algorithm \ref{alg:cap}.

\begin{algorithm}[!t]
    \linespread{1.3}\selectfont
    \caption{Optimization of MRSS}\label{alg:cap}
    \begin{algorithmic}
    \Require $\bZ_{it}$, $\bX_{it}$, $\ba_{it}$
    \State Obtaining initial values $\bLambda^0$, $\bbeta^0$, $\bT^0$, $\bc^0$, $\bQ^0$ by assuming $p(\bZ_{it}|\balpha_{it})$ follows a multivariate normal distribution.
    \State Calculate $\log\hat L^{(0)}(\bpsi)$ via equation \eqref{eqn:ind}, \eqref{eqn:all}.
    \While{$\log \hat L^{(l+1)}(\bpsi)-\log\hat L^{(l)}(\bpsi)>tol$}
    \For{$i=1:c$}
        \State Estimating $\bLambda$, $\bbeta$ by maximizing $\log\hat L^{(l)}$.
        \State Estimating $\bT$ by maximizing $\log\hat L^{(l)}$.
        \State Estimating $\bc$ by maximizing $\log\hat L^{(l)}$.
        \State Estimating $\bT$ by maximizing $\log\hat L^{(l)}$.
    \EndFor
        \State Estimating $\bH$ (if applicable), $\bQ$ by maximizing $\log\hat L^{(l)}$.
        \State Estimating $\bT$ by maximizing $\log\hat L^{(l)}$.
        \State  Calculate $\log\hat L^{(l+1)}(\bpsi)$ via equation \eqref{eqn:ind}, \eqref{eqn:all}.
    \EndWhile
    \end{algorithmic}
\end{algorithm}

The fitted MRSS model can provide forecasts separately for each subject and these forecasts can be used to predict personalized treatment effects for a patient at a future time point. For inference on the estimated parameters, we adopt the likelihood ratio test (LRT) to obtain $p$-values \citep{gamerman2013non}. Lastly, we use AIC to select the dimension of the latent states \citep{kitagawa1987non, bengtsson2006improved}.

\section{Simulation}
In the first part of the simulation study, we compare our proposed method (MRSS) to a generalized linear mixed effect model (GLMM) to assess the estimation for fixed effect parameters ($\bbeta$). In the second part, we compare MRSS to a vector autoregressive model (VAR) to assess the prediction performance. 

\subsection{Simulation Model and Comparison Methods}
We simulate a total of $N$ ($N=20, 40$ or $60$) independent subjects. For each subject, we generate a three-dimensional synthetic time series data $(\bY^{(1)}, \bY^{(2)}, \bY^{(3)})$ with $T$ ($T=15, 30$ or $60$) time points. Let $(Y^{(1)}_{t}, Y^{(2)}_{t}, Y^{(3)}_{t})$ be the $t$th observation ($t=1,\cdots, T$) and the three components follow a binomial distribution $Bin(1, \operatorname{expit}(\mu^{(1)}_{t}))$, a Poisson distribution $Pois(\operatorname{exp}(\mu^{(2)}_{t}))$, and a normal distribution $N(\mu^{(3)}_{t},1)$, respectively.

We assume that the parameters $(\mu^{(1)}_{t},\mu^{(2)}_{t},\mu^{(3)}_{t})^T$ in the observation equation have the following structure:
$    (\mu^{(1)}_{t},\mu^{(2)}_{t},\mu^{(3)}_{t})^T=\blambda_t \balpha_t+X_1+2X_2+0.03 t,$
where $X_1$, $X_2$ are time-invariant fixed effects for each subject and $X_1\sim N(-5,2)$, $X_2\sim Bin(4,0.5)$; $\blambda_t$ is a $3$ by $2$ time-varying loading matrix; $\balpha_t$ is a two-dimensional latent state vector. Specifically, the latent state vector $\balpha_t=(\alpha^{(1)}_t, \alpha^{(2)}_t)^T \in\mathbb{R}^2$ has an AR(1) structure
$\balpha_t = \left({0.6 \atop 0} {0 \atop 0.8}\right)\balpha_{t-1} + \left({1.2 \atop 2}\right)+\beps_t$ with
initial value $\balpha_1 = (10,10)^T$ and white noise $\beps_t\sim N\left(\boldsymbol{0},  \left({1 \atop 0} {0 \atop 1.5}\right)\right)$. The loading matrix $
\blambda_t=\left({-0.5a_t \atop 0.1} {0.2a_t \atop 0.2} {-a_t \atop 1}\right)^T$
depends on a time-varying treatment indicator $a_t$, 
where $a_1, a_2,\cdots,a_{T}$ independently follow $Bin(1, p)$ ($p=0, 0.15, 0.3, 0.45, 0.6, 0.75$ or $0.9$) indicating the time points that $\alpha^{(1)}_t$ has an effect on the observation. 

In the first set of experiments, we compare with GLMM to examine the estimation of the fixed effect of $X_1$ and $X_2$. Because the usual GLMM does not analyze different types of outcomes jointly, we 
fit three separate GLMMs to $\bY^{(1)}, \bY^{(2)}$ and $\bY^{(3)}$ and set the response type as `binomial', `Poisson' and `Gaussian'. In each model, we include $X_1$ and $X_2$ as fixed effect terms, a random slope of $a_t$, and a random intercept for each subject as random effects. {For the proposed method, we include two latent states as the true model and use all $T$ time points to estimate parameters. In addition, we explore the cases when a wrong number of states is used in MRSS.}

In the second set of experiments, we focus on the time series model to examine prediction performance. The first $5/6$th time points are used for estimating models, and the rest of the $1/6$th time points are used to quantify the out-of-sample prediction performance. 
We compare to a vector autoregressive model (VAR) in fitting the time series data. For each subject, the VAR model is specified as 
$    \tilde{\boldsymbol{Y}}_t=\boldsymbol{A} \tilde{\boldsymbol{Y}}_{t-1} + \boldsymbol{C}a_t + \boldsymbol{u}_t,$
where response $\tilde{\boldsymbol{Y}}_t=(\tilde Y^{(1)}_{t}, \tilde Y^{(2)}_{t}, \tilde Y^{(3)}_{t})^T=(Y^{(1)}_{t}, \log(Y^{(2)}_{t}), Y^{(3)}_{t})^T$; 
and $\boldsymbol{u_t}$ is a 3-dimensional process with $\operatorname{E}(\boldsymbol{u_t}) = 0$ and time-invariant positive definite covariance matrix $\operatorname{E}(\boldsymbol{u}_t \boldsymbol{u}_t^T) = \Sigma_{\boldsymbol{u}}$ (white noise). Since the VAR model can only be applied to a single subject's time series but there are $N$ independent subjects in the simulated data, we compare two variants of the VAR model: 1) individual-VAR: we fit $N$ separate VAR models in total, one for each individual; 2) pooled-VAR: we average  the estimated $\hat{\boldsymbol{A}}$ and $\hat{\boldsymbol{C}}$ from the individual-VAR models and apply that set of parameters to all subjects for prediction. 
 
\subsection{Results}
All results are based on $200$ repetitions.
In the first set of simulations, we compare the performance in estimating fixed effects ($\bbeta$).
In \textbf{Setting 1}, we vary the sample size ($N$) from $20$ to $60$ and fix the time series length ($T$) at $30$ and the expectation of $a_t$ ($p$) at $0.3$. In \textbf{Setting 2}, we vary time series length ($T$) from $15$ to $60$ and fix the sample size ($N$) at $40$ and the expectation of $a_t$ ($p$) at $0.3$. In \textbf{Setting 3}, we vary the expectation of $a_t$ ($p$) from $0$ to $0.9$, and fix the sample size ($N$) at $40$ and the time series length ($T$) at $30$. {In all settings, $99\%$ of repetitions achieves convergence. Supplementary Appendix Figure~\ref{wbfig:conv} shows an example of the likelihood change in $200$ repetitions during the optimization process ($N=40$, $T=30$, $p=0.45$). Most models converge in around $100$ iterations.}

\begin{figure}[!t]
    \parbox{0.55\textwidth}{
        \textbf{(a.1)} Response $\boldsymbol{Y}^{(1)}$; Varying sample size ($N$)\\
        \includegraphics[clip, trim=16pt 15pt 0 50pt, width=0.5\textwidth]{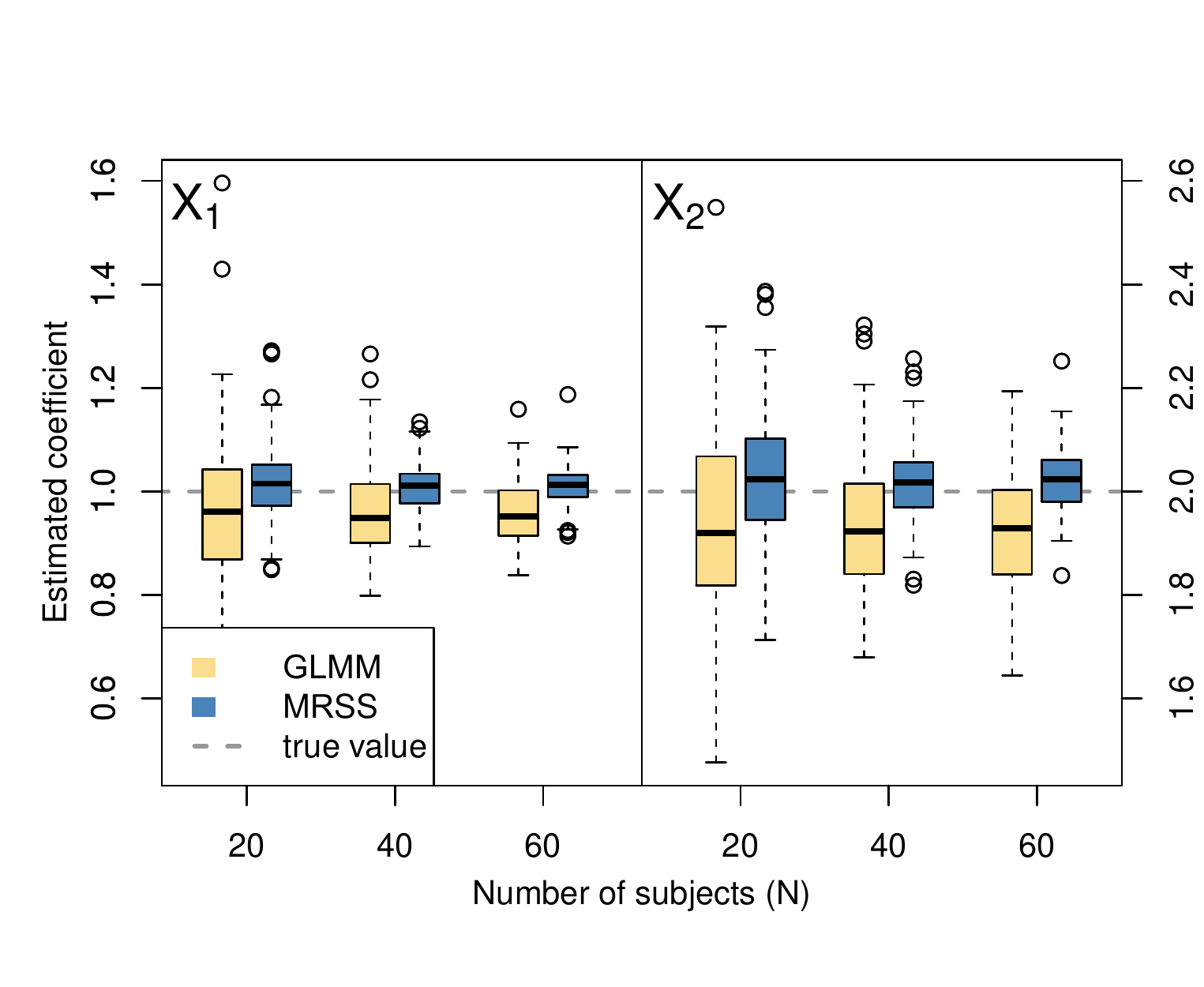}
    }
    \parbox{0.5\textwidth}{
        \textbf{(a.2)} Response $\boldsymbol{Y}^{(3)}$; Varying sample size ($N$)\\
        \includegraphics[clip, trim=16pt 15pt 0 50pt, width=0.5\textwidth]{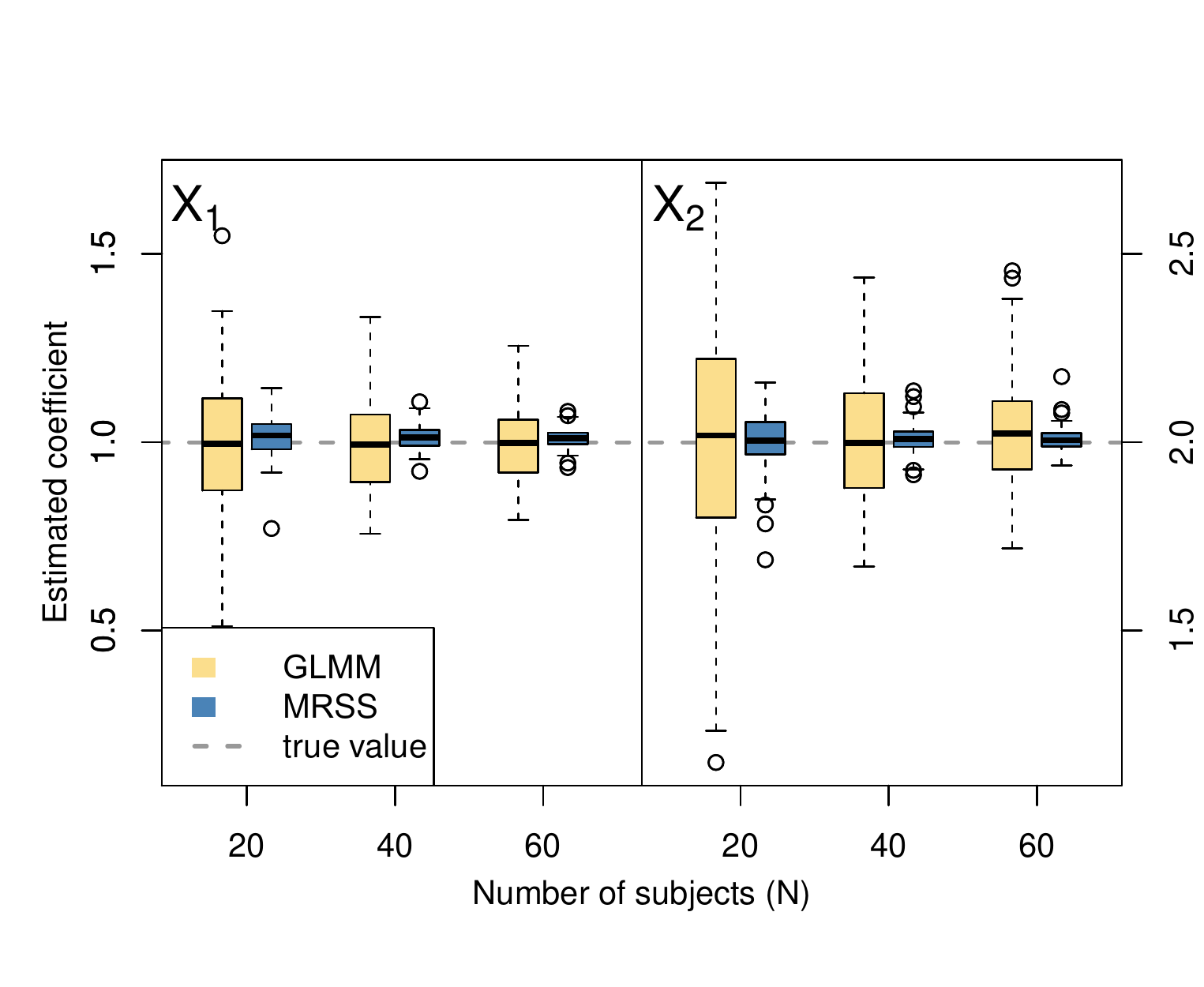}
    }
    \parbox{0.55\textwidth}{
        \textbf{(b.1)} Response $\boldsymbol{Y}^{(1)}$; Varying time length ($T$)\\
        \includegraphics[clip, trim=16pt 15pt 0 50pt, width=0.5\textwidth]{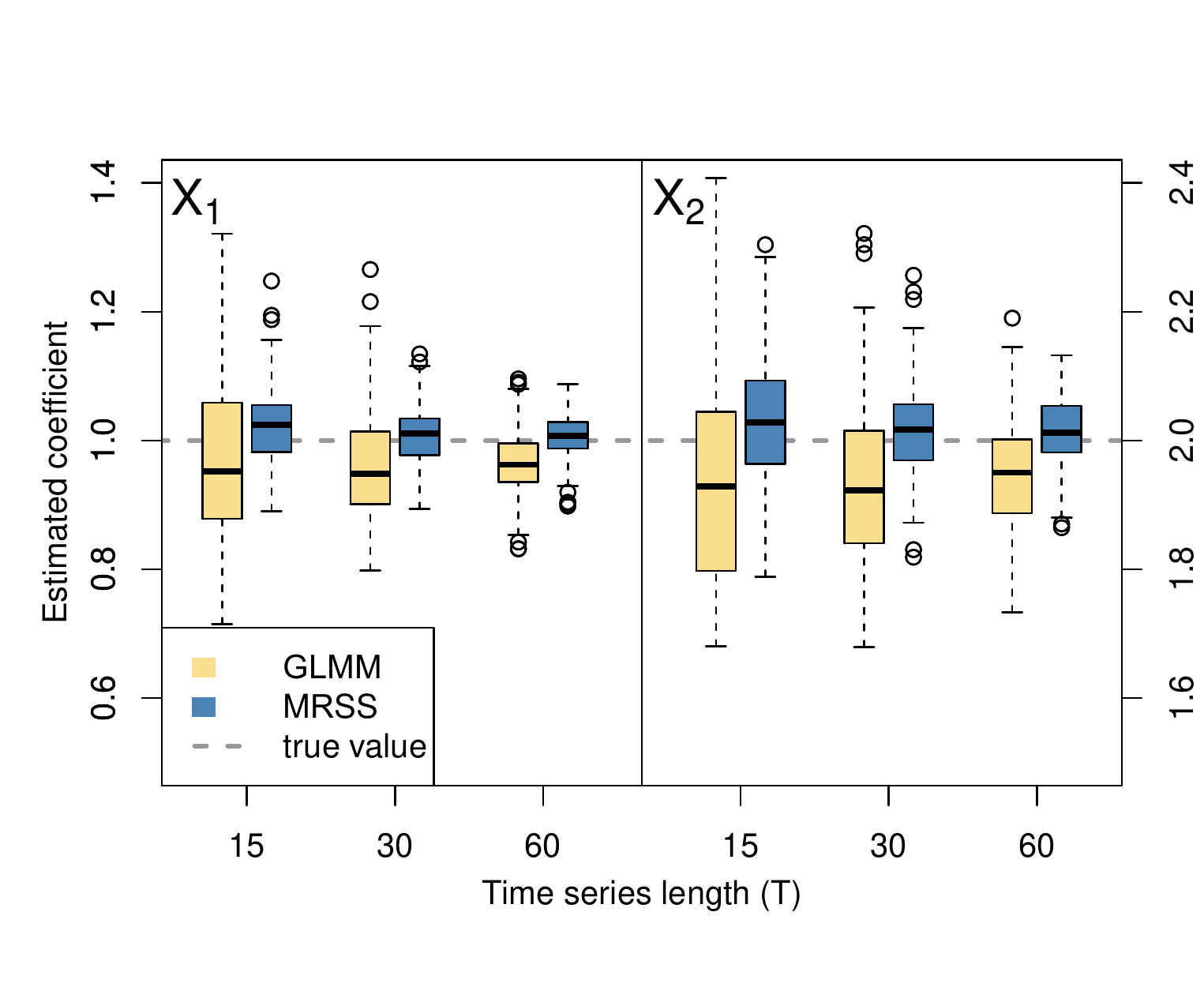}
    }
    \parbox{0.5\textwidth}{
        \textbf{(b.2)} Response $\boldsymbol{Y}^{(3)}$; Varying time length ($T$)\\
        \includegraphics[clip, trim=16pt 15pt 0 50pt, width=0.5\textwidth]{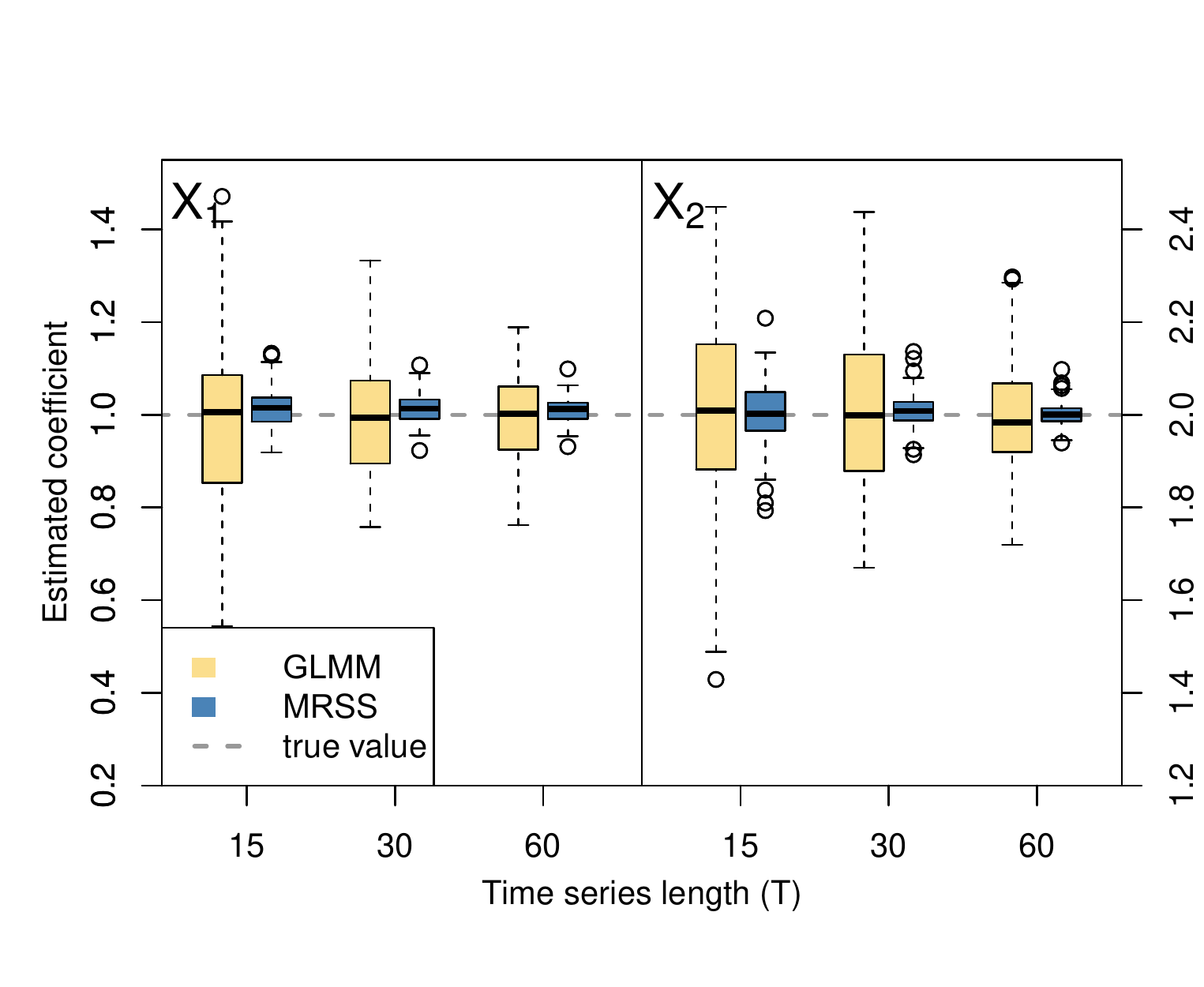}
    }
    \parbox{0.55\textwidth}{
        \textbf{(c.1)} Response $\boldsymbol{Y}^{(1)}$; Varying $p$\\
        \includegraphics[clip, trim=16pt 30pt 0 50pt, width=0.5\textwidth]{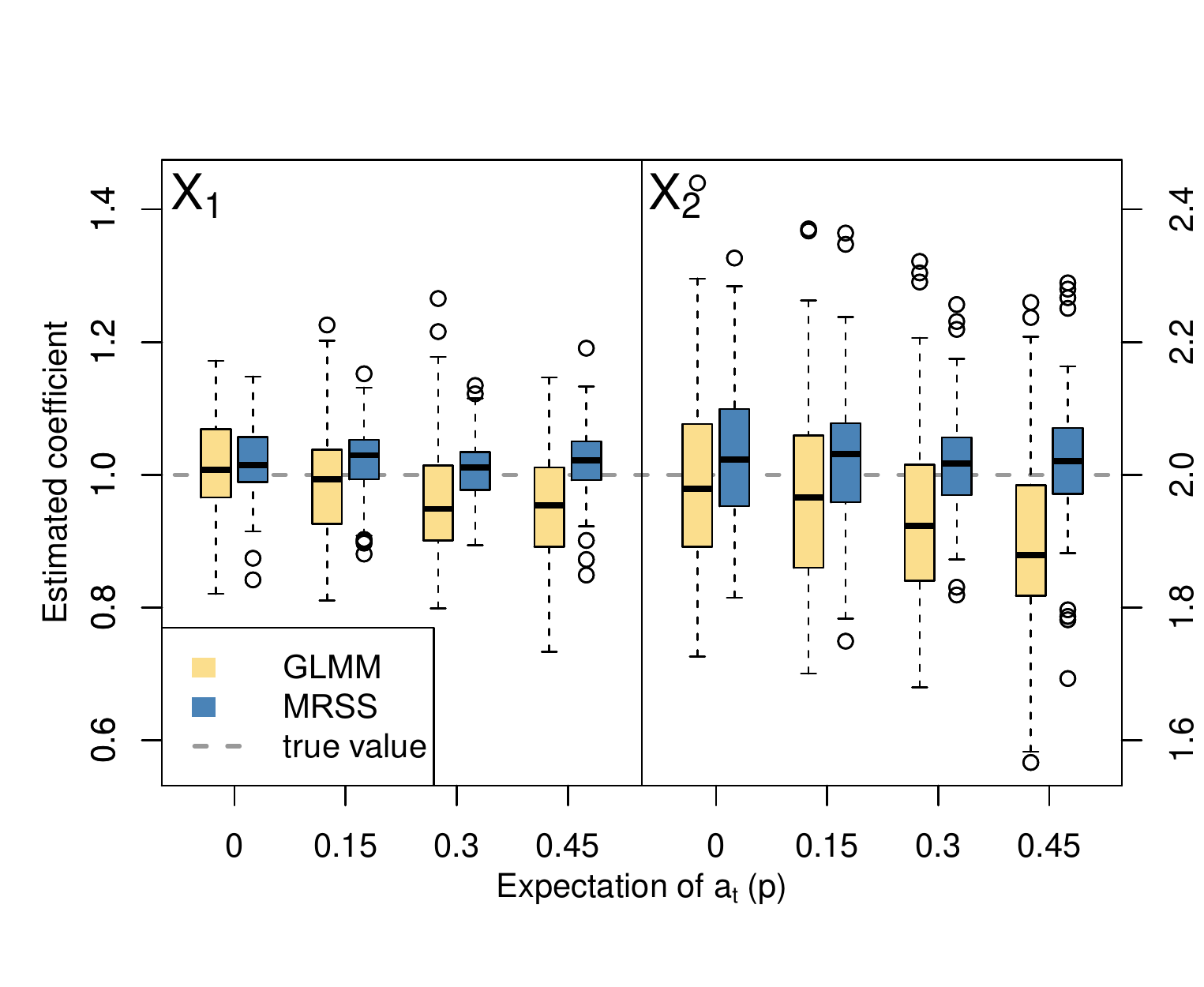}
    }
    \parbox{0.5\textwidth}{
        \textbf{(c.2)} Response $\boldsymbol{Y}^{(3)}$; Varying $p$\\
        \includegraphics[clip, trim=16pt 30pt 0 50pt, width=0.5\textwidth]{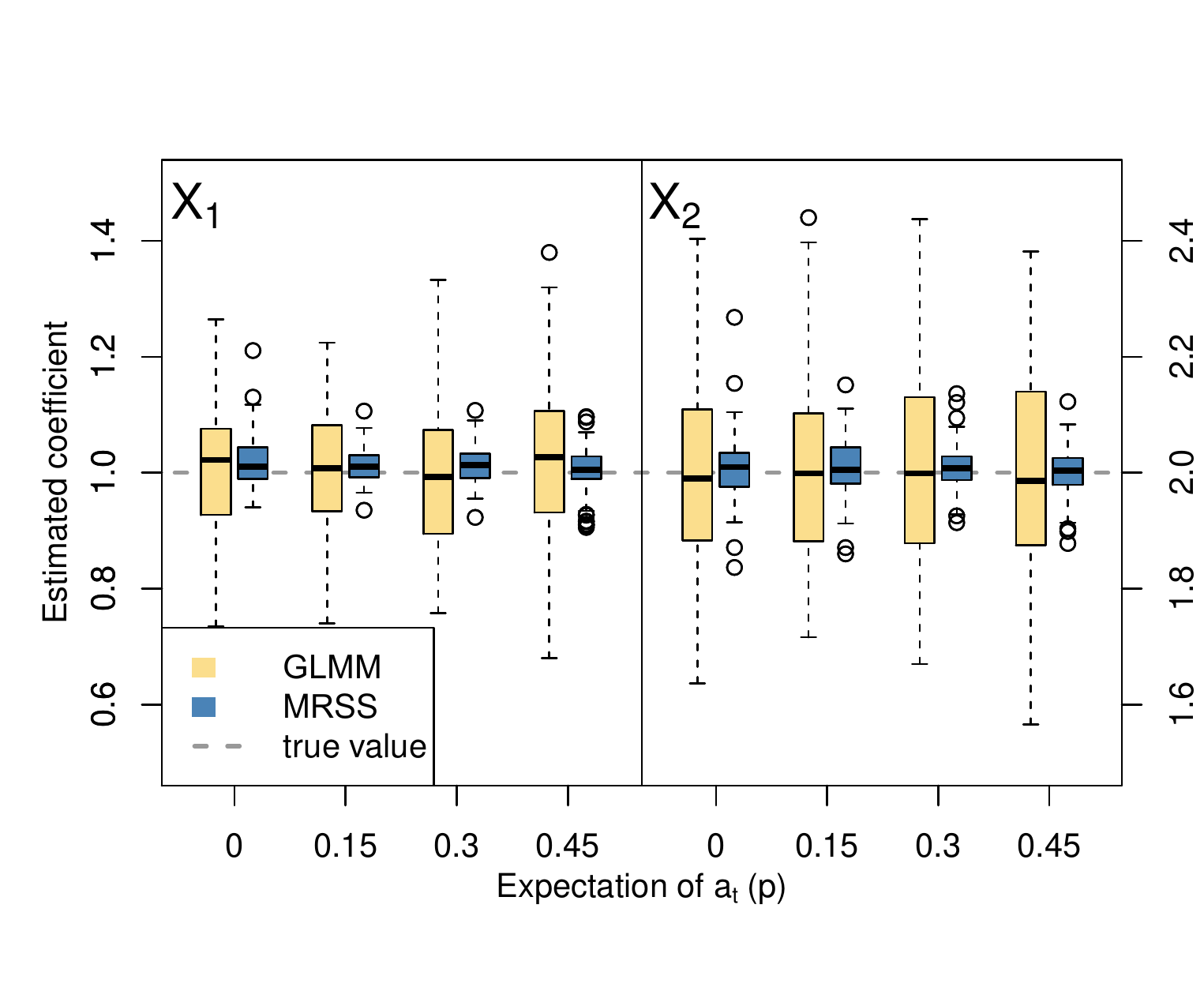}
    }
    \caption{Estimated coefficients from response $\boldsymbol{Y}^{(1)}$ or $\boldsymbol{Y}^{(3)}$ under different simulation settings. In (a.1), (a.2), the sample size ($N$) varies;  In (b.1), (b.2), the time series length ($T$) varies;  In (c.1), (c.2), the expectation of $a_t$ ($p$) varies. Estimated coefficients from response $\boldsymbol{Y}^{(2)}$ are in Supplementary Appendix Figure~\ref{wbfig:coefficients}.}\label{fig:coefficients}
\end{figure}

Figure~\ref{fig:coefficients} (a.1) and (b.1) show the results for the binomial response $\boldsymbol{Y}^{(1)}$. The proposed method provides unbiased coefficient estimates, but the estimation of both coefficients in GLMM is biased, and such bias is not alleviated by increasing the sample size or the number of observations. This bias might be due to the misspecification of the random effects structure in the GLMM.  Figures~\ref{fig:coefficients}~(a.2) and (b.2) show that both methods are unbiased for the normally distributed response $\boldsymbol{Y}^{(3)}$. In both settings, except for the bias, the variances of the estimated coefficients are also much smaller in the proposed method. Supplementary Appendix Table~\ref{wbtab:sample_size} and \ref{wbtab:time_length} present all MSEs, and the proposed method is shown to be more efficient in all the scenarios. 

Next, we change the expectation of $a_t$ ($p$) in $\bLambda_t$ from $0$ to $0.45$. When $p=0$, 
the observed responses only depend on one latent state $a^{(2)}_t$. While $p>0$, both latent states play a role in the observation equation. As $p$ increases, the proportion of effect from $a^{(1)}_t$ increases. In Figure~\ref{fig:coefficients} (c.1), when $p$ increases from $0$, the bias of GLMM also increases, which reveals that GLMM cannot correctly handle multiple latent processes even though the indicator variable $a_t$ is included in the model. The proposed model is not affected by varying $p$. Supplementary Appendix Table~\ref{wbtab:a} contains more comprehensive results where $q$ ranges from $0$ to $0.9$. The MSE of GLMM increases with $p$ while MRSS does not show a difference. 

{As a sensitivity analysis, we further fit several MRSS models with wrong numbers of states in Supplementary Appendix Table~\ref{wbtab:num}. The sample size ($N$) is fixed at $40$, time series length ($T$) is $30$, and the expectation of $a_t$ ($p$) is $0.3$. We find that the performance hardly changes when the model contains one or two unnecessary latent states. However, if we miss any of the treatment effect state $\bv$ or  health  state $\bb$, the estimation becomes biased with larger variances. Therefore, we recommend fitting slightly larger models to examine the model sensitivity and adopting AIC to determine the number of latent states as described in Section~\ref{sec:estimation}.}

\begin{figure}[!ht]
    \parbox{0.55\textwidth}{
        \textbf{(a)} In-sample\\
        \includegraphics[clip, trim=0pt 15pt 0 50pt, width=0.5\textwidth]{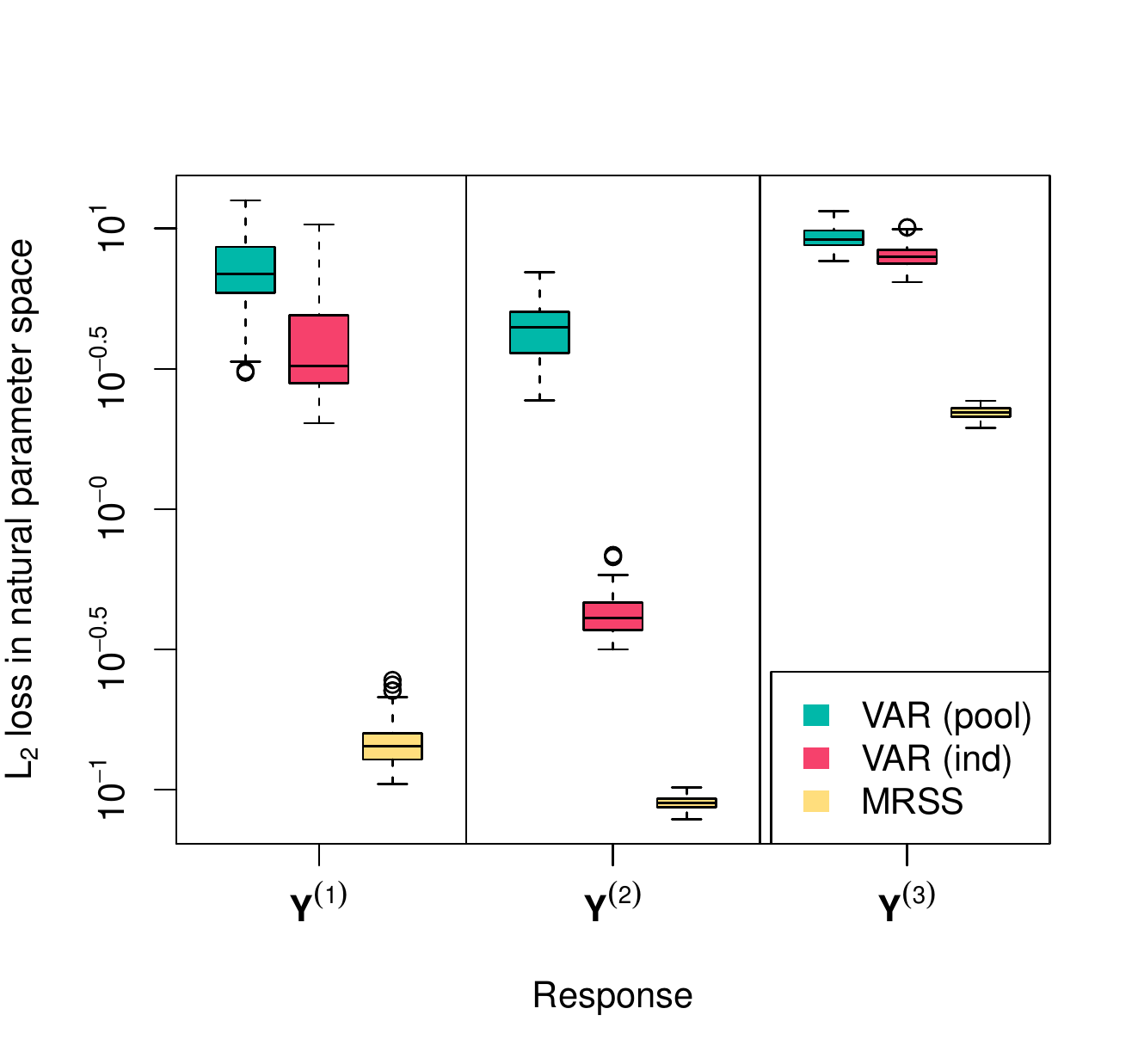}
    }
    \parbox{0.5\textwidth}{
        \textbf{(b)} Out-of-sample\\
        \includegraphics[clip, trim=0pt 15pt 0 50pt, width=0.5\textwidth]{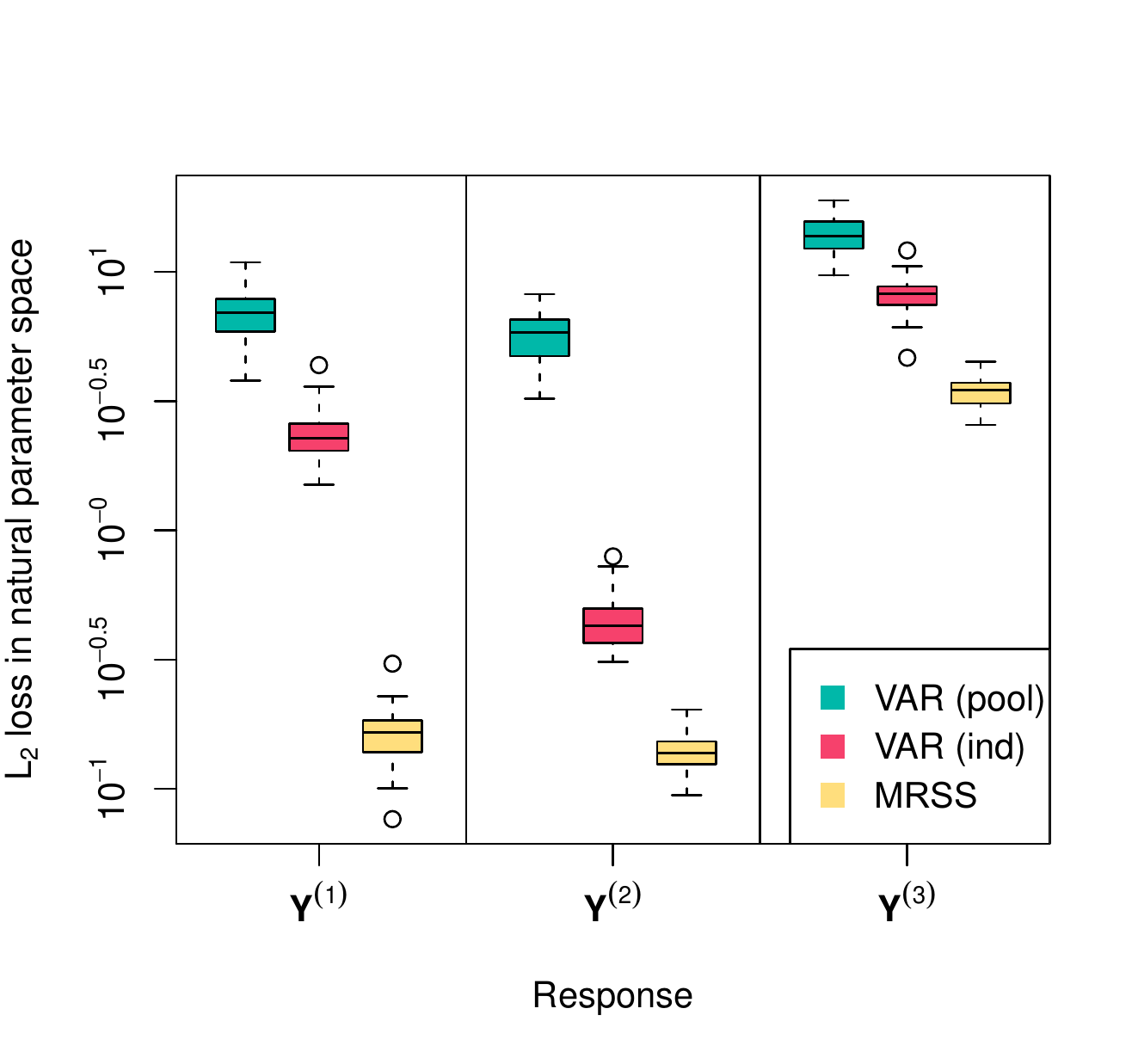}
    }
    \caption{Mean prediction errors in simulation studies. (a) Mean training errors of the first $25$ time points.  We used the first $25$ time points to train the model so that this is training error. (b) Mean testing prediction errors of the last $5$ time points.}\label{fig:prediction}
\end{figure}

In the second set of simulations, we compare the prediction accuracy with the setting $N=40$, $T=30$, $p=0.3$. We use the first $25$ time points to train the model and then we predict $\boldsymbol{Y}_t$ based on $\boldsymbol{Y}_1$ to $\boldsymbol{Y}_{t-1}$ for all $t > 1$. We define the average MSE as the squared error in the natural parameter space averaged across the training time points.
Similarly, we  define the out-of-sample prediction error as the average MSE from time points $t=26$ to $t=30$ (see Supplementary \ref{wbsec:prediction} for details). Figure~\ref{fig:prediction} shows the mean prediction error for $40$ subjects in $200$ repetitions. The pooled-VAR model performs the worst, which demonstrates that naively taking the average value of estimated parameters will not improve the model's performance. The proposed method outperforms the individual-VAR. One  reason is that MRSS estimates much fewer parameters than individual-VAR, since the latter fits $40$ separate models in total. In addition, individual-VAR does not properly handle the non-Gaussian distributions. 

\begin{figure}[!t]
    \parbox{0.55\textwidth}{
        \textbf{(a)} \\
        \includegraphics[clip, trim=0pt 15pt 28pt 50pt, width=0.5\textwidth]{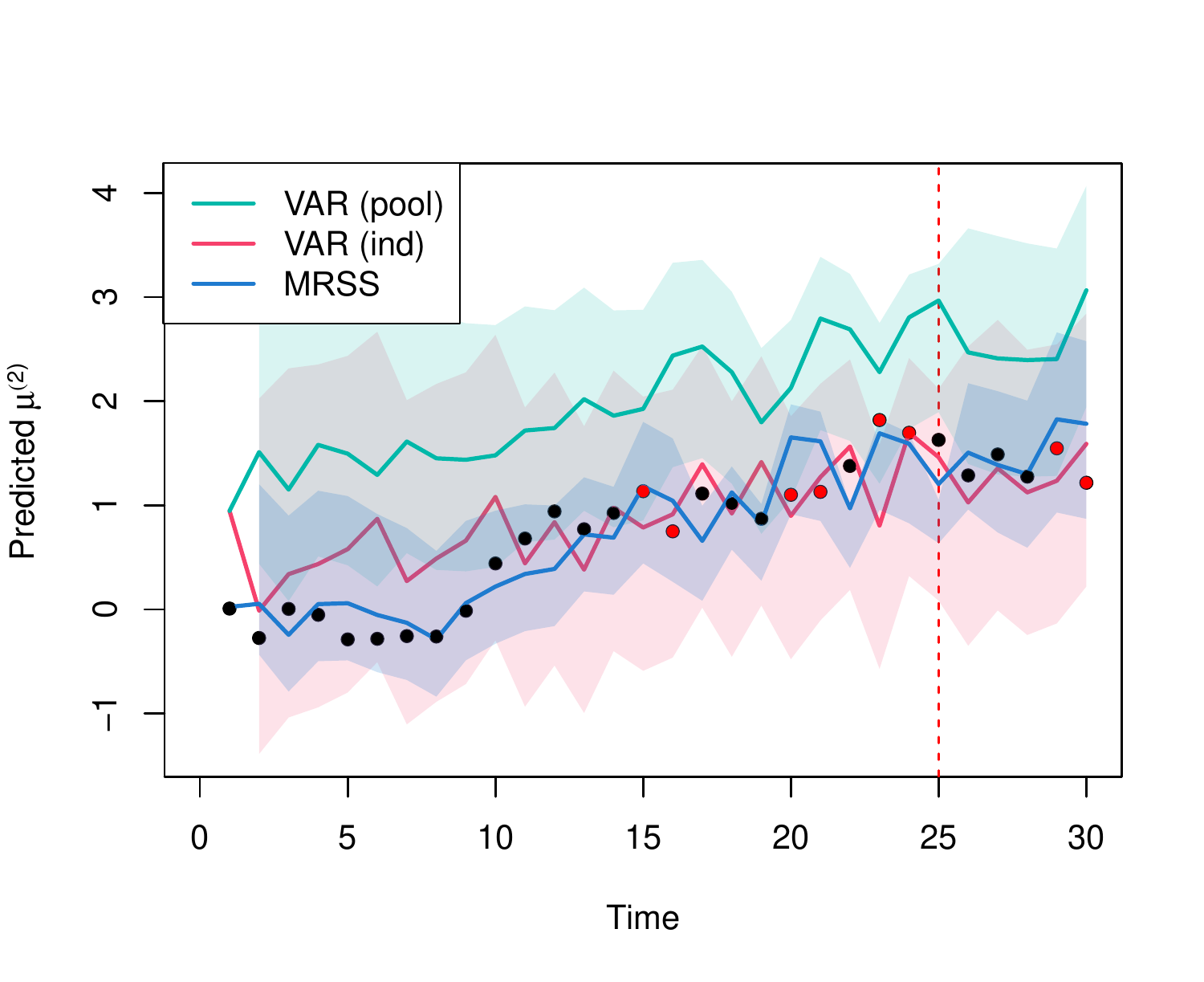}
    }
    \parbox{0.5\textwidth}{
        \textbf{(b)} \\
        \includegraphics[clip, trim=0pt 15pt 28pt 50pt, width=0.5\textwidth]{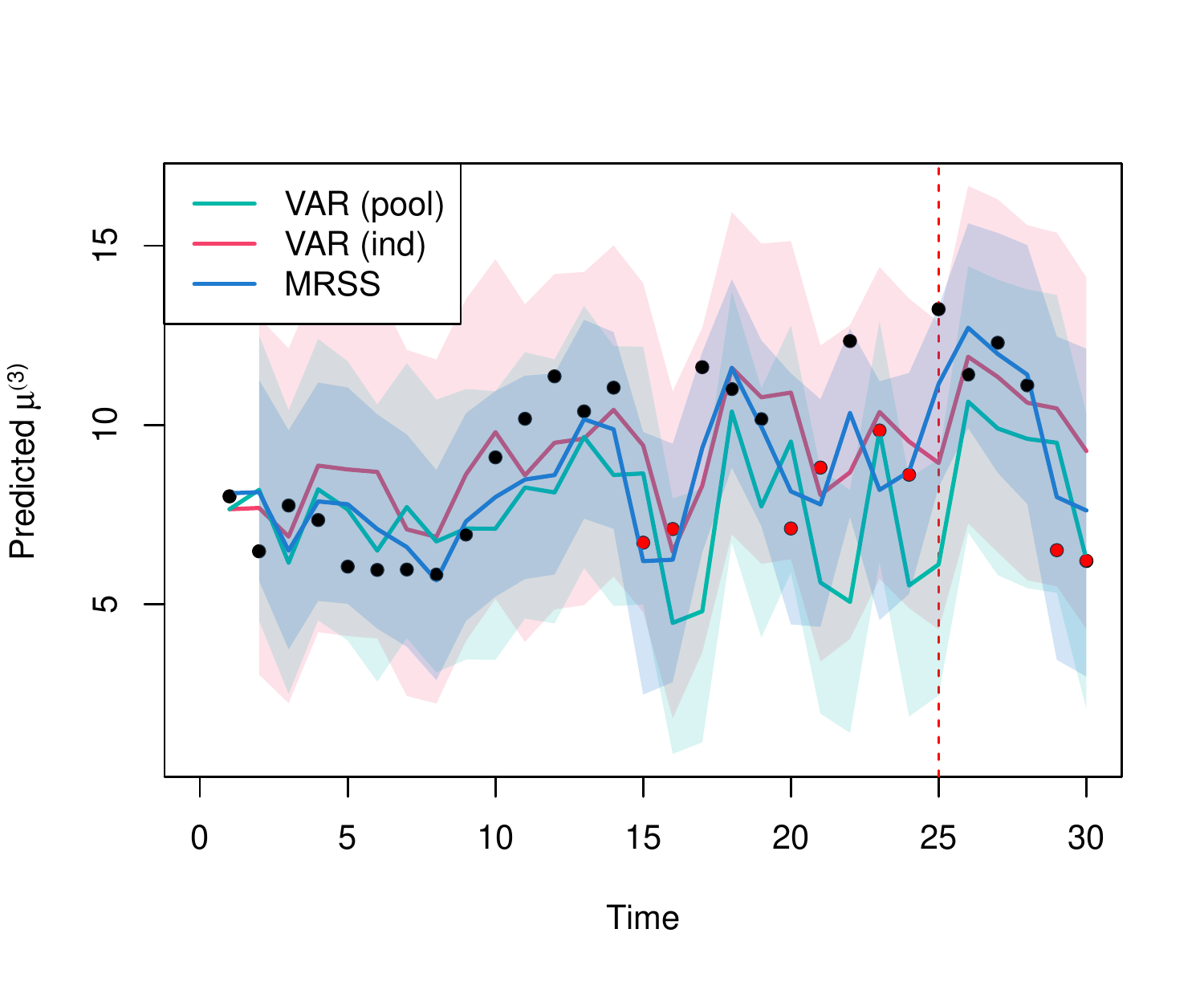}
    }
    \caption{Predicted trajectories of a typical subject in simulation studies. Black and red dots are true values when treatment $a_t=0$ and $a_t=1$. (a) Predicted $\log(Y^{(2)})$ in VAR or predicted $\mu^{(2)}$ in  MRSS. (b) Predicted $Y^{(3)}$ in VAR or predicted $\mu^{(3)}$ in MRSS.}\label{fig:example}
\end{figure}

Figure~\ref{fig:example} shows the predicted trajectories ($Y^{(2)}$ and $Y^{(3)}$) for a typical subject. Black and red dots are true values when $a_t=0$ or $a_t=1$.  The pooled-VAR provides  biased prediction on $Y^{(2)}$, but the bias is smaller on $Y^{(3)}$. In Figure~\ref{fig:example}(a), MRSS has better predictive performance than individual-VAR, and its confidence band is much narrower. In Figure~\ref{fig:example}(b), red dots are away from black dots with lower values. Visual inspections reveal that MRSS is better at capturing responses when  $a_t=0$ or $a_t=1$. Supplementary Appendix Figure~\ref{wbfig:y1pre} to Appendix Figure~\ref{wbfig:y3pre} show more randomly selected subjects, and similar conclusions hold.

\section{Application to mPower}
The mPower study \citep{bot2016mpower}
is an observational smartphone-based study that aims to evaluate the feasibility of remotely collecting frequent information in PD patients about their daily changes in symptom severity and their sensitivity to medication treatment. A total of $6,805$ participants completed the enrollment survey, among which $1,087$ were self-identified as having a professional diagnosis of PD while $5,581$ were not ($137$ opted not to answer the question).  The detailed study design is described by \citet{bot2016mpower}. In Supplementary Figure~\ref{wbfig:pipe}, we show how the digital phenotypes were collected. Briefly, a participant was instructed to perform various activity tasks, including voice, walking, speeded tapping, and a memory game on a smartphone across three to six months.

\subsection{Data Preprocessing}
For each task performed on the mobile device, we extract a series of digital phenotypes based on prior literature and treat them as the response variables \citep{snyder2020mhealthtools, sieberts2021crowdsourcing}. These features were shown to be predictive of the severity of PD symptoms \citep{tsanas2012accurate} and thus included as digital phenotypes of interest in our joint model. The complete list of phenotypes is in Supplementary \ref{wbsec:preprocessing}. In this analysis, the main ``motor'' model is a multi-dimensional model that consists of $3$ features from the walking modality, $3$ from the tapping modality, and $2$ measurement process indicators of each modality. Additionally, we also build a ``memory'' model (a one-dimensional model including a key score of memory performance and its measurement process) and a ``voice'' model (a multi-dimensional model including $8$ voice features extracted from sound recordings and the measurement process).  We briefly introduce each phenotype modality in the following.

\noindent\textbf{Walking Modality:}
The walking tasks assess participants' gait and balance. Participants were asked to walk 20 steps in a straight line (walking stage), turn around, and stand still for 30 seconds (resting stage).  
We  obtain the readouts of the accelerometer in $X$, $Y$, $Z$ directions and also define the combined reading as 
$
	C_1 = \sqrt{X^2+Y^2+Z^2},$ and 
	$C_2=\sqrt{(X-\bar X)^2+(Y-\bar Y)^2+(Z-\bar Z)^2}.
$
The following extracted features are most predictive of PD status and thus used in our analysis \citep{snyder2020mhealthtools}: median value of walking stage $C_1$; the standard deviation of resting stage $C_2$; estimates of the scaling exponent in detrended fluctuation analysis of walking stage $Y$ (conducting first order cumulative summations on the original data before performing a DFA). A healthier subject tends to walk faster (larger `median $C_1$'), stand more steadily (smaller `sd $C_2$'), and has fewer long-range power-law autocorrelations in accelerometer readout (smaller `dfa $Y$').

\noindent\textbf{Tapping Modality:}
Finger tapping can be used to measure bradykinesia, one of the hallmarks of PD symptoms. In the tapping task, participants were instructed to lay their phone on a flat surface and use two fingers of the same hand to alternatively tap two stationary points on the screen for 20 seconds.  The software recorded  where and when the participant tapped on the phone screen. 
We used the standard deviation, interquartile range of the time interval of two consecutive taps, and the number of tappings in $20$ seconds as outcome variables. A healthier subject taps the screen at a faster and more stable speed (larger `number taps' and smaller `sd tap inter', `range tap inter').

\noindent\textbf{Memory Modality:}
In memory tasks, participants observed a grid of flowers that illuminated one at a time and were then asked to replicate the illumination pattern by touching the flowers in the same order. A game score was calculated by the game software. A higher score means better memory strength.

\noindent\textbf{Voice Modality:} Speech disorders have been linked to PD and there is strong supporting evidence of degrading performance in voice with PD progression \citep{ho2008better, skodda2009progression}. In the mPower study, participants recorded sustained phonation by saying `Aaaaah' into the microphone at a steady volume for up to 10 seconds. We aimed to extract distinguishing sound features for PD as the orchestrated muscle movements are involved in voice production. We select eight features that best distinguish PD patients and controls from each voice recording as responses in our model: Shimmer\textsubscript{PQ3} (amplitude differences using a $3$ sample window of fundamental frequency $F_0$ estimates), GQ\textsubscript{std\_cycle\_closed} (standard deviation of the glottal quotient for the duration where vocal folds are in collision), GNE\textsubscript{mean} (mean value of the glottal to noise excitation), MFCC\textsubscript{mean\_1st\_coef} (the mean value of 1st Mel Frequency cepstral coefficient), 
Log Entropy\textsubscript{det\_1st\_coef} (log entropy of detail coefficients at 1st level (of total $10$ levels) in Daubechies 8 wavelet decomposition for $F_0$), Log Entropy\textsubscript{app\_1st\_coef} (log entropy of approximate coefficients at 1st level), Log Entropy\textsubscript{det\_LT\_1st\_coef} (log entropy of detail coefficients at 1st level for log-transformed $F_0$), Log Entropy\textsubscript{app\_LT\_1st\_coef} (log entropy of approximate coefficients at 1st level for log-transformed $F_0$). A healthier subject has larger Shimmer\textsubscript{PQ3}, MFCC\textsubscript{mean\_1st\_coef} and smaller GQ\textsubscript{std\_cycle\_closed}, GNE\textsubscript{mean}. 
The wavelet coefficients measure the resemblance indices between the selected wavelet and the signal at each level. A healthier subject can sustain a vowel with a smaller deviation from exact periodicity and thus has smaller log entropy of these coefficients. 

\subsection{Model Building and Results}
\subsubsection{Motor Model}
We jointly model $6$ phenotypes in two modalities (walking and tapping) and the corresponding measurement indicators (measurements taken before or after Levodopa treatment). Response $\bA_{it}\in \mathbb{R}^2$ is the measurement indicators for modality walking and tapping (A\_walking, A\_tapping), and they are modeled as binomial distributions. Response $\bY_{it}\in \mathbb{R}^6$ includes $6$ phenotypes from two modalities: `median of C\textsubscript{1} (resting)', `Sd of C\textsubscript{2}', `Dfa Y' from walking modality, and `Range of tapping time interval', `number of taps', `Sd of tapping time interval' from tapping modality. {The values in $\bY_{it}$ are standardized and assumed to be normally distributed. This distributional assumption can be checked with fitted residuals. 
Subjects took measurements every 1.34 (for walking modality) or 1.75 (for tapping modality) days and were followed up to six months. Considering the `early-morning akinesia' effect \citep{marsden1976off}, we let $t=1,2,\cdots, m_i$ represent semi-daily transitions so that we can differentiate measurements in the morning or the afternoon. }

We focus on estimating personalized treatment effects to determine how each individual responded to the medication treatment. Since mPower is an observational study, the observed association may be due to confounding. For example, an earlier analysis of mPower study showed that circadian rhythms/daily routine
activities might be responsible for patients' poorer performance in the morning \citep{neto2017analysis}. Thus, we distinguish morning versus afternoon or evening measurements and model them using separate processes to parse out the circadian effect from the medication effect. This leads to four latent states: $v^{(1)}_{it}$ and $v^{(2)}_{it}$ reflect the health status; $b^{(1)}_{it}$ is the treatment effect of Levodopa; and $b^{(2)}_{it}$ is a latent state to model the `early-morning akinesia' effect \citep{marsden1976off}. The transition matrix of the four states is defined as
    \begin{align}
        \begin{pmatrix}
            b^{(1)}_{it}\\b^{(2)}_{it}\\v^{(1)}_{it}\\v^{(2)}_{it}
        \end{pmatrix} = \bT \begin{pmatrix}
            b^{(1)}_{i,t-1}\\b^{(2)}_{i,t-1}\\v^{(1)}_{i,t-1}\\v^{(2)}_{i,t-1}
        \end{pmatrix} + \begin{pmatrix}
            \varepsilon_{b^{(1)}}\\\varepsilon_{b^{(2)}}\\\varepsilon_{v^{(1)}}\\\varepsilon_{v^{(2)}}
        \end{pmatrix},
    \end{align}
where $\bT\in\mathbb{R}^{4\times 4}$ is a diagonal transition matrix and it takes different values for PD patients taking Levodopa,  PD patients not taking Levodopa, and healthy controls; $(  \varepsilon_{b^{(1)}}, \varepsilon_{b^{(2)}}, \varepsilon_{v^{(1)}}, \varepsilon_{v^{(2)}})^T\sim N_{4}(\boldsymbol{\mu}_\varepsilon, \bSigma_\varepsilon)$ and $\varepsilon_{v^{(1)}} \indep \varepsilon_{v^{(2)}}$.

For PD patients taking Levodopa, 
the observation equation is as follows:
    \begin{eqnarray}
    \binom{\bA_{it}}{\bY_{it}}\sim p\left(\binom{\bA_{it}}{\bY_{it}}\middle|\begin{pmatrix}
        \textbf{0} & \blambda_{Ab^{(2)}}\operatorname{diag}(\boldsymbol{M}_{it}) & \blambda_{Av^{(1)}} & \blambda_{Av^{(2)}}\\
        \blambda_{Yb^{(1)}}\operatorname{diag}(\ba_{it}) & \blambda_{Yb^{(2)}}\operatorname{diag}(\boldsymbol{M}_{it}) & \blambda_{Yv^{(1)}}&\blambda_{Yv^{(2)}}
    \end{pmatrix} \begin{pmatrix}
        b^{(1)}_{it}\\b^{(2)}_{it}\\v^{(1)}_{it}\\v^{(2)}_{it}
    \end{pmatrix}+\bbeta\bX_{it}  \right),
    \end{eqnarray}
where $\ba_{it}\in\mathbb{R}^6$ is the treatment indicator for the corresponding $6$ phenotypes;  $\boldsymbol{M}_{it}\in\mathbb{R}^6$ is the indicator of whether the treatment is taken in the early morning for the corresponding $6$ phenotypes; $\blambda_{Ab^{(2)}},\blambda_{Av^{(1)}},\blambda_{Av^{(2)}},\blambda_{Yb^{(1)}},\blambda_{Yb^{(2)}}, \blambda_{Yv^{(1)}},\blambda_{Yv^{(2)}}$ are the corresponding loading parameters. 
The covariates $\Xb_{it}$ include `age', `gender', `smoking history' (years of smoking before the study), `PD history' (years from diagnosis of PD), and `deep brain stimulation' (whether a patient had deep brain stimulation or not). 

For PD patients who didn't take treatment, treatment effect state $b_{it}^{(1)}$ and response $\bA_{it}$ do not exist, which leads to the following model structure:
\begin{eqnarray}
   {\bY_{it}}\sim p\left({\bY_{it}}\middle|\begin{pmatrix}
   \blambda_{Yb^{(2)}}\operatorname{diag}(\boldsymbol{M}_{it}) & \blambda_{Yv^{(1)}}&\blambda_{Yv^{(2)}}
   \end{pmatrix} \begin{pmatrix}
      b^{(2)}_{it}\\v^{(1)}_{it}\\v^{(2)}_{it}
   \end{pmatrix}+\bbeta\bX_{it}^{}  \right).
\end{eqnarray}

For healthy controls, we only include one health status state $v^{(1)}_{it}$ and covariates $\bX_{it}$ excluding `PD history' and `deep brain stimulation'.
The observation model is as follows:
   \begin{eqnarray}
   {\bY_{it}}\sim p\left({\bY_{it}}\middle|\begin{pmatrix}
    \blambda_{Yb^{(2)}}\operatorname{diag}(\boldsymbol{M}_{it}) &\blambda_{Yv^{(1)}}
    \end{pmatrix} 
    \begin{pmatrix}
    b^{(2)}_{it}\\v^{(1)}_{it}
   \end{pmatrix}+\bbeta^{} \bX_{it}^{} \right).
   \end{eqnarray}

We also compare with GLMMs and individual-VARs (fit $N$ separate VAR models in total, one for each individual). 
For binomial responses `A\_walking', `A\_tapping', the GLMM includes `age', `gender', 'PD' (patient or control), `medication' (whether taking Levodopa as treatment), `smoking history', `PD history', `deep brain stimulation', and `morning' (whether the measurement is taken in the early morning) as fixed effect terms and a random intercept for each subject. 
For the six Gaussian digital phenotypes, the GLMM includes the same predictors, a random intercept, and an additional indicator variable (`A\_walking' or `A\_tapping') to quantify whether one measurement is taken immediately after the treatment or not.
In individual-VAR, we fit $8$ responses in one model. The covariates are the same as GLMM, except that `A\_walking' and `A\_tapping' are ruled out since they appear in responses. 


The estimated loading matrix $\blambda$ and $\bbeta$ in the proposed MRSS and the estimated coefficients of GLMM are presented in Supplementary Appendix Table~\ref{wbtab:estimated}, \ref{wbtab:estglmm}. 
In the GLMM model, measurement indicators `A\_walking' and `A\_tapping' are not significantly associated with any covariates in `age', `gender', `smoking history', `PD history', `deep brain stimulation' or `morning'. In contrast, the MRSS detects that `A\_walking' and `A\_tapping' are significantly related to the time state $b^{(2)}$, the health states $v^{(1)}$ and $v^{(2)}$, and age. This indicates that older patients tend to take measurements immediately after treatment, and such behavior is also more likely to occur in the morning or when a patient's health status deteriorates. 

\begin{figure}[!b]
	\parbox{.5\textwidth}{
	\textbf{(a)}\\	
\includegraphics[width=0.5\textwidth]{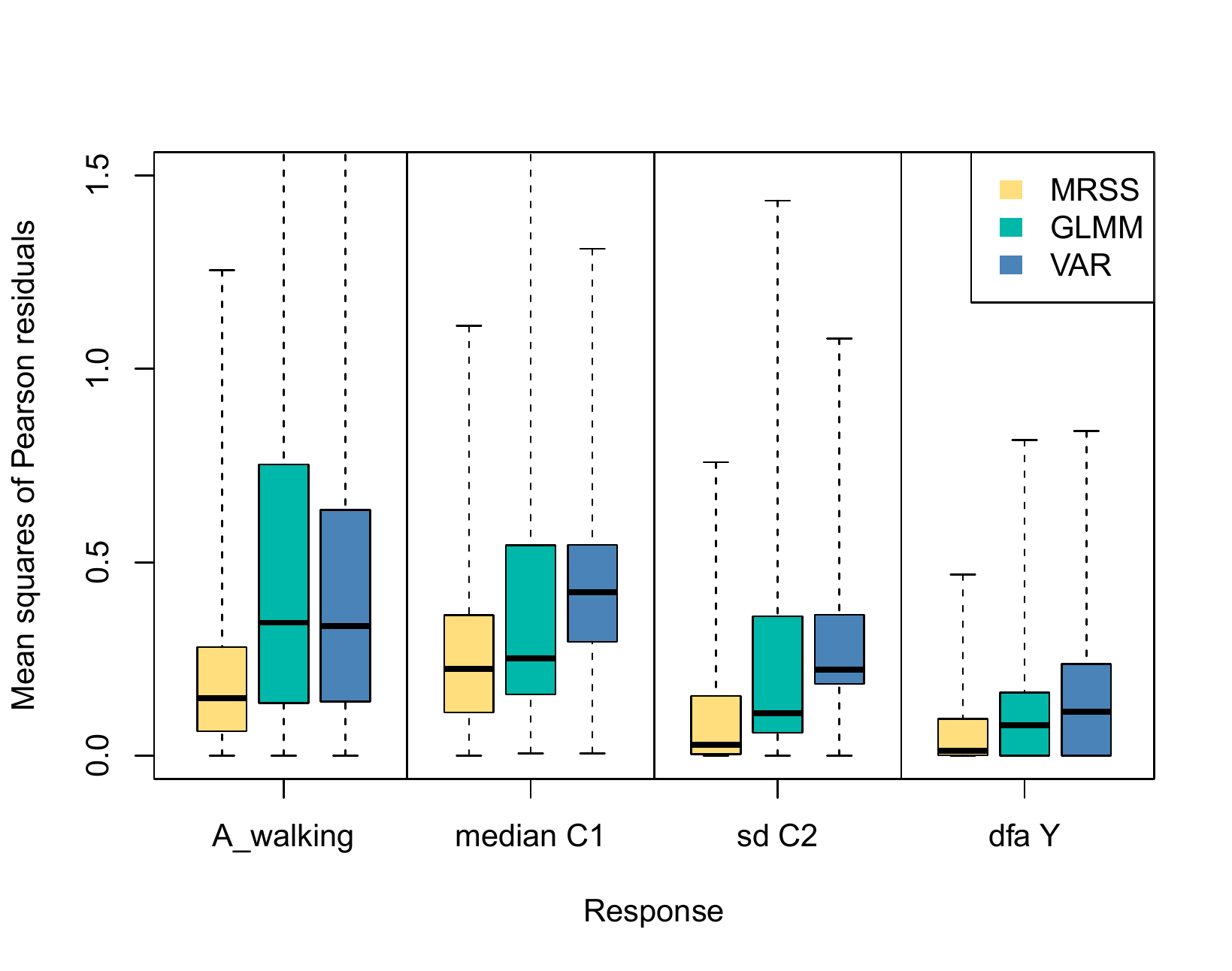}}
\parbox{.5\textwidth}{
	\textbf{(b)}\\
		\includegraphics[width=0.5\textwidth]{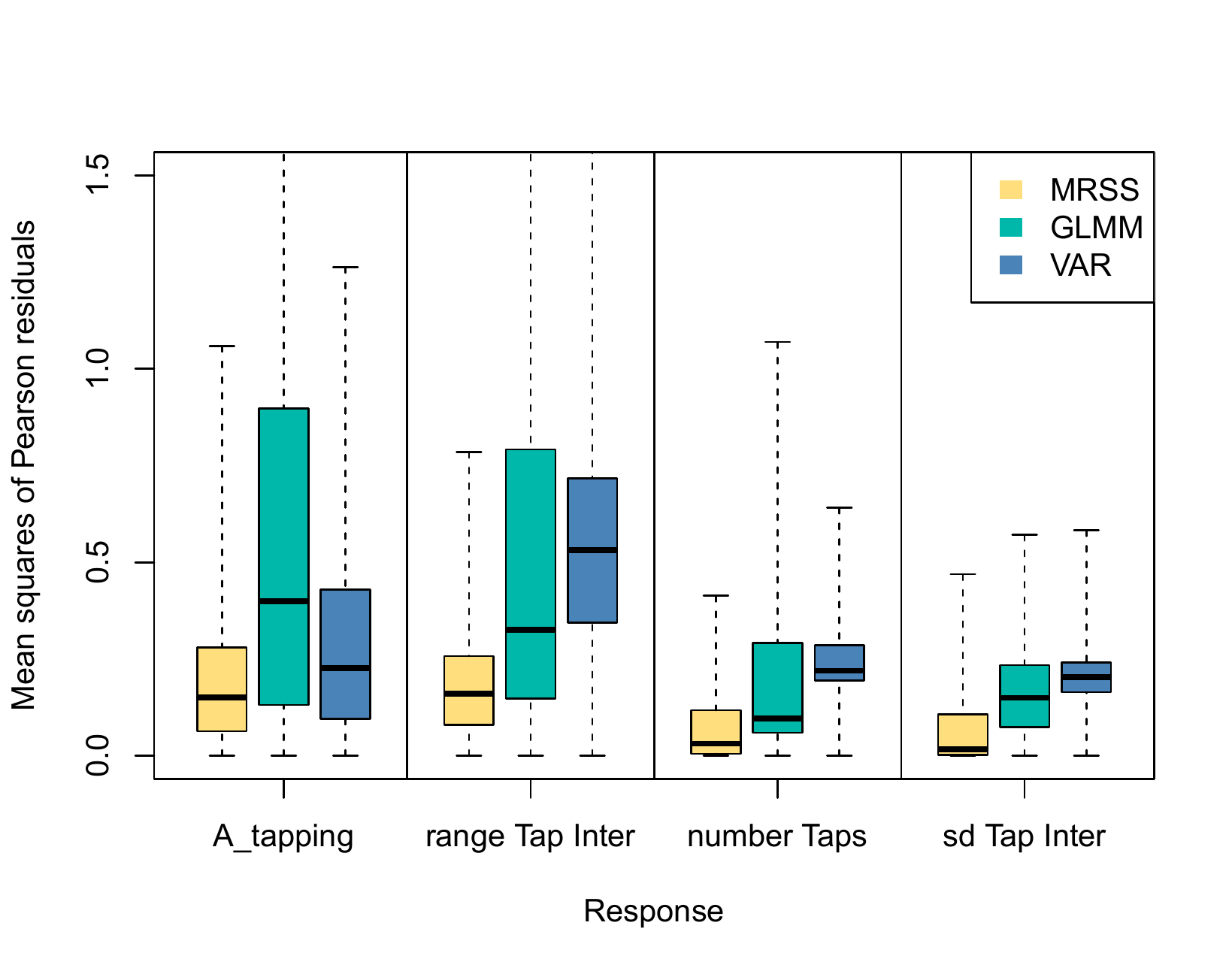}}
        \caption{The out-of-sample prediction error of each response. The boxplot shows the mean squares of the Pearson residuals of $5$ predicted time points.}\label{fig:pred}
\end{figure}

For the three phenotypes in the walking modality, MRSS finds that the phenotypes are significantly associated with the time state $b^{(2)}$ and health status states $v^{(1)}$. The performance on tapping is generally worse in the early morning, which is consistent with previous literature \citep{bhidayasiri2008motor}. The three phenotypes in tapping modality are significantly affected by the short-term treatment state $b^{(1)}$ and health status state $v^{(1)}$ and $v^{(2)}$. These results confirm the well-known short-term effect of Levodopa. In addition, all phenotypes are significantly associated with `age', which GLMM does not detect. The long-term treatment effect is reflected in the difference in the health status states in each group. The estimated expectations of health status states $v^{(1)}$, $v^{(2)}$ are $-1.15$ and $-0.89$ for PD patients with treatment, and are $-0.90$ and $-0.11$ for PD patients without treatment. We multiplied the loading matrix to the expected latent state values and observed better average digital phenotype responses for Levodopa receivers, which reflects that Levodopa, on average, provides a long-term effect to slow the disease progression. 

\begin{figure}[!b]
	\parbox{.5\textwidth}{
	\textbf{(a)}\\	
\includegraphics[width=0.5\textwidth]{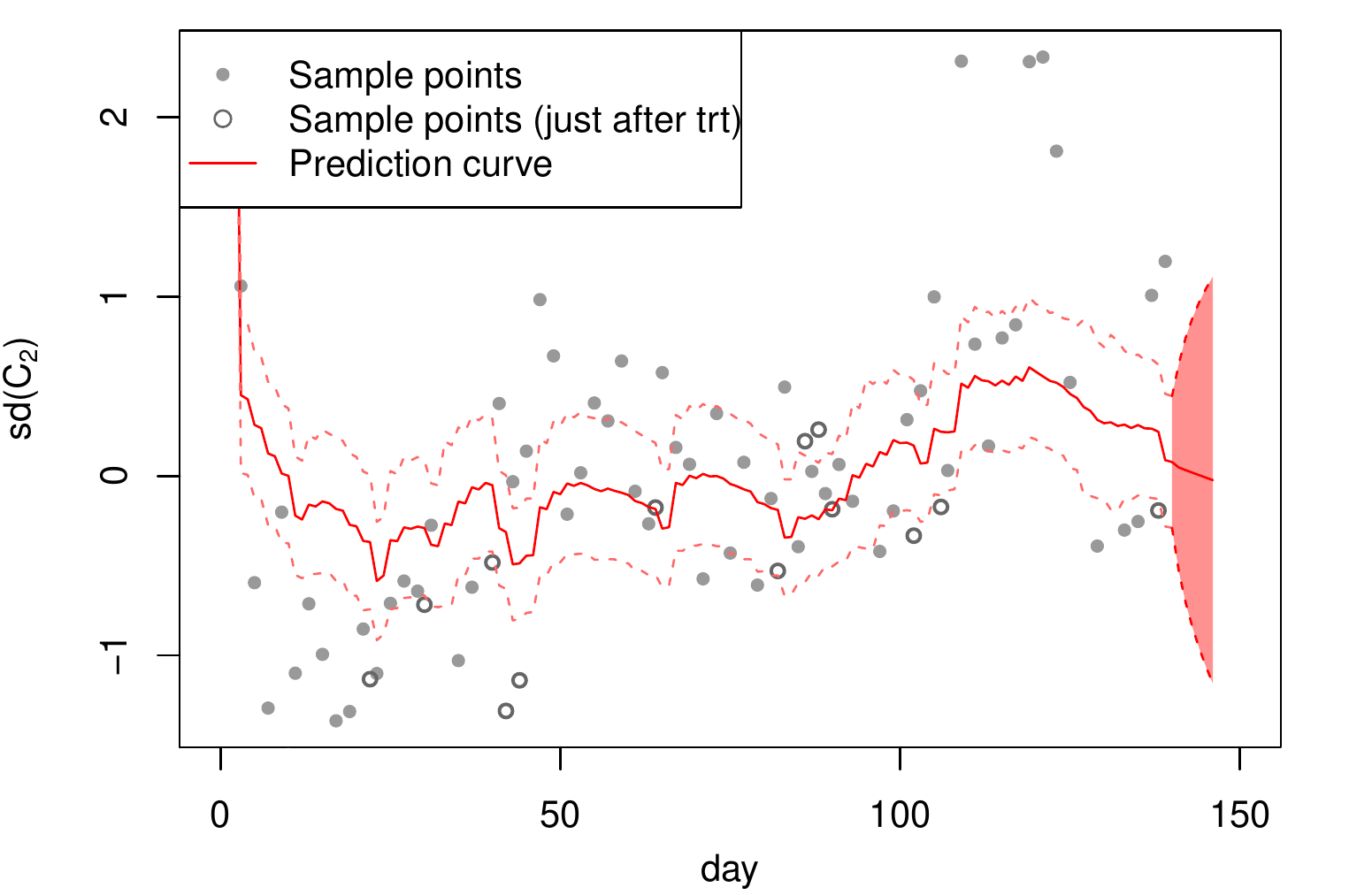}}
\parbox{.5\textwidth}{
	\textbf{(b)}\\
		\includegraphics[width=0.5\textwidth]{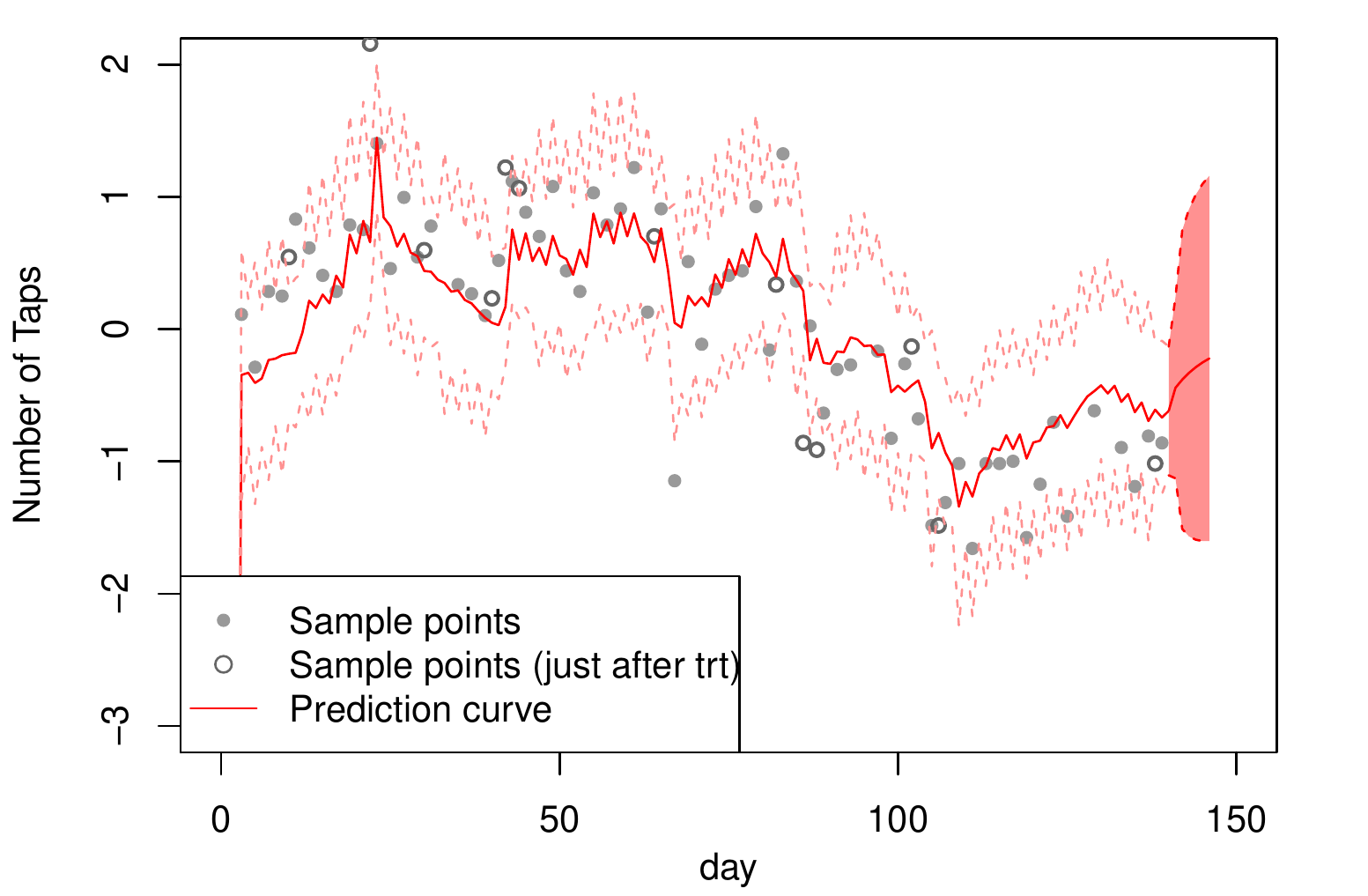}}
        \caption{Predicted trajectories of patients in the real world mPower study. (a) One-step ahead prediction on the walking modality (standard deviation of $C_2$). (b) One-step ahead prediction on the tapping modality (number of taps in $20s$).}\label{fig:sdc2}
\end{figure}

\begin{figure}[!t]
    \centering	\includegraphics[width=0.9\textwidth]{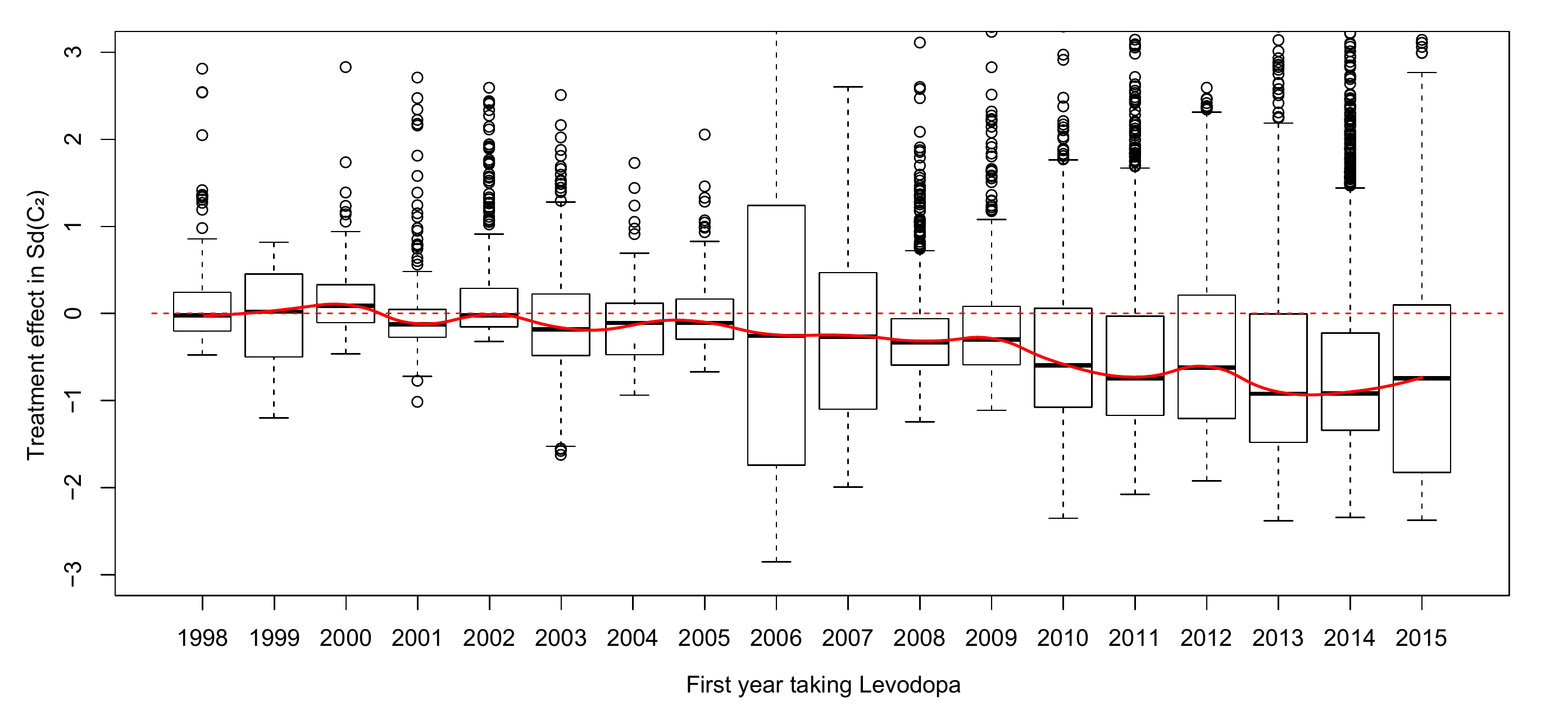}
    \caption{Predicted Levodopa treatment effect in year $2016$. The year in the horizontal axis represents the first year when a patient started taking the treatment. The vertical axis shows the decrease in Sd(C\textsubscript{2}). The lower value indicates a better treatment effect. }\label{fig:trt}
\end{figure}

Finally, we make an out-of-sample prediction for each subject to predict the last $5$ time points and calculate the mean squares of the Pearson residuals of each response in Figure~\ref{fig:pred}. The proposed method achieves much lower  prediction errors without lower variability. We also demonstrate the one-step-ahead prediction of `Sd of C\textsubscript{2}' and `number of taps' with an example patient in Figure~\ref{fig:sdc2}. This patient's health condition worsened within 120 days when the walking trembling increased and the number of screen taps decreased. However, after day 120, his/her symptoms were controlled, and the one-step-ahead prediction indicates a good prognosis. In addition, we predict the response `Sd of C\textsubscript{2}' for each patient in year $2016$ if he/she measures immediately after a dose of Levodopa and if he/she stops taking the treatment. The difference between the two predicted values can be regarded as the predicted treatment effect. We do this to all subjects in the study and group the results by the first year they started taking Levodopa, as shown in Figure~\ref{fig:trt}. We find the treatment effect wears off for most subjects who took Levodopa for more than $10$ years (the first year before $2006$), and better control in orthostatic tremor for subjects who just started taking the drug. 

\subsubsection{Memory Model \& Voice Model}
Similarly, we jointly model the memory game score and its corresponding measurement indicators. We include three latent states: $v^{(1)}_{it}$ reflects the health status; $b^{(1)}_{it}$ is the treatment effect; and $b^{(2)}_{it}$ is the time state to indicate early morning. The model details are in Supplementary \ref{wbsec:memory}. The estimated parameters are in Supplementary Appendix Table~\ref{wbtab:estimated_game}. Memory deterioration is mainly significantly associated with increased age. Also, we see that Levodopa improves cognitive function in patients. The `early morning' effect does not play an important role in memory test results.  

The voice model includes eight selected features and the corresponding measurement indicators. There are five latent states: $v^{(1)}_{it}$, $v^{(2)}_{it}$ reflect the health status; $b^{(1)}_{it}$, $b^{(2)}_{it}$ are the treatment effects; and $b^{(3)}_{it}$ is the `early morning' state. The model details are in Supplementary \ref{wbsec:voice}, and the estimated parameters are in Supplementary Appendix Table~\ref{wbtab:estimated_voice}. As expected, voice features are always affected by gender since males and females sound greatly different. At least one treatment state is associated with each response, which implies Levodopa improves preparatory motor set related activity and alleviates hypophonia. Besides, the `early morning' is not observed for voice features. 

\section{Discussion}
This paper proposes an MRSS model to jointly model mixed-type digital phenotypes collected from multiple modalities and their measurement processes using latent processes. The proposed MRSS has several advantages: {it can accommodate intensively measured time series data without requiring balanced measurements (patients were measured at different time points since baseline); the MRSS constructs a unified framework by using the exponential distribution family to model a wide range of outcome measures and the outcome correlations between different distributions (e.g., binary, Poisson, Gaussian, etc) are modeled through the dependence on a few common latent states.} The proposed MRSS model allows  inferring patients' heterogeneous latent health states from the observed multi-modal phenotypes while accounting for informative measurement through the joint modeling framework.
Individualized treatment effects (e.g., the short-term and long-term Levodopa treatment effect) are reflected in the latent health state, and each observed outcome can be inferred. The MRSS also allows a forecast of future treatment responses, given the patient's past treatment history and measurements. 

We use importance sampling and Laplace approximation to estimate parameters for non-Gaussian phenotypes. Simulation studies and real data analysis demonstrate that MRSS can handle complex, intensively measured time series and improve estimation accuracy and efficiency than existing methods. Our analysis results demonstrate the feasibility of using remote sensors to track PD patients' symptoms and capture short-term and long-term dopamine therapy treatment effects.

There are some extensions of our method worth further exploration. {We considered discrete time state space models here. In practice, a subject can take multiple measurements daily; thus, an MRSS model on the continuous-time scale may be considered.  With the continuous-time MRSS, additional periodic state variables for the early morning effect or other seasonal effects may be incorporated. 
More efficient ways available to construct the importance sampling density can be considered. For example, \citet{chan2016observed} and \citet{chan2018bayesian} directly
compute the Gaussian approximations based on fast band matrix routines to reduce the computational burden. 
Our real data application models the short-term treatment effect with a linear time state. However, a `delayed on' phenomenon may be present \citep{marsden1976off} and the treatment effect can wear off gradually, in which case, a non-linear transition model can be useful. Additionally, when phenotypes are non-stationary, we may allow the transition matrix to vary with time, and estimation can proceed using a sliding-windows approach. When substantial heterogeneity is present across subjects, we may consider including random effects in system matrices $\blambda$ and $\bT$ to account for between-individual variations beyond fixed effects \citep{liu2011mixed}. }

\bigskip
\begin{center}
{\large\bf SUPPLEMENTARY MATERIAL}
\end{center}

\begin{description}

\item[Supplementary:] Details on  Gaussian and Non-Gaussian state space model, and additional simulations and application section results. 


\end{description}
\section*{REFERENCE}
Reference is at the end of this document.

\renewcommand*\footnoterule{}
\renewcommand{\figurename}{Appendix Figure.}
\renewcommand{\tablename}{Appendix Table.}
\renewcommand*{\thesection}{Appendix~\Alph{section}}
\title{\vspace{-60pt}\bf Supplementary Materials for ``Mixed-Response State-Space Model for Analyzing Multi-Dimensional Digital Phenotypes''}
\date{}

\maketitle
\setlength{\abovedisplayskip}{3pt}
\setlength{\belowdisplayskip}{3pt}
\setlength{\abovedisplayshortskip}{3pt}
\setlength{\belowdisplayshortskip}{3pt} 
\setcounter{section}{0}
\setcounter{figure}{0}
\setcounter{table}{0}

\section{Computational Algorithms for State Space Models}\label{wesec:model}
In this section, we give a detailed account of the computational algorithms of the proposed mixed-response state-space model (MRSS). In \ref{wesec:model}.1, we describe  the Gaussian state space model which gives the initial values for the MRSS in Section 2 of the main paper. In \ref{wesec:model}.2, we describe general algorithms for MRSS that includes non-Gaussian outcomes.

\subsection{Gaussian State Space Model}
When the observation distribution $p$ is a multivariate normal distribution with mean   $\blambda_{it}\balpha_{it}+\bd_{it}$ and variance $\bH$, the model \eqref{eqn:nmodel} in Section~\ref{sec:estimation} becomes a standard $p$-dimensional linear Gaussian state-space model ($p=m_A+m_Y$). We will omit the subscript $i$ when describing the model for $i$th subject in the following sections.
\begin{equation}\label{wbeqn:model}
\begin{alignedat}{2}
    \bZ_{t} &= \blambda_t\balpha_{t}+\bd_t+\beps_t, & & \beps_t \sim \mathrm{N}_p\left(\boldsymbol{0}, \bH\right),  \quad t=1, \ldots, m \\
    \balpha_{t+1} &=\bT \balpha_{t}+\bc+\bfeta_t, \qquad && \bfeta_t  \sim \mathrm{N}_w\left(\boldsymbol{0}, \bQ\right),\\
    &&& \balpha_1 \sim \operatorname{N}_w(\ba_1, \bP_1)
\end{alignedat}  
\end{equation}
where $\bZ_{t}$ is a $p \times 1$ vector of observations called the observation vector and $\balpha_{t}$ is an unobserved $w \times 1$ vector called the state vector ($p=m_A+m_Y$, $w=m_b+m_v$). The first equation of \eqref{wbeqn:model}
is called the observation equation and the second is called the state equation. The error terms $\beps$ and $\bfeta$ are assumed to be serially independent and independent of each other at all time points.  
matrices $\blambda_t, \bd_t, \bT,\bc, \bH$ and $\bQ$ are  assumed to be known. We discuss their estimation in Section A.2. 

Fitting parameters in a Gaussian state space model is the maximum likelihood. The derivation of the likelihood function would rely on recursively using the Kalman Filter and Kalman smoother described here. Denote the set of observations $\bZ_1,\cdots ,\bZ_{t}$ by $\bZ_{(t)}$. The Kalman filter is a recursion for calculating $\ba_{t|t} = \e(\balpha_t|\bZ_{(t)})$,
$\ba_{t+1} = \e(\balpha_{t+1}|\bZ_{(t)})$, $\bP_{t|t} = \var(\balpha_t|\bZ_{(t)}$) and $\bP_{t+1} = \var(\balpha_{t+1}|\bZ_{(t)})$ given $\ba_t$ and $\bP_t$. Specifically, we have \citep{harvey1989time}
\begin{equation}\label{eqn:kalman}
    \begin{aligned}
      \bv_{t} &=\bZ_{t}-\blambda_t \ba_{t}-\bd_{t}, & \bF_{t} &=\blambda_t \bP_{t} \blambda_t^{T}+\bH, \\
      \ba_{t \mid t} &=\ba_{t}+\bP_{t} \blambda_t^{T} \bF_{t}^{-1} \bv_{t}, & \bP_{t \mid t} &=\bP_{t}-\bP_{t} \blambda_t^{T} \bF_{t}^{-1} \blambda_t \bP_{t}, \\
      \ba_{t+1} &=\bT \ba_{t \mid t}+\bc, & \bP_{t+1} &=\bT \bP_{t \mid t} \bT^{T}+ \bQ ^{T}
      \end{aligned}
\end{equation}
for $t=1,\cdots,m$.

The smoother is a recursion for calculating the conditional mean $\hat \balpha_t=\e(\balpha_t|\bZ_{(m)})$ and conditional variance matrix $\bV_t=\var(\balpha_t|\bZ_{(m)})$ for $t=m-1,\cdots ,1$ given full sample $\bZ_{(m)}$. Specifically \citep{de1989smoothing},
\begin{equation}\label{eqn:smoother}
  \begin{aligned}
    \br_{t-1} &=\blambda_t^T\bF_t^{-1}\bv_t+\bL_t^T\br_t, & \bN_{t-1} &=\blambda_t^T \bF_{t}^{-1} \blambda_t^T+\bL_t^T\bN_t\bL_t, \\
    \hat\balpha_t &=\ba_t+\bP_t\br_{t-1}, & \bV_t &=\bP_t-\bP_t\bN_{t-1}\bP_t 
    \end{aligned}
\end{equation}
for $t=m-1,\cdots, 1$ initialized with $\br=\boldsymbol{0}$ and $\bN=\boldsymbol{0}$. We
interpret $\bL^T\cdots \bL^T_{n-1}$ as $\bI$ when $t=n$ and as $\bL^T_{n-1}$ when $t=n-1$.




\subsubsection{Likelihood Function and Estimation for Gaussian State Space Model}\label{wbsec:mle}
We first assume the initial state vector $\balpha_1$ has density $N(\ba_1,\bP_1)$ where $\ba_1$ and $\bP_1$ are known. The log likelihood of $\bZ_{(m)}$ is 
\begin{align*}
  \log L(\bZ_{(m)})&=\sum_{t=1}^m \log p(\bZ_t|\bZ_{(t-1)})
\end{align*}
where $p(\bZ_1|\bZ_{(0)})=p(\bZ_1)$. Putting $\bv_{t} =\bZ_{t}-\blambda_t \ba_{t}-\bd_{t}$, $\bF_{t} =\blambda_t \bP_{t} \blambda_t^{T}+\bH$ and substituting $N_p(\blambda_t\ba_t+ \bd_t, \bF_t)$ for $p(\bZ_{t}|\bZ_{(t-1)})$, we obtain:
\begin{align*}
  \log L(\bZ_{(m)})&=-\frac{mp}{2}\log(2\pi)-\frac{1}{2}\sum_{t=1}^{m}\left(\log(|\bF_t|)+\bv^T_t\bF^{-1}\bv_t\right)
\end{align*}
The quantities $\bv_t$ and $\bF_t$ are calculated routinely by the Kalman filter 
so $\log L(\bZ_{(m)})$ is easily computed from the Kalman filter output.

When $\ba_1$ and $\bP_1$ in the initial state vector $\balpha_1$ density $N(\ba_1,\bP_1)$ are unknown, we can assume $\balpha\sim N(\b0, \kappa \bP_{\infty}+\bP_*)$ where $\bP_{\infty}$ is a diagonal matrix with ones
on the diagonal and $\kappa\to \infty$. Following \citet{de1991diffuse}, the log likelihood (denoted as $L_d$) now becomes:
\begin{align*}
  \log L_{d}\left(\bZ_{(m)}\right)=\lim _{\kappa \rightarrow \infty}\left[\log L\left(\bZ_{(m)}\right)+\frac{q}{2} \log \kappa\right].
\end{align*}

Let $\bpsi$ denote the vector of all unknown parameters. We will estimate $\bpsi$ by maximum likelihood. In the diffuse case we shall take it for granted that for models of interest, estimates of $\bpsi$ obtained by maximizing $\log L(\bZ_{(m)}|\bpsi)$ for fixed $\kappa$ converge to the estimates obtained by maximizing the diffuse log-likelihood $\log L_d(\bZ_{(m)}|\bpsi)$ as $\kappa \to \infty$. In order to estimate parameters, one can use Newton's method to solve the equation:
\begin{align*}
  \frac{\partial\log L(\bZ_{(m)}|\bpsi)}{\partial \bpsi}=0.
\end{align*}

\subsubsection{Matrix Form of Gaussian State Space Model}\label{wbsec:matrix}
Finally, we introduce the matrix form of the Gaussian state space model. These results are useful when we generalize the model to non-Gaussian outcomes.
The Gaussian state space model \eqref{wbeqn:model} can be represented in a matrix form. The observation equation takes the form:
\begin{align}\label{eqn:mat_obs}
  \bZ_{(m)}=\blambda^* \balpha^* +\bd^*+ \beps^*, \qquad \beps^*\sim N(\b0, \bH^*)
\end{align}
where
\begin{gather*}
  \bZ_{(m)}=\left(\begin{array}{c}
  \bZ_{1} \\
  \vdots \\
  \bZ_{m}
  \end{array}\right), \quad \blambda^*=\left(\begin{array}{llll}
  \blambda_{1} & & \b0 & \b0 \\
  & \ddots & & \vdots \\
  \b0 & & \blambda_{m} & \b0
  \end{array}\right), \quad \balpha^*=\left(\begin{array}{c}
  \balpha_{1} \\
  \vdots \\
  \balpha_{m} \\
  \balpha_{m+1}
  \end{array}\right),
  \end{gather*}
  \begin{gather*}
  \bd^*=\left(\begin{array}{c}
    \bd_{1} \\
    \vdots \\
    \bd_{m}
    \end{array}\right), \quad
  \beps^*=\left(\begin{array}{c}
  \beps_{1} \\
  \vdots \\
  \beps_{m}
  \end{array}\right), \quad \bH^*=\left(\begin{array}{ccc}
  \bH & & \b0 \\
  & \ddots & \\
  \b0 & & \bH
  \end{array}\right)
\end{gather*}

The state equation takes the form:
\begin{align}\label{eqn:mat_state}
  \balpha^*=\bT^*(\balpha_1^*+\bc^*+\bR^*\bfeta^*), \qquad \bfeta^*\sim N(\b0, \bQ^*),
\end{align}
where 
\begin{gather*}
  \bT^*=\left(\begin{array}{rrrrrrr}
    \bI & \b0 & \b0 & \b0 & & \b0 & \b0 \\
    \bT & \bI & \b0 & \b0 & & \b0 & \b0 \\
    \bT^2 & \bT & \bI & \b0 & & \b0 & \b0 \\
    \bT^3 & \bT^2 & \bT & \bI & & \b0 & \b0 \\
    & & & & \ddots & & \vdots \\
    \bT^{m-1}  & \bT^{m-2} & \bT^{m-3}  & \bT^{m-4}  & & \bI & \b0 \\
    \bT^{m}  & \bT^{m-1}  & \bT^{m-2} & \bT^{m-3}  & \cdots & \bT & \bI
    \end{array}\right),
    \end{gather*}
    \begin{gather*}
  \balpha_{1}^{*}=\left(\begin{array}{c}
  \balpha_{1} \\
  \b0 \\
  \b0 \\
  \vdots \\
  \b0
  \end{array}\right), \quad \bR^*=\left(\begin{array}{cccc}
  \b0 & \b0 & \cdots & \b0 \\
  \bI & \b0 & & \b0 \\
  \b0 & \bI & & \b0 \\
  & & \ddots & \vdots \\
  \b0 & \b0 & \cdots & \bI
  \end{array}\right) ,\quad  \bc^*=\left(\begin{array}{c}
    \b0\\
    \bc \\
    \vdots \\
    \bc
    \end{array}\right),
      \end{gather*}
    \begin{gather*}
  \bfeta^*=\left(\begin{array}{c}
  \bfeta_{1} \\
  \vdots \\
  \bfeta_{m}
  \end{array}\right), \quad \bQ^*=\left(\begin{array}{ccc}
  \bQ & & \b0 \\
  & \ddots & \\
  \b0 & & \bQ
  \end{array}\right).
\end{gather*}
It follows that $\e(\balpha_1^*)=\ba_1^*$, $\var(\balpha_1^*)=\bP_1^*$ with
\begin{gather*}
  \ba_{1}^{*}=\left(\begin{array}{c}
    \ba_{1} \\
    \b0 \\
    \b0 \\
    \vdots \\
    \b0
    \end{array}\right), \quad \bP_{1}^{*}=\left(\begin{array}{ccccc}
    \bP_{1} & \b0 & \b0 & \cdots & \b0 \\
    \b0 & \b0 & \b0 & & \b0 \\
    \b0 & \b0 & \b0 & & \b0 \\
    \vdots & & & \ddots & \\
    \b0 & \b0 & \b0 & & \b0
    \end{array}\right) 
\end{gather*}

Furthermore, we show that the observation vectors $\bZ_{(t)}$ are linear functions of the initial state vector $\balpha_1^*$ and the disturbance vectors $\beps^*$ and $\bfeta^*$  since it follows by substitution of \eqref{eqn:mat_state} into \eqref{eqn:mat_obs} that
\begin{align*}
  \bZ_{(m)}=\blambda^* \bT^*(\balpha_1^*+\bc^*+\bR^*\bfeta^*) +\bd^*+ \beps^*
\end{align*}
Marginally, we have:
\begin{gather*}
  \e(\balpha^*)=\ba^*=\bT^*\ba_1^*+\bT^*\bc^*, \quad \var(\balpha^*)=\bV^*=\bT^*(\bP_1^*+\bR^*\bQ^*\bR^{*T})\bT^{*T}\\
  \e(\bZ_{(m)})=\bmu^*=\blambda^*\bT^*(\ba_1^*+\bc^*)+\bd^*, \quad \var(\bZ_{(m)})=\bOmega^*=\blambda^*\bT^*(\bP_1^*+\bR^*\bQ^*\bR^{*T})\bT^{*T}\blambda^{*T}+\bH^*
\end{gather*}
Therefore, the corresponding density functions are:
\begin{align}
  \log p(\balpha^*)&=Const-\frac{1}{2}\log|\bV^*|-\frac{1}{2}(\balpha^*-\ba^*)^T\bV^{*-1}(\balpha^*-\ba^*)\\
  \log p(\btheta^*)&=Const-\frac{1}{2}\log|\bSigma^*|-\frac{1}{2}(\btheta^*-\bmu^*)^T\bSigma^{*-1}(\btheta^*-\bmu^*)\label{wbeqn:density}\\
  \log p(\bZ_{(m)})&=Const-\frac{1}{2}\log|\bOmega^*|-\frac{1}{2}(\bZ_{(m)}-\bmu^*)^T\bOmega^{*-1}(\bZ_{(m)}-\bmu^*)\label{wbeqn:like}\\
  \log p(\bZ_{(m)}|\balpha^*)&=Const-\frac{1}{2}\log|\bH^*|-\frac{1}{2}(\bZ_{(m)}-\btheta^*)^T\bH^{*-1}(\bZ_{(m)}-\btheta^*)\label{wbeqn:105}\\
  \log p(\balpha^*, \bZ_{(m)})&=\log p(\balpha^*)+\log p(\bZ_{(m)}|\balpha^*)
\end{align}
  where $\btheta^*=\blambda^*\balpha^*+\bd^*$ is called as the signal, and $\bSigma^*=\var(\btheta^*)=\blambda^*\bV^*\blambda^{*T}$.

Applying the conditional Gaussian distribution theorem, we obtain the filtering/smoothing formula in matrix form.  For Gaussian State Space model \eqref{wbeqn:model}, the conditional mean and variance of the signal $\btheta^*=\blambda^*\balpha^*+\bd^*$ given all observations $\bZ_{(m)}$ is:
\begin{align}
  \hat{\btheta^*}&=\e(\btheta^*|\bZ_{(m)})=\bmu^*+\bSigma^*\bOmega^{*-1}(\bZ_{(m)}-\bmu^*)
  =(\bSigma^{*-1}+\bH^{*-1})^{-1}(\bSigma^{*-1}\bmu^*+\bH^{*-1}\bZ_{(m)}),\label{eqn:mat_filter}\\
  \text{and}\notag\\
  &\var(\btheta^*|\bZ_{(m)})=\bSigma^*-\bSigma^*\bOmega^{*-1}\bSigma^*.
\end{align}

In addition, the state filter/smooth is:
$$\hat\balpha=\e(\balpha|\bZ_{(m)})=\bT\balpha_1^*+\bT\bc^*+\bT(\bP_1^*+\bR\bQ\bR^T)\bT^T\blambda^T(\bOmega^{-1}(\bZ_{(m)}-\bmu)).$$

\subsection{Non-Gaussian State Space Model}
Now suppose we have non-Gaussian observations:
\begin{equation}
  \begin{alignedat}{2}
      \bZ_{t} &= p(\bZ_{t}|\blambda_t\balpha_{t}+\bd_t),  \qquad & & \\
      \balpha_{t+1} &=\bT \balpha_{t}+\bc+ \bfeta_{t}, \qquad && \bfeta_{t}  \sim \mathrm{N}_w\left(\boldsymbol{0}, \bQ\right),  \quad t=1, \ldots, m,\\
      &&& \balpha_1 \sim \operatorname{N}_w(\ba_1, \bP_1)
  \end{alignedat}  
  \end{equation}
The signal $\btheta_t=\blambda_t\balpha_{t}+\bd_t$. 

\subsubsection{Mode Approximation of Non-Gaussian Responses}\label{wbsec:approx}
When $\bZ_t|\btheta_t$ is not normally distributed, it is not easy to estimate the smoothing states $\e(\btheta^*|\bZ_{(m)})$ directly. Instead, here we aim to estimate the mode for the smoothed density where the signal is linear and Gaussian. 
Therefore, we express the smoothed log density by
\begin{align}\label{eqn:smoothp}
  \log p(\btheta^*|\bZ_{(m)}) = \log p(\bZ_{(m)}|\btheta^*) + \log p(\btheta^*)-\log p(\bZ_{(m)})
\end{align}
and maximize this expression with respect to $\btheta^*$ numerically using the Newton-Raphson method. The expression of $\log p(\btheta^*)$ is \eqref{wbeqn:density} since $\btheta^*$ is linear and Gaussian.  
The first and second derivatives of $\log p(\btheta^*|\bZ_{(m)})$ are:
\begin{align*}
  \frac{\partial\log p(\btheta^*|\bZ_{(m)})}{\partial\btheta^*}&=\frac{\partial\log p(\bZ_{(m)}|\btheta^*)}{\partial\btheta^*}-\bSigma^{*-1}(\btheta^*-\bmu^*)\\
  \frac{\partial^2\log p(\btheta^*|\bZ_{(m)})}{\partial\btheta^{*2}}&=\frac{\partial^2\log p(\bZ_{(m)}|\btheta^*)}{\partial\btheta^{*2}}-\bSigma^{*-1}
\end{align*}
For a given guess of the mode, say $\tilde \btheta^{*(i)}$, a new estimation of the mode $\btheta^*$ is given by
\begin{align}
  \tilde \btheta^{*(i+1)}&=\tilde \btheta^{*(i)}-\left(\frac{\partial^2\log p(\bZ_{(m)}|\btheta^*)}{\partial\btheta^{*2}}\middle|_{\btheta^*=\tilde\btheta^{*(i)}}-\bSigma^{*-1}\right)^{-1}\left(\frac{\partial\log p(\bZ_{(m)}|\btheta^*)}{\partial \btheta^*}\middle|_{\btheta^*=\tilde\btheta^{*(i)}}-\bSigma^{*-1}(\tilde\btheta^{*(i)}-\bmu^*)\right)\\
  &=(\bSigma^{*-1}+\bA^{-1})^{-1}(\bA^{-1}\btheta^{\dagger(i)}+\bSigma^{*-1}\bmu^*)\label{eqn:max}
\end{align}
where 
\begin{align*}
  \bA^{-1} = - \frac{\partial^2\log p(\bZ_{(m)}|\btheta^*)}{\partial\btheta^{*2}} \bigg|_{\btheta^*=\tilde \btheta^{*(i)}}, \quad \btheta^{\dagger(i)}=\tilde \btheta^{*(i)}+\bA \frac{\partial\log p(\bZ_{(m)}|\btheta^*)}{\partial\btheta^*}\bigg|_{\btheta^*=\tilde \btheta^{*(i)}}
\end{align*}

This optimization can alternatively be derived by matching the first and second derivatives of the smoothing density $\log p(\btheta^*|\bZ_{(m)})$ with a smoothed approximating density of the linear Gaussian model $\log g(\btheta^*|\bZ_{(m)})$. Similar to \eqref{eqn:smoothp}, the log density $\log g(\btheta^*|\bZ_{(m)})$ can be decomposed as:
\begin{align}
  \log g(\btheta^*|\bZ_{(m)}) = \log g(\bZ_{(m)}|\btheta^*) + \log g(\btheta^*)-\log g(\bZ_{(m)})
\end{align}
where $\log g(\bZ_{(m)}|\btheta^*)$ is the log density of the observation and is defined by \eqref{wbeqn:105}; $\log g(\btheta^*)$ is given by \eqref{wbeqn:density}; $\log g(\bZ_{(m)})$ is the log density of observation and does not depend on $\btheta^*$. 

We want to match the first and second derivatives of the densities $g(\btheta^*|\bZ_{(m)})$ with $p(\btheta^*|\bZ_{(m)})$ with respect to $\btheta^*$. This is equivalent to matching the two derivatives of  $g(\bZ_{(m)}|\btheta^*)$ with $p(\bZ_{(m)}|\btheta^*)$:
\begin{align*}
  \frac{\partial\log g(\bZ_{(m)}|\btheta^*)}{\partial\btheta^*}&=\bH^{*-1}(\bZ_{(m)}-\btheta^*)\approx\frac{\partial\log p(\bZ_{(m)}|\btheta^*)}{\partial\btheta^*}\\
  \frac{\partial^2\log g(\bZ_{(m)}|\btheta^*)}{\partial\btheta^{*2}}&=-\bH^{*-1}\approx\frac{\partial^2\log p(\bZ_{(m)}|\btheta^*)}{\partial\btheta^{*2}}
\end{align*}
We solve the equations by iteration.
\begin{align*}
  \bZ_{(m)}\approx\btheta^*+\bH^{*-1}\frac{\partial\log p(\bZ_{(m)}|\btheta^*)}{\partial\btheta^*},\qquad
  \bH^*\approx-\left\{\frac{\partial^2\log p(\bZ_{(m)}|\btheta^*)}{\partial\btheta^{*2}}\right\}^{-1}
\end{align*}

Therefore, for a given $\hat\btheta^{*(i)}$, we regard $\left(\tilde \btheta^{*(i)}+\bA \frac{\partial\log p(\bZ_{(m)}|\btheta^*)}{\partial\btheta^*}\Big|_{\btheta^*=\tilde \btheta^{*(i)}}\right)$ as the response ``$\bZ_{(m)}$'' which follows a normal distribution $g$ with observation disturbance variance matrix $-\left\{\frac{\partial^2\log p(\bZ_{(m)}|\btheta^*)}{\partial\btheta^{*2}}\right\}^{-1}$ and apply Kalman filter/smoother \eqref{eqn:mat_filter} to update $\hat \btheta^{*(i+1)}$. The process leads to an iteration
from which the final linearized model is the linear Gaussian model with the same
conditional mode of $\btheta^*$ given $\bZ_{(m)}$ as the non-Gaussian nonlinear model.
In short, the expectation of approximating Gaussian densities $g(\btheta^*|\bZ_{(m)})$ that matches the first and second derivatives with $p(\btheta^*|\bZ_{(m)})$ wrt $\btheta^*$ is equivalent to the mode of  $p(\btheta^*|\bZ_{(m)})$ by maximizing the smoothed density of the signal $p(\btheta^*|\bZ_{(m)})$ wrt $\btheta^*$.

In a similar manner, we are able to conduct state filtering/smoothing and disturbance filtering/smoothing by estimating the conditional mode of $\balpha^*$ and $\bfeta^*$.  The mode is obtained by repeated linearization of the smoothed density around a trial estimate of its mode. The linearization is based on an approximating linear model and allows the use of the Kalman filter and is smoother.


Section~\ref{sec:estimation} in the main manuscript describes the likelihood computation and optimization for the non-Gaussian state space model. In addition, when $\ba_1$ and $\bP_1$ in the initial state vector $\balpha_1$ density $N(\ba_1,\bP_1)$ are unknown, \citet{durbin2012time} introduced the diffuse initialization in Chapter 11.4.4. They showed that the main results in Section~\ref{wbsec:approx} still hold with some modifications.

\section{Definition of Prediction Errors in Simulation Study}\label{wbsec:prediction}
For the VAR model, we define the within-sample prediction error as follows:
\begin{align*}
\text{$\boldsymbol{Y}^{(1)}$ (Binomial): }&\sum_{t=2}^{25} \left(\operatorname{logit}(\hat{\tilde Y}^{(1)}_t) - \mu^{(1)}_t\right)^2/24,\\
\text{$\boldsymbol{Y}^{(2)}$ (Poisson): }&\sum_{t=2}^{25} \left(\hat{\tilde Y}^{(2)}_t - \mu^{(2)}_t\right)^2/24,\\
\text{$\boldsymbol{Y}^{(3)}$ (Gaussian): }&\sum_{t=2}^{25} \left(\hat{\tilde Y}^{(3)}_t - \mu^{(3)}_t\right)^2/24.
\end{align*}
We measure the loss in the natural parameter space. 
Note that for $\boldsymbol{Y}^{(2)}$, we already model it in the logarithmic scale as mentioned. For the proposed method, we define the within-sample prediction error as follows:
\begin{align*}
\text{$\boldsymbol{Y}^{(1)}$ (Binomial): }&\sum_{t=2}^{25} \left(\hat{\mu}^{(1)}_t - \mu^{(1)}_t\right)^2/24,\\
\text{$\boldsymbol{Y}^{(2)}$ (Poisson): }&\sum_{t=2}^{25} \left(\hat{\mu}^{(2)}_t - \mu^{(2)}_t\right)^2/24,\\
\text{$\boldsymbol{Y}^{(3)}$ (Gaussian): }&\sum_{t=2}^{25} \left(\hat{\mu}^{(3)}_t- \mu^{(3)}_t\right)^2/24,
\end{align*} 
where $\hat{\mu}^{(1)}, \hat{\mu}^{(2)}, \hat{\mu}^{(3)}$ are predicted parameters in the exponential family. Similarly, we could define the out-of-sample prediction error of the last $5$ time points for VAR and MSS models.

\section{Feature Extraction in mPower Study}\label{wbsec:preprocessing}
A complete list of extracted phenotypes: \\
\noindent\textbf{Walking Modality:}
The walking activity evaluates participants' gait and balance. Participants are asked to walk 20 steps in a straight line (waling stage), turn around, and stand still for 30 seconds (resting stage).  In the walking stage, we can obtain the readouts of the accelerometer in $X$, $Y$, $Z$ directions. We can also define the combined reading as 
$
	C_1 = \sqrt{X^2+Y^2+Z^2},$ and 
	$C_2=\sqrt{(X-\bar X)^2+(Y-\bar Y)^2+(Z-\bar Z)^2}.
$
For each of five time series, the following features are extracted as suggested by \citet{snyder2020mhealthtools}: 1) Mean, standard deviation, mode, skewness, kurtosis, median,  first/third quartile, interquartile range, coefficient of variation; 2) Autocorrelation at lag one; 3) Detrended fluctuation analysis of the acceleration series; 4) Teager-kaiser energy operator of the acceleration series; 5) Features from Least-squares spectral analysis (frequency, maximum power, etc). In the resting stage, we extract the above features 1), 2), and 3) from readouts $C_1$ and $C_2$.

\noindent\textbf{Voice Modality:} Speech disorders have been linked to PD and there is strong supporting evidence of degrading performance in voice with PD progression \citep{ho2008better, skodda2009progression}. In the study, participants recorded sustained phonation by saying `Aaaaah' into the microphone at a steady volume for up to 10 s. We aim to infer properties of the speech signals, extracting useful distinguishing features which are altered as the orchestrated muscle movements involved in voice production. We extract  more than three hundred features ($339$) from each voice recording as follows: 
1)  Various pitch detection algorithms that quantify the cycle-to-cycle changes in fundamental frequency \citep{tsanas2011nonlinear, camacho2008sawtooth}. 
2) Glottal Quotient (GQ) calculated by the DYPSA algorithm that is similar to  pitch detection, taking into account the length of time that vocal folds are apart (glottis is open) or in collision (glottis is closed) \citep{naylor2006estimation}. 
3) Various shimmer measurements that quantify the cycle-to-cycle changes in amplitude \citep{tsanas2011nonlinear}. 
4) Harmonics-to-Noise Ratio (HNR) quantifies the ratio of actual signal information over noise \citep{boersma2006praat}.
5) The Glottal-to-Noise Excitation Ratio (GNE) aims to express the amount of noise in the speech signal \citep{michaelis1997glottal}. 
6) Coefficients of Linear Predicting Coding (LPCC) that indicates the accuracy of prediction of the current data sample as a function of the preceding samples \citep{chatfield2019analysis}. 
7)  Mel-Frequency Cepstral Coefficients (MFCC) that are the Mel frequency scales in combination with cepstral analysis and often serve as the reference standard feature for speaker identification and automatic speech recognition \citep{mermelstein1976distance, VOICEBOX}. 
8) Recurrence Period Density Entropy (RPDE) that addresses the ability of the vocal folds to sustain stable vocal fold oscillation \citep{kantz2004nonlinear}. 
9) Wavelet measures that are used in the study of fractal properties and self-similarity of signals, and characteristics of speech signals. They are widely used in developing dysphonia measures \citep{tsanas2010new}. 
10) Some variants and combinations of the aforementioned measures \citep{tsanas2009accurate}.

\noindent\textbf{Tapping:}
Finger tapping can be used to measure bradykinesia, one of the hallmarks of PD symptoms. In the tapping task, participants were instructed to lay their phone on a flat surface and to use two fingers on the same hand to alternatively tap two stationary points on the screen for 20 seconds.  The software recorded  where and when the participant tapped on the iPhone screen. 
Based on these records, we extracted the following features. For the time between two tappings, we calculated 1) mean, standard deviation, skewness, kurtosis, median,  first/third quartile, interquartile range, coefficient of variation, 2) autocorrelation (lag=1), and 3) Teager-kaiser energy operator. For the tapping position drift from the center, we calculated the mean, standard deviation, skewness, kurtosis, median,  interquartile range, and coefficient of variation. We also recorded the number of tappings in $20$s.

\section{Modeling Details in mPower Study}
\subsection{Memory Model}\label{wbsec:memory}
We model the memory game score and its corresponding measurement indicators (measurement taken before or after Levodopa treatment) jointly. Response $\bA_{it}\in \mathbb{R}$ is the measurement indicator for the phenotype (A\_memory) and is modeled as a binomial distribution. Response $\bY_{it}\in \mathbb{R}$ is the game score. The values in $\bY_{it}$ are standardized and assumed to be normally distributed. We let $t=1,2,\cdots, m_i$ represent semi-daily transitions.

There are three latent states: $v^{(1)}_{it}$ reflects the health status; $b^{(1)}_{it}$ is the treatment effect of Levodopa; and $b^{(2)}_{it}$ is a latent state to model the `early-morning akinesia' effect \citep{marsden1976off}. The transition matrix of the four states is defined as
    \begin{align}
        \begin{pmatrix}
            b^{(1)}_{it}\\b^{(2)}_{it}\\v^{(1)}_{it}
        \end{pmatrix} = \bT \begin{pmatrix}
            b^{(1)}_{i,t-1}\\b^{(2)}_{i,t-1}\\v^{(1)}_{i,t-1}
        \end{pmatrix} + \begin{pmatrix}
            \varepsilon_{b^{(1)}}\\\varepsilon_{b^{(2)}}\\\varepsilon_{v^{(1)}}
        \end{pmatrix},
    \end{align}
where $\bT\in\mathbb{R}^{3\times 3}$ is a diagonal transition matrix and it takes different values for PD patients taking Levodopa,  PD patients not taking Levodopa, and healthy controls; $(  \varepsilon_{b^{(1)}}, \varepsilon_{b^{(2)}}, \varepsilon_{v^{(1)}})^T\sim N_{3}(\boldsymbol{\mu}_\varepsilon, \bSigma_\varepsilon)$.

For PD patients taking Levodopa, 
the observation equation is as follows:
    \begin{eqnarray*}
    \binom{\bA_{it}}{\bY_{it}}\sim p\left(\binom{\bA_{it}}{\bY_{it}}\middle|\begin{pmatrix}
        \textbf{0} & \blambda_{Ab^{(2)}}\operatorname{diag}(\boldsymbol{M}_{it}) & \blambda_{Av^{(1)}}  \\
        \blambda_{Yb^{(1)}}\operatorname{diag}(\ba_{it}) & \blambda_{Yb^{(2)}}\operatorname{diag}(\boldsymbol{M}_{it}) & \blambda_{Yv^{(1)}}
    \end{pmatrix} \begin{pmatrix}
        b^{(1)}_{it}\\b^{(2)}_{it}\\v^{(1)}_{it}
    \end{pmatrix}+\bbeta\bX_{it}  \right),
    \end{eqnarray*}
where $\ba_{it}\in\mathbb{R}$ is the treatment indicator;  $\boldsymbol{M}_{it}\in\mathbb{R}$ is the indicator of whether the treatment is taken in the early morning; $\blambda_{Ab^{(2)}},\blambda_{Av^{(1)}},\blambda_{Yb^{(1)}},\blambda_{Yb^{(2)}}, \blambda_{Yv^{(1)}}$ are the corresponding loading parameters. 
The covariates $\Xb_{it}$ include `age', `gender', `smoking history' (years of smoking before the study), `PD history' (years from diagnosis of PD), and `deep brain stimulation' (whether a patient had deep brain stimulation or not). 

For PD patients who didn't take treatment, treatment effect state $b_{it}^{(1)}$ does not affect their measurements, and response $\bA_{it}$ is not needed for modeling, which leads to the following model structure:
\begin{eqnarray*}
   {\bY_{it}}\sim p\left({\bY_{it}}\middle|\begin{pmatrix}
   \blambda_{Yb^{(2)}}\operatorname{diag}(\boldsymbol{M}_{it}) & \blambda_{Yv^{(1)}}
   \end{pmatrix} \begin{pmatrix}
      b^{(2)}_{it}\\v^{(1)}_{it}
   \end{pmatrix}+\bbeta\bX_{it}  \right).
\end{eqnarray*}

For healthy controls, the structure of the observation model is the same as PD patients without treatments, but covariate $\Xb_{it}$ only contains `age', `gender', and `smoking history' (years of smoking before the study). 

\subsection{Voice Model}\label{wbsec:voice}
We model eight sound features and their corresponding measurement indicators (measurements taken before or after Levodopa treatment) jointly. Response $\bA_{it}\in \mathbb{R}$ is the measurement indicator for the phenotype (A\_voice) and is modeled as a binomial distribution. Response $\bY_{it}\in \mathbb{R}^8$ is the sound feature. The values in $\bY_{it}$ are standardized and assumed to be normally distributed.  We let $t=1,2,\cdots, m_i$ represent semi-daily transitions.

There are five latent states: $v^{(1)}_{it}$, $v^{(2)}_{it}$ reflect the health status; $b^{(1)}_{it}$, $b^{(2)}_{it}$ are the treatment effects, and $b^{(3)}_{it}$ is a latent state to model the `early-morning akinesia' effect \citep{marsden1976off}. The transition matrix of the four states is defined as
    \begin{align}
        \begin{pmatrix}
            b^{(1)}_{it}\\b^{(2)}_{it}\\b^{(3)}_{it}\\v^{(1)}_{it}\\v^{(2)}_{it}
        \end{pmatrix} = \bT \begin{pmatrix}
          b^{(1)}_{i,t-1}\\b^{(2)}_{i,t-1}\\b^{(3)}_{i,t-1}\\v^{(1)}_{i,t-1}\\v^{(2)}_{i,t-1}
        \end{pmatrix} + \begin{pmatrix}
            \varepsilon_{b^{(1)}}\\\varepsilon_{b^{(2)}}\\\varepsilon_{b^{(3)}}\\
            \varepsilon_{v^{(1)}} \\\varepsilon_{v^{(2)}}
        \end{pmatrix},
    \end{align}
where $\bT\in\mathbb{R}^{5\times 5}$ is a diagonal transition matrix and it takes different values for PD patients taking Levodopa,  PD patients not taking Levodopa, and healthy controls; $(  \varepsilon_{b^{(1)}}, \varepsilon_{b^{(2)}}, \varepsilon_{b^{(3)}},\varepsilon_{v^{(1)}},\varepsilon_{v^{(2)}})^T\sim N_{5}(\boldsymbol{\mu}_\varepsilon, \bSigma_\varepsilon)$ and $\varepsilon_{b^{(1)}} \indep \varepsilon_{b^{(2)}}$, $\varepsilon_{v^{(1)}} \indep \varepsilon_{v^{(2)}}$.

For PD patients taking Levodopa, 
the observation equation is as follows:
    \begin{eqnarray*}
    \binom{\bA_{it}}{\bY_{it}}\sim p\left(\binom{\bA_{it}}{\bY_{it}}\middle|\begin{pmatrix}
        \textbf{0} & \blambda_{Ab^{(3)}}\operatorname{diag}(\boldsymbol{M}_{it}) & \blambda_{Av^{(1)}} & \blambda_{Av^{(2)}}\\
        \blambda_{Yb^{(1,2)}}\operatorname{diag}(\ba_{it}) & \blambda_{Yb^{(3)}}\operatorname{diag}(\boldsymbol{M}_{it}) & \blambda_{Yv^{(1)}}&\blambda_{Yv^{(2)}}
    \end{pmatrix} \begin{pmatrix}
        b^{(1)}_{it}\\b^{(2)}_{it}\\b^{(3)}_{it}\\v^{(1)}_{it}\\v^{(2)}_{it}
    \end{pmatrix}+\bbeta\bX_{it}  \right),
    \end{eqnarray*}
where $\ba_{it}\in\mathbb{R}^8$ is the treatment indicator for the corresponding $8$ phenotypes;  $\boldsymbol{M}_{it}\in\mathbb{R}^8$ is the indicator of whether the treatment is taken in the early morning for the corresponding $8$ phenotypes; $\blambda_{Ab^{(3)}},\blambda_{Av^{(1)}},\blambda_{Av^{(2)}},\blambda_{Yb^{(1,2)}},\blambda_{Yb^{(3)}}, \blambda_{Yv^{(1)}},\blambda_{Yv^{(2)}}$ are the corresponding loading parameters. 
The covariates $\Xb_{it}$ include `age', `gender', `smoking history' (years of smoking before the study), `PD history' (years from diagnosis of PD), and `deep brain stimulation' (whether a patient had deep brain stimulation or not). 

For PD patients who didn't take treatment, treatment effect states $b_{it}^{(1)}$, $b_{it}^{(2)}$ and response $\bA_{it}$ do not exist, which leads to the following model structure:
\begin{eqnarray*}
   {\bY_{it}}\sim p\left({\bY_{it}}\middle|\begin{pmatrix}
   \blambda_{Yb^{(3)}}\operatorname{diag}(\boldsymbol{M}_{it}) & \blambda_{Yv^{(1)}}&\blambda_{Yv^{(2)}}
   \end{pmatrix} \begin{pmatrix}
      b^{(3)}_{it}\\v^{(1)}_{it}\\v^{(2)}_{it}
   \end{pmatrix}+\bbeta\bX_{it}^{}  \right).
\end{eqnarray*}

For healthy controls, we only include one health status state $v^{(1)}_{it}$ and covariates $\bX_{it}$ excluding `PD history' and `deep brain stimulation'.
The observation model is as follows:
   \begin{eqnarray*}
   {\bY_{it}}\sim p\left({\bY_{it}}\middle|\begin{pmatrix}
    \blambda_{Yb^{(3)}}\operatorname{diag}(\boldsymbol{M}_{it}) &\blambda_{Yv^{(1)}}
    \end{pmatrix} 
    \begin{pmatrix}
    b^{(3)}_{it}\\v^{(1)}_{it}
   \end{pmatrix}+\bbeta^{} \bX_{it}^{} \right).
   \end{eqnarray*}

\newpage
\def\spacingset#1{\renewcommand{\baselinestretch}%
{#1}\small\normalsize} \spacingset{1}
\section*{Appendix Figures \& Tables}

\begin{figure}[!ht]
  \centering	\includegraphics[width=1\textwidth]{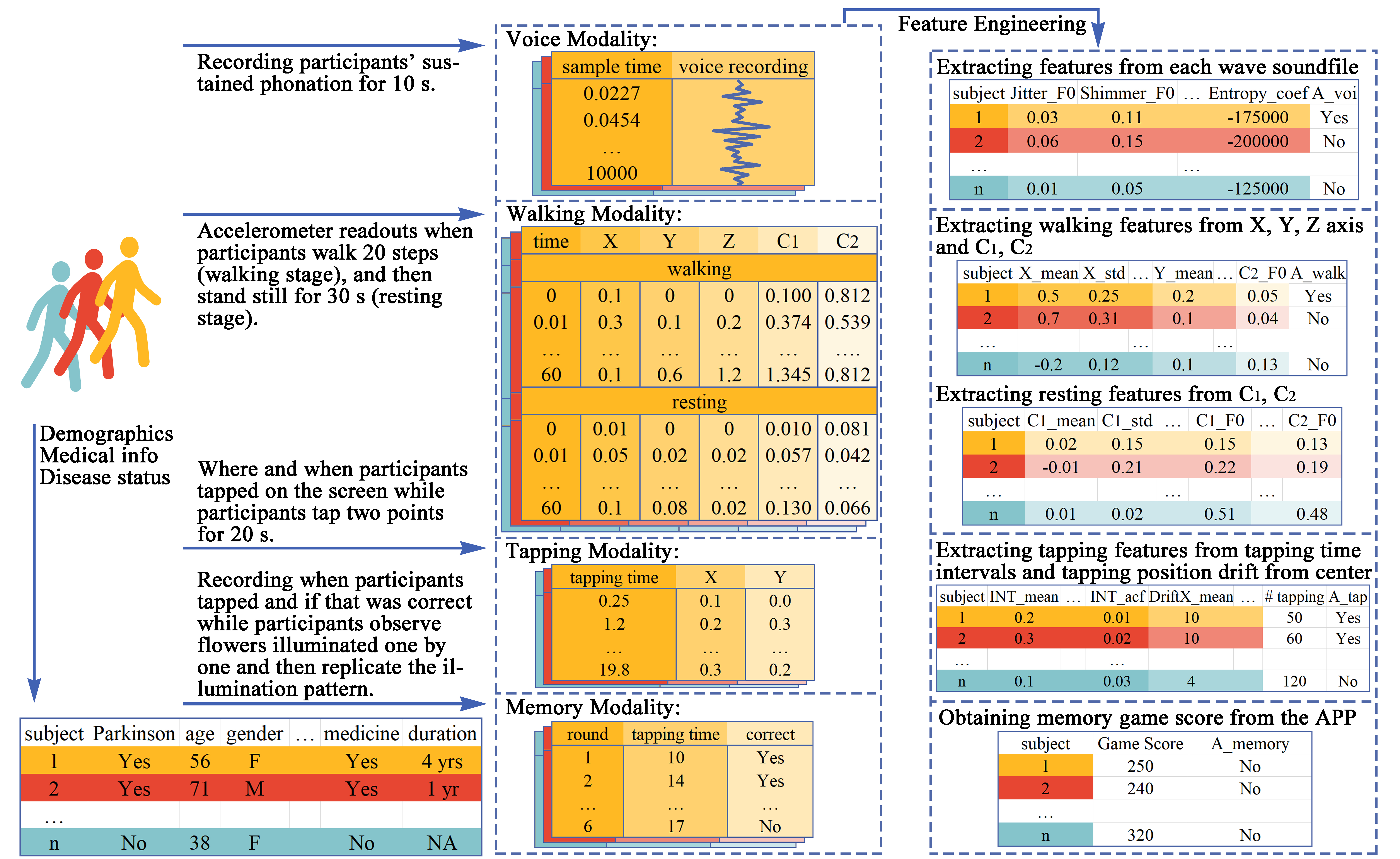}
  \caption{Remote data collection and feature engineering process in mPower study.}\label{wbfig:pipe}
\end{figure}

\begin{figure}[!ht]
  \centering\includegraphics[width=0.7\textwidth]{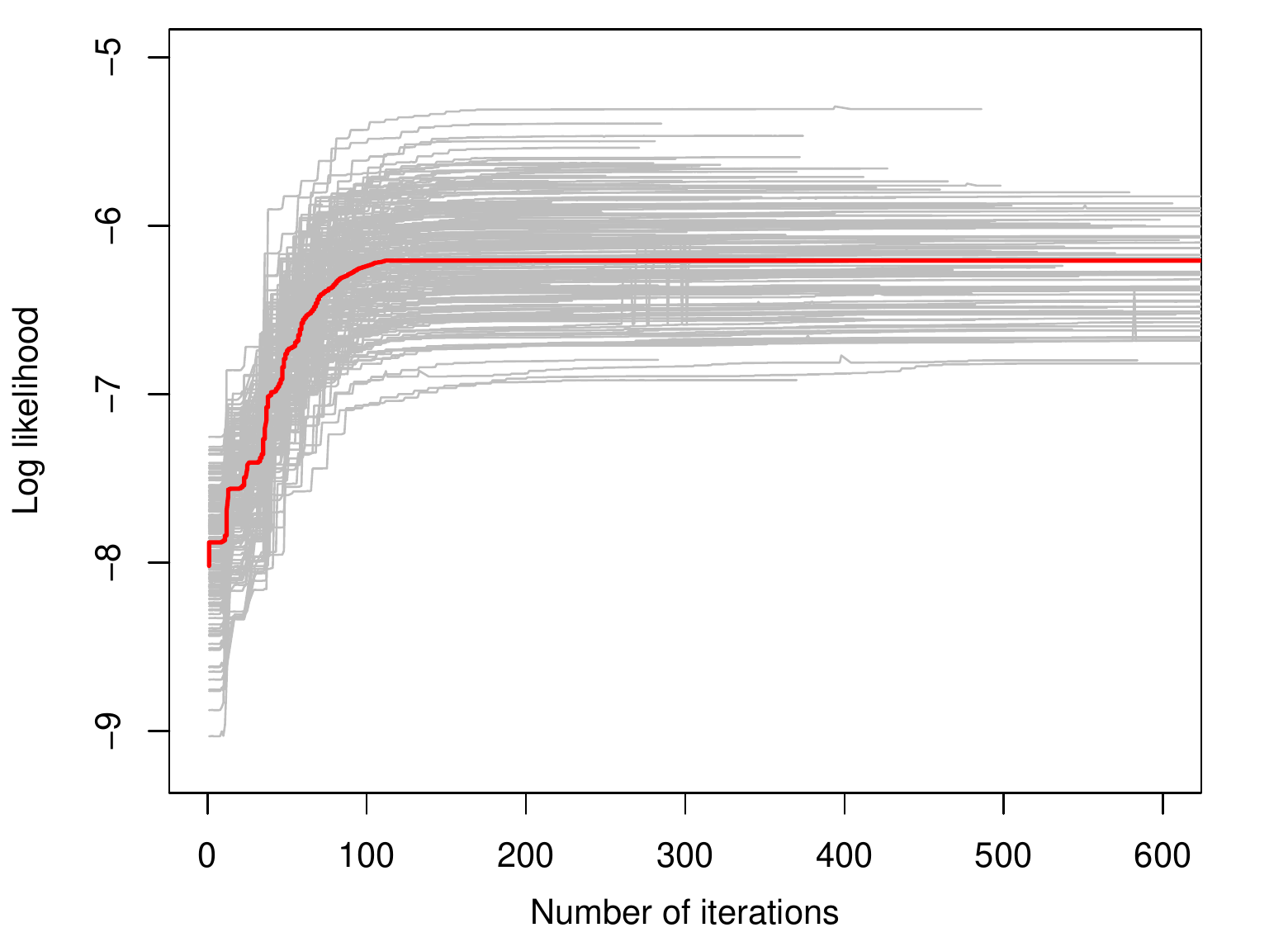}
  \caption{Likelihood change of $200$ repetitions in the simulation study ($N=40$, $T=30$, $p=0.45$). }\label{wbfig:conv}
\end{figure}

\begin{figure}[!ht]
    \parbox{0.55\textwidth}{
        \textbf{(a)} Response $\boldsymbol{Y}^{(2)}$; Varying sample size (N)\\
        \includegraphics[clip, trim=16pt 16pt 0 50pt, width=0.5\textwidth]{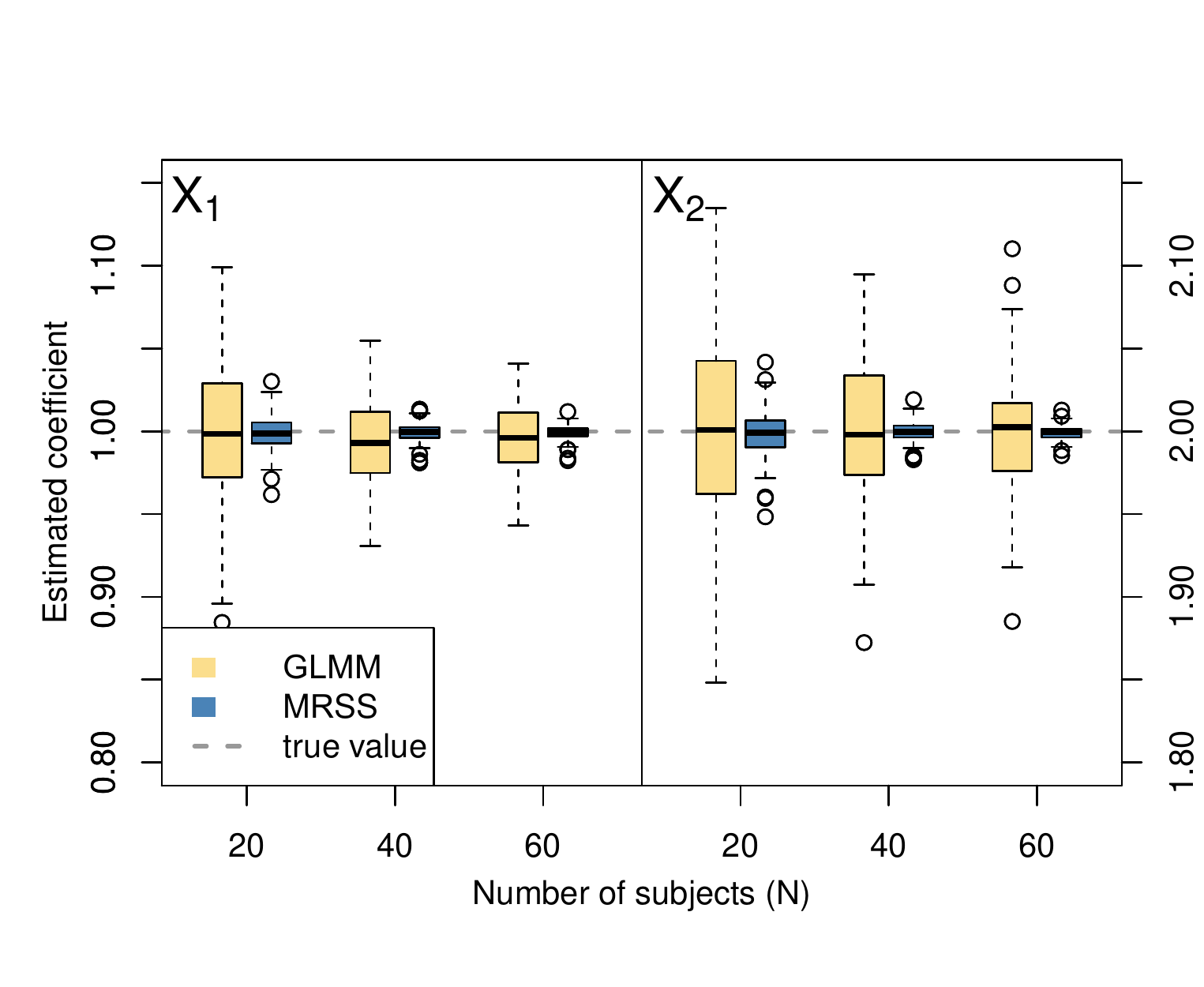}
    }
    \parbox{0.5\textwidth}{
        \textbf{(b)} Response $\boldsymbol{Y}^{(2)}$; Varying time length (T)\\
        \includegraphics[clip, trim=16pt 16pt 0 50pt, width=0.5\textwidth]{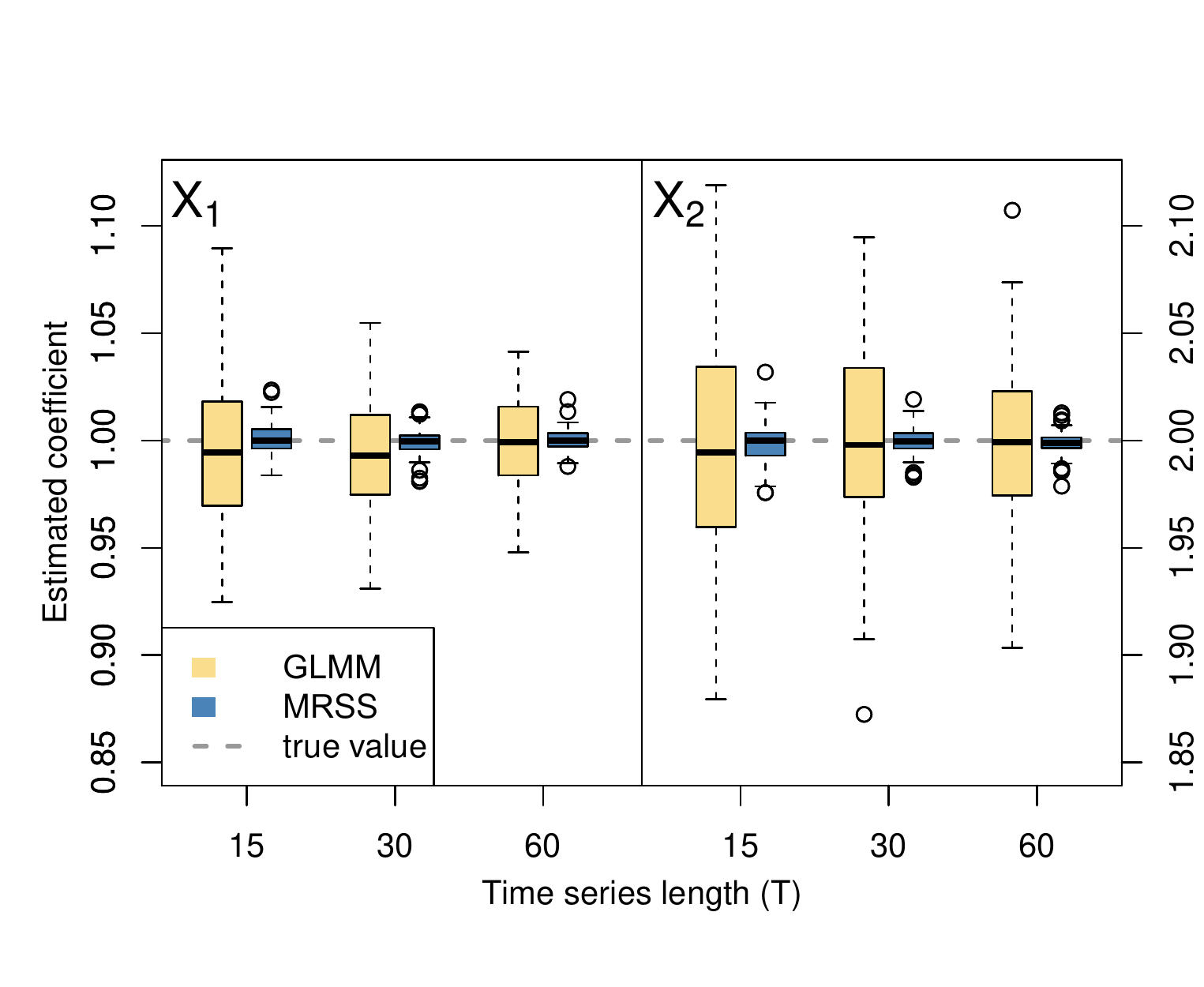}
    }
    \begin{center}
        \parbox{0.55\textwidth}{
            \textbf{(c)} Response $\boldsymbol{Y}^{(2)}$; Varying $p$\\
            \includegraphics[clip, trim=16pt 30pt 0 50pt, width=0.5\textwidth]{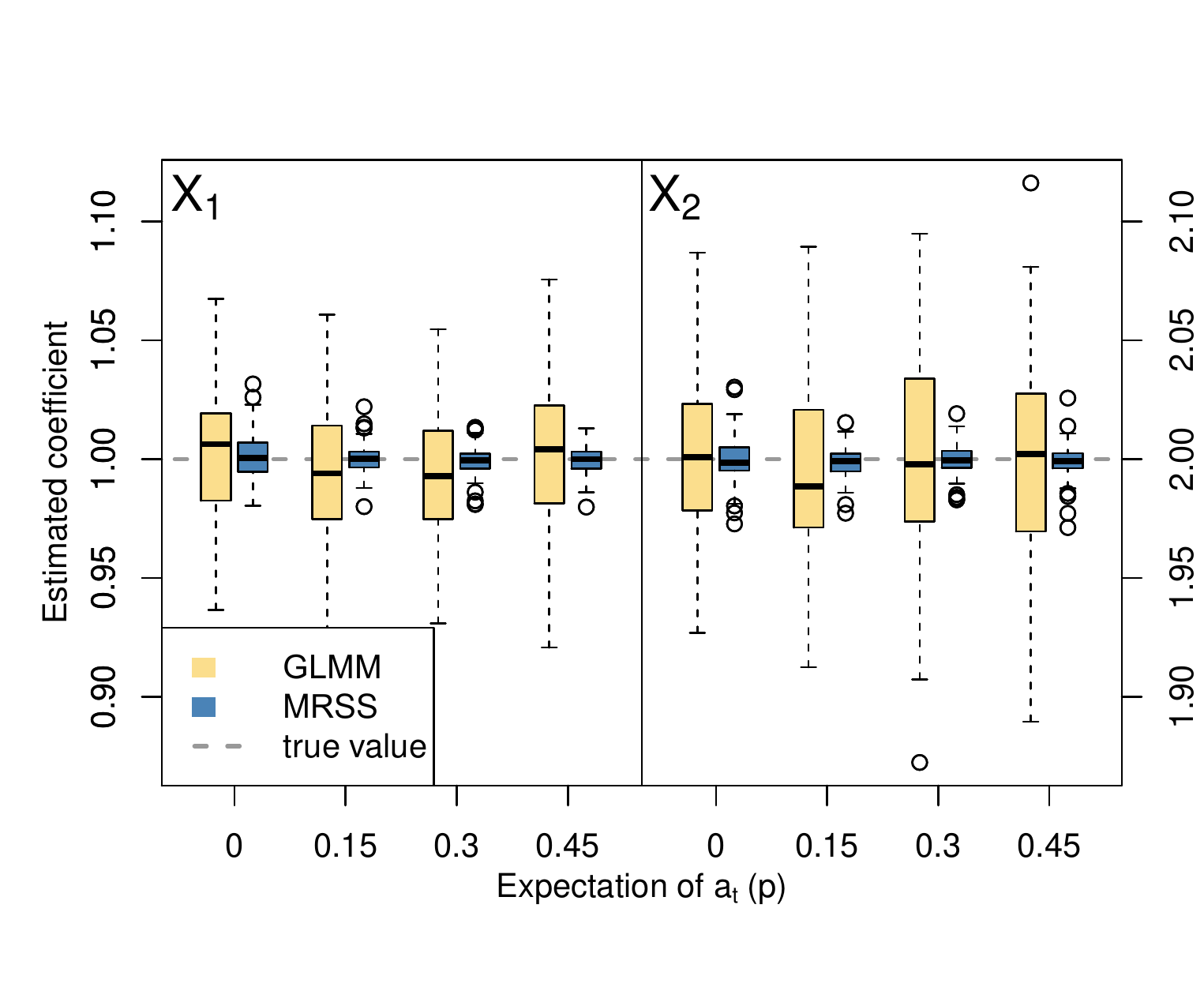}
        }
    \end{center}
    \caption{Estimated coefficients from response $\boldsymbol{Y}^{(2)}$ under different settings. In (a), the sample size ($N$) varies;  In (b), the time series length ($T$) varies;  In (c), the expectation of $a_t$ ($p$) varies. }\label{wbfig:coefficients}
\end{figure}

\begin{figure}
    \hspace{-0.05\textwidth}
    \includegraphics[width=1.1\textwidth]{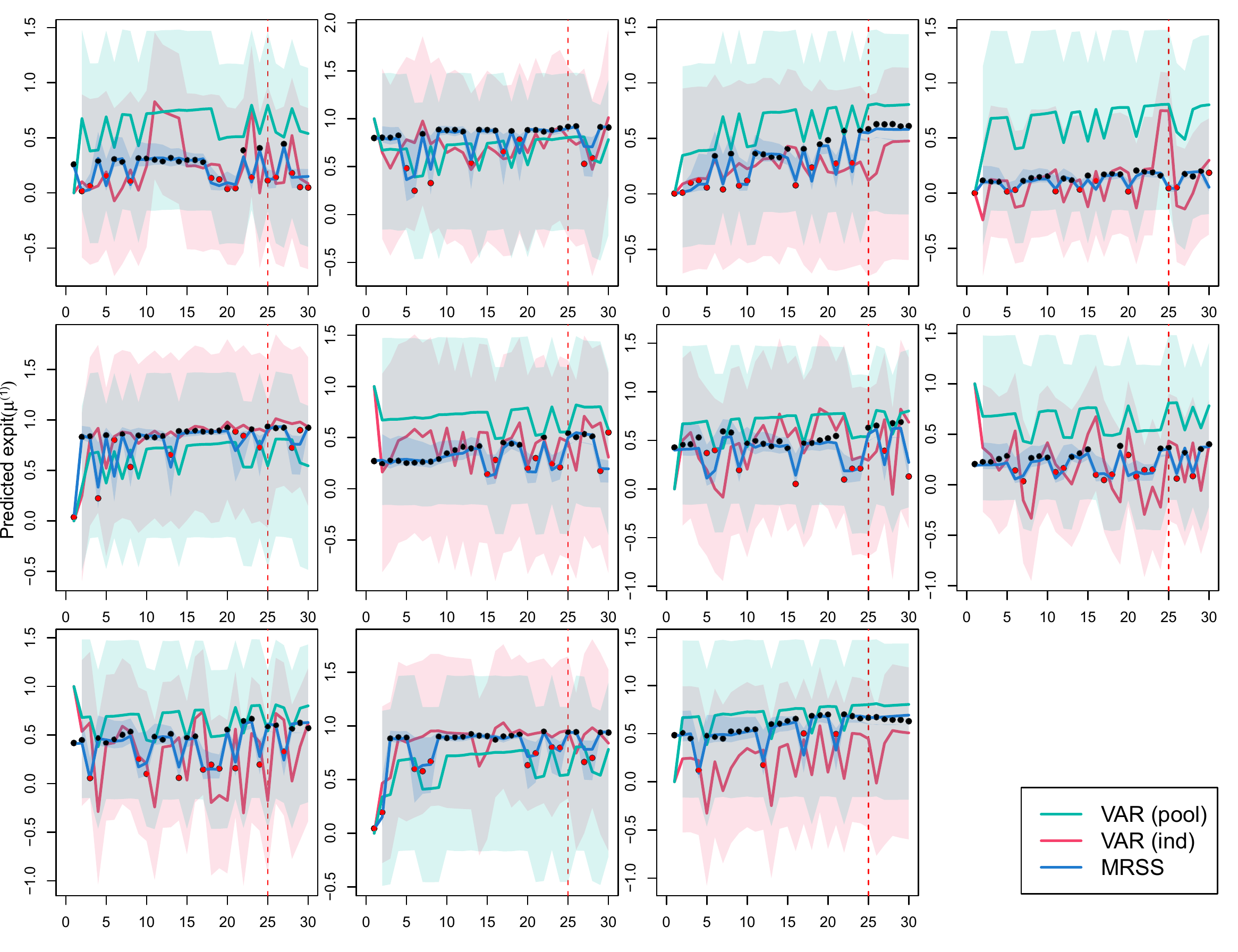}
    \caption{Predicted trajectories ($\operatorname{expit(\mu^{(1)})}$) of $11$ randomly selected subjects. Black and red dots are true values when $a_t=0$ and $a_t=1$. }\label{wbfig:y1pre}
    \end{figure}

\begin{figure}[!ht]
\hspace{-0.05\textwidth}
\includegraphics[width=1.1\textwidth]{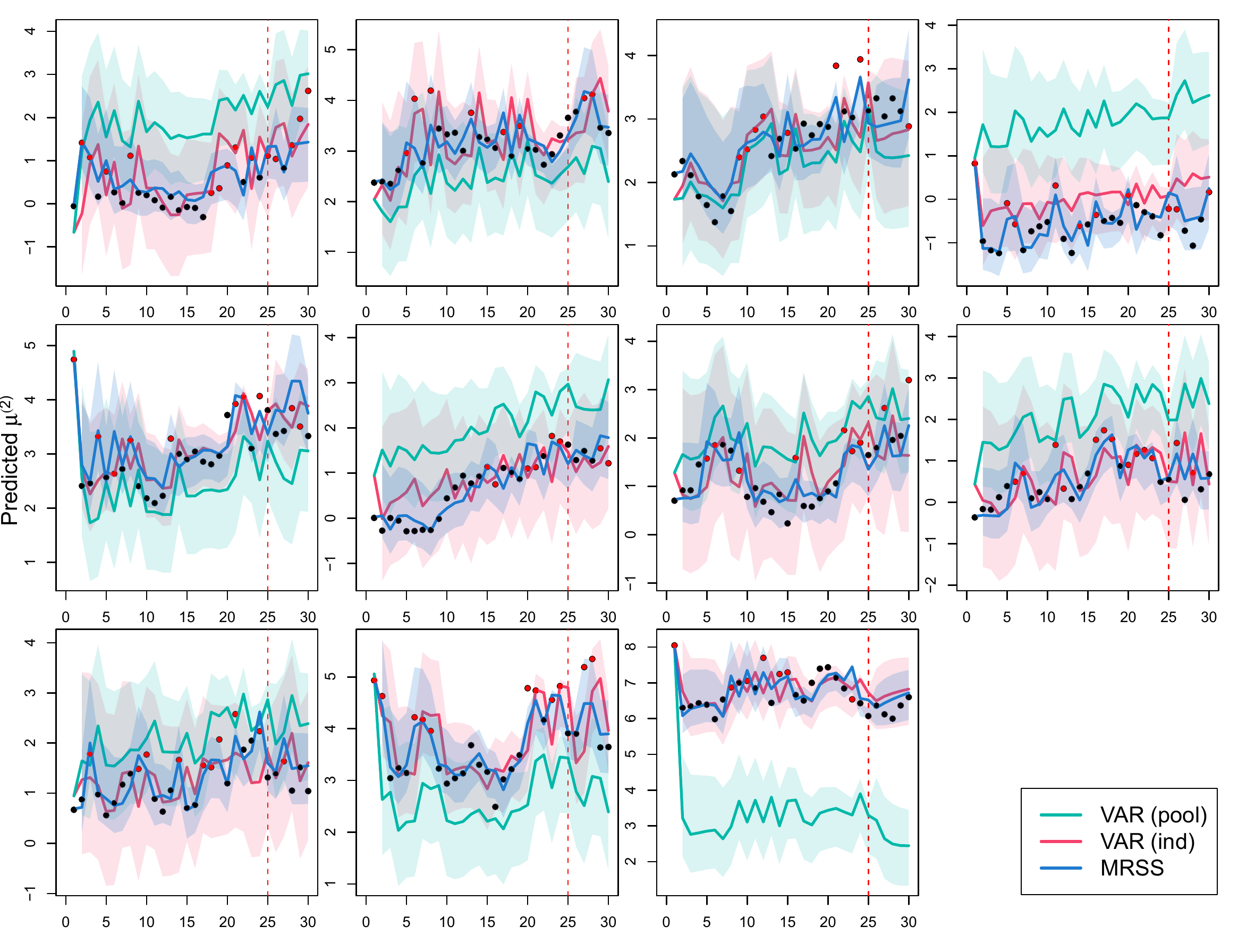}
\caption{Predicted trajectories ($\mu^{(2)}$) of $11$ randomly selected subjects. Black and red dots are true values when $a_t=0$ and $a_t=1$. }\label{wbfig:y2pre}
\end{figure}

\begin{figure}[!ht]
    \hspace{-0.05\textwidth}
    \includegraphics[width=1.1\textwidth]{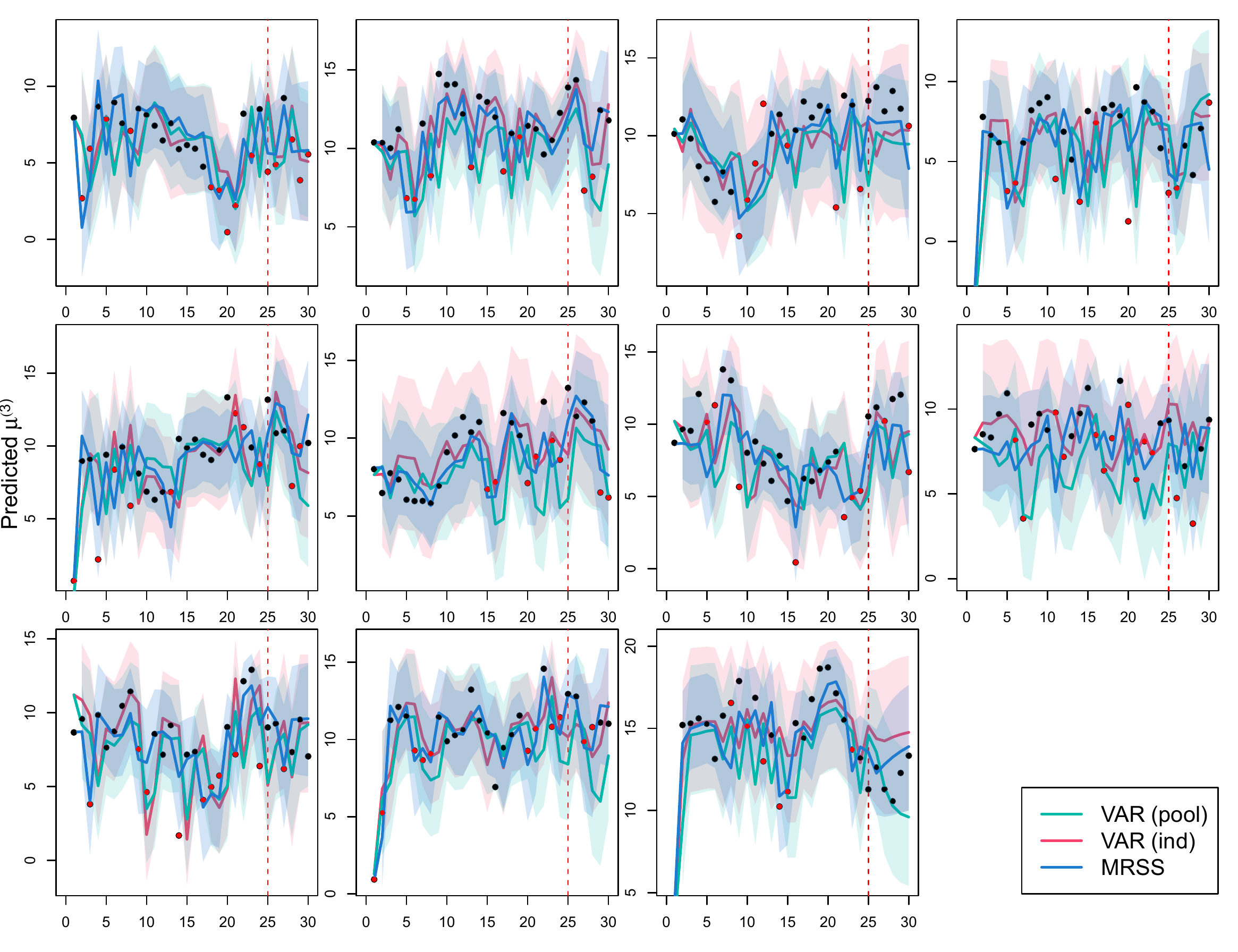}
    \caption{Predicted trajectories ($\mu^{(3)}$) of $11$ randomly selected subjects. Black and red dots are true values when $a_t=0$ and $a_t=1$. }\label{wbfig:y3pre}
\end{figure}

\clearpage
\begin{table}
\caption{MSE ($\times 10^{-2}$) of estimated coefficient with different sample sizes ($N$)}\label{wbtab:sample_size}
\newcolumntype{Y}{>{\centering\arraybackslash}X}
\hspace{-0.05\textwidth}
\begin{tabularx}{1.12\textwidth}{ccYYccYc}
    \toprule
    \multirow{2}[2]{*}{Method} & \multirow{2}[2]{*}{\makecell[b]{Sample\\size\,($N$)}} & \multicolumn{2}{c}{$\boldsymbol{Y}^{(1)}$} & \multicolumn{2}{c}{$\boldsymbol{Y}^{(2)}$} & \multicolumn{2}{c}{$\boldsymbol{Y}^{(3)}$} \\
    \cmidrule{3-8}
      &   & $X_1$ & $X_2$ & $X_1$ & $X_2$ &  $X_1$ & $X_2$ \\
      \midrule
    \multirow{3}[0]{*}{GLMM} & 20 & 1.89 (4.33) & 3.87 (5.28) & 0.20 (0.27) & 0.41 (0.60) & 3.36 (5.05) & 8.40$\!$ (13.04) \\
      & 40 & 0.91 (1.14) & 2.23 (2.68) & 0.07 (0.09) & 0.18 (0.24) & 1.36 (1.70) & 2.68 (3.08) \\
      & 60 & 0.56 (0.66) & 1.77 (2.16) & 0.04 (0.05) & 0.13 (0.23) & 0.91 (1.16) & 2.08 (3.47) \\
      \cmidrule{1-2}
    \multirow{3}[0]{*}{MRSS} & 20 & 0.66 (1.32) & 1.75 (2.86) & 0.01 (0.02) & 0.02 (0.04) & 0.31 (0.64) & 0.54 (1.18) \\
      & 40 & 0.24 (0.35) & 0.65 (1.13) & $<$0.01 (0.01) & $<$0.01 (0.01) & 0.12 (0.19) & 0.15 (0.28) \\
      & 60 & 0.16 (0.38) & 0.44 (0.80) & $<$0.01 (0.00) & $<$0.01 (0.00) & 0.07 (0.12) & 0.10 (0.32) \\
      \bottomrule
    \end{tabularx}
\end{table}

\begin{table}
\caption{MSE ($\times 10^{-2}$) of estimated coefficient with different time series lengths ($T$)}\label{wbtab:time_length}
\newcolumntype{Y}{>{\centering\arraybackslash}X}
    \hspace{-0.05\textwidth}
\begin{tabularx}{1.139\textwidth}{ccYYccYY}
    \toprule
    \multirow{2}[2]{*}{Method} & \multirow{2}[2]{*}{\makecell[b]{Time\\length\,($T$)}} & \multicolumn{2}{c}{$\boldsymbol{Y}^{(1)}$} & \multicolumn{2}{c}{$\boldsymbol{Y}^{(2)}$} & \multicolumn{2}{c}{$\boldsymbol{Y}^{(3)}$} \\
    \cmidrule{3-8}
      &   & $X_1$ & $X_2$ & $X_1$& $X_2$ & $X_1$ & $X_2$\\
    \midrule
    \multirow{3}[0]{*}{GLMM} & 15 & 1.43 (1.66) & 3.33 (3.96) & 0.13 (0.19) & 0.29 (0.35) & 2.31 (4.03) & 4.75 (7.07) \\
      & 30 & 0.91 (1.14) & 2.23 (2.68) & 0.07 (0.09) & 0.18 (0.24) & 1.36 (1.70) & 2.68 (3.08) \\
      & 60 & 0.43 (0.60) & 1.11 (1.54) & 0.05 (0.06) & 0.12 (0.17) & 0.80 (1.01) & 1.45 (2.14) \\
      \cmidrule{1-2}
    \multirow{3}[0]{*}{MRSS} & 15 & 0.52 (0.90) & 1.34 (2.83) & 0.01 (0.01) & 0.01 (0.02) & 0.21 (0.35) & 0.45 (0.79) \\
      & 30 & 0.24 (0.35) & 0.65 (1.13) & $<$0.01 (0.01) & $<$0.01 (0.01) & 0.12 (0.19) & 0.15 (0.28) \\
      & 60 & 0.15 (0.23) & 0.32 (0.42) & $<$0.01 (0.00) & $<$0.01 (0.01) & 0.08 (0.13) & 0.07 (0.13) \\
    \bottomrule
\end{tabularx}%
\end{table}

\begin{table}
\caption{MSE ($\times 10^{-2}$) of estimated coefficient with different expectation of $a_t$ ($p$)}\label{wbtab:a}
\newcolumntype{Y}{>{\centering\arraybackslash}X}
\hspace{-0.05\textwidth}
\begin{tabularx}{1.1\textwidth}{ccYYccYY}
    \toprule
\multirow{2}[2]{*}{Method} & \multirow{2}[2]{*}{$p$} & \multicolumn{2}{c}{$\boldsymbol{Y}^{(1)}$} & \multicolumn{2}{c}{$\boldsymbol{Y}^{(2)}$} & \multicolumn{2}{c}{$\boldsymbol{Y}^{(3)}$} \\
\cmidrule{3-8}
  &   & $X_1$ & $X_2$ & $X_1$ & $X_2$ &  $X_1$ & $X_2$ \\
  \midrule
\multirow{7}[0]{*}{GLMM} & 0 & 0.56 (0.72) & 1.83 (2.58) & 0.08 (0.11) & 0.13 (0.18) & 1.43 (1.78) & 2.52 (3.26) \\
  & 15 & 0.63 (0.91) & 2.18 (2.80) & 0.07 (0.10) & 0.15 (0.19) & 1.31 (1.85) & 2.53 (3.34) \\
  & 30 & 0.91 (1.14) & 2.23 (2.68) & 0.07 (0.09) & 0.18 (0.24) & 1.36 (1.70) & 2.68 (3.08) \\
  & 45 & 0.92 (1.22) & 2.95 (3.78) & 0.09 (0.13) & 0.17 (0.23) & 1.70 (2.51) & 2.87 (3.51) \\
  & 60 & 1.29 (1.45) & 3.88 (3.43) & 0.08 (0.11) & 0.19 (0.24) & 1.47 (1.78) & 3.71 (4.65) \\
  & 75 & 1.75 (1.70) & 5.88 (4.98) & 0.08 (0.11) & 0.16 (0.25) & 1.51 (2.05) & 3.11 (4.79) \\
  & 90 & 1.68 (2.00) & 6.68 (5.71) & 0.07 (0.11) & 0.17 (0.23) & 1.91 (2.85) & 4.06 (7.77) \\
  \cmidrule{1-2}
\multirow{7}[0]{*}{MRSS} & 0 & 0.35 (0.51) & 1.26 (1.84) & 0.01 (0.02) & 0.01 (0.02) & 0.22 (0.51) & 0.30 (0.82) \\
  & 15 & 0.29 (0.40) & 1.04 (2.01) & $<$0.01 (0.01) & $<$0.01 (0.01) & 0.10 (0.17) & 0.24 (0.39) \\
  & 30 & 0.24 (0.35) & 0.65 (1.13) & $<$0.01 (0.01) & $<$0.01 (0.01) & 0.12 (0.19) & 0.15 (0.28) \\
  & 45 & 0.28 (0.51) & 0.95 (1.83) & $<$0.01 (0.01) & $<$0.01 (0.01) & 0.16 (0.25) & 0.15 (0.28) \\
  & 60 & 0.19 (0.30) & 0.59 (0.76) & $<$0.01 (0.01) & $<$0.01 (0.01) & 0.09 (0.14) & 0.25 (0.41) \\
  & 75 & 0.26 (0.54) & 0.70 (1.14) & $<$0.01 (0.00) & $<$0.01 (0.01) & 0.12 (0.17) & 0.27 (0.39) \\
  & 90 & 0.26 (0.57) & 0.69 (1.18) & $<$0.01 (0.01) & 0.01 (0.02) & 0.23 (0.59) & 0.38 (0.70) \\
  \bottomrule
\end{tabularx}%
\end{table}

\begin{table}
\caption{MSE ($\times 10^{-2}$) of estimated coefficient with different number of states in the MRSS model}\label{wbtab:num}
\newcolumntype{Y}{>{\centering\arraybackslash}X}
\hspace{-0.05\textwidth}
\begin{tabularx}{1.1\textwidth}{cYYcccccc}
\toprule
\multicolumn{3}{c}{\multirow{2}[4]{*}{Model}} & \multicolumn{2}{c}{$\bY^{(1)}$} & \multicolumn{2}{c}{$\bY^{(2)}$} & \multicolumn{2}{c}{$\bY^{(3)}$} \\
\cmidrule{4-9}\multicolumn{3}{c}{}  & $X_1$    & $X_2$    & $X_1$    & $X_2$    & $X_1$    & $X_2$ \\
\midrule
\multicolumn{3}{l}{GLMM (ref)} & 0.91 (1.14) & 2.23 (2.68) & 0.07 (0.09) & 0.18 (0.24) & 1.36 (1.70) & 2.68 (3.08) \\
\cmidrule{1-3}
\multirow{7}[1]{*}{MRSS} &  $\# \bb$  &  $\# \bv$ &       &       &       &       &       &  \\
      & 1     & 1     & 0.24 (0.35) & 0.65 (1.13) & $<$0.01 (0.01) & $<$0.01 (0.01) & 0.12 (0.19) & 0.15 (0.28) \\
      & 0     & 1     & 15.1 (16.1) & 196 (387) & 0.56 (0.68) & 6.44 (4.32) & 20.3 (26.3) & 62.2 (73.4) \\
      & 1     & 0     & 174 (554) & 537 (1536) & 2.95 (2.41) & 23.7 (37.5) & 340 (851) & 153 (325) \\
      & 1     & 2     & 0.56 (0.97) & 1.06 (2.55) & 0.01 (0.01) & 0.01 (0.02) & 0.13 (0.37) & 0.26 (0.59) \\
      & 2     & 1     & 0.66 (0.90) & 1.74 (3.51) & 0.09 (0.03) & 0.11 (0.21) & 0.14 (0.46) & 0.44 (0.96) \\
      & 2     & 2     & 0.77 (1.00) & 1.76 (4.42) & 0.10 (0.09) & 0.20 (0.26) & 0.13 (0.48) & 0.40 (1.57) \\
\bottomrule
\end{tabularx}\\

Note: 1) GLMM is included for reference purposes.\\
\phantom{Note: }2) MRSS model with $1$ $\bb$ and $1$ $\bv$ is the true model.
\end{table}

\begin{landscape}
\thispagestyle{empty}
\begin{table}[!h]
\caption{Estimated loading matrix and coefficients of MRSS model (Memory Model)}\label{wbtab:estimated_game}
\hspace{-0.03\textwidth}
\begin{tabular}{ccccccccc}
    \toprule
    & \multicolumn{3}{c}{loading $\lambda$} & \multicolumn{5}{c}{coefficients $\beta$} \\
    \cmidrule(r){2-4}\cmidrule(l){5-9}
  Response & $b^{(1)}$ & $b^{(2)}$ & $v^{(1)}$ & age & gender & smoke & PD history & brain stimu \\
  \midrule
A\_memory &   & 0.18 [0.922] & \textcolor[rgb]{ 1,  0,  0}{2.56 [0.002]} & \textcolor[rgb]{ 1,  0,  0}{0.46 [0.030]} & -0.03 [0.768] & 0.00 [0.693] & 0.00 [0.189] & 0.00 [0.615] \\
score & \textcolor[rgb]{ 1,  0,  0}{0.48 [0.003]} & 0.13 [0.465] & \textcolor[rgb]{ 1,  0,  0}{1.26 [0.048]} & \textcolor[rgb]{ 1,  0,  0}{-0.35 [0.033]} & 0.02 [0.399] & 0.00 [0.819] & \textcolor[rgb]{ 1,  0,  0}{0.32 [0.020]} & 0.01 [0.754] \\
  \bottomrule
\end{tabular}\\

Note: 1) Values in square brackets are corresponding p-values. \\
\phantom{Note: }2) Values in {\color{red}red} are significant estimates ($p<0.05$).
\end{table}

\begin{table}[!h]
\caption{Estimated loading matrix and coefficients of MRSS model (Voice Model)}\label{wbtab:estimated_voice}
\hspace{-0.154\textwidth}
\begin{tabular}{ccccccccccc}
    \toprule
    & \multicolumn{5}{c}{loading $\lambda$} & \multicolumn{5}{c}{coefficients $\beta$} \\
    \cmidrule(r){2-6}\cmidrule(l){7-11}
  Response & $b^{(1)}$ & $b^{(2)}$ &$b^{(3)}$& $v^{(1)}$ & $v^{(2)}$ & age & gender & smoke & PD history & brain stimu \\
  \midrule
A\_voice &   &   & \textcolor[rgb]{ 1,  0,  0}{0.47 [0.002]} & \textcolor[rgb]{ 1,  0,  0}{2.60 [0.000]} & \textcolor[rgb]{ 1,  0,  0}{-0.64 [0.000]} & \textcolor[rgb]{ 1,  0,  0}{-0.11 [0.012]} & \textcolor[rgb]{ 1,  0,  0}{-0.32 [0.007]} & \textcolor[rgb]{ 1,  0,  0}{0.11 [0.017]} & \textcolor[rgb]{ 1,  0,  0}{0.06 [0.020]} & \textcolor[rgb]{ 1,  0,  0}{-0.35 [0.000]} \\
\textcolor[rgb]{ .267,  .447,  .769}{V27} & \textcolor[rgb]{ 1,  0,  0}{-0.11 [0.000]} & 0.00 [0.771]& 0.01 [0.658] & 0.10 [0.699]& \textcolor[rgb]{ 1,  0,  0}{-0.21 [0.000]} & -0.03 [0.197] & \textcolor[rgb]{ 1,  0,  0}{0.52 [0.024]} & 0.03 [0.464]& 0.09 [0.645]& 0.00 [0.930] \\
V51 & \textcolor[rgb]{ 1,  0,  0}{0.08 [0.039]} & 0.01 [0.650] & 0.00 [0.982] & \textcolor[rgb]{ 1,  0,  0}{-0.13 [0.018]} & 0.01 [0.782] & 0.05 [0.277] & \textcolor[rgb]{ 1,  0,  0}{-0.40 [0.001]} & 0.03 [0.819] & 0.17 [0.440] & 0.02 [0.608] \\
V52 & \textcolor[rgb]{ 1,  0,  0}{-0.07 [0.004]} & -0.02 [0.507] & 0.01 [0.452]& \textcolor[rgb]{ 1,  0,  0}{-0.09 [0.009]} & 0.00 [0.737] & 0.02 [0.324]& \textcolor[rgb]{ 1,  0,  0}{-0.56 [0.000]} & -0.06 [0.431] & 0.02 [0.493] & \textcolor[rgb]{ 1,  0,  0}{-0.16 [0.022]} \\
\textcolor[rgb]{ .267,  .447,  .769}{V73} & \textcolor[rgb]{ 1,  0,  0}{0.10 [0.016]} & 0.06 [0.122] & 0.00 [0.862] & \textcolor[rgb]{ 1,  0,  0}{-0.05 [0.029]} & \textcolor[rgb]{ 1,  0,  0}{-0.09 [0.019]} & -0.01 [0.405] & \textcolor[rgb]{ 1,  0,  0}{0.22 [0.000]} & -0.03 [0.434] & -0.10 [0.454]& 0.09 [0.713] \\
V176 &0.11 [0.077] & \textcolor[rgb]{ 1,  0,  0}{0.10 [0.017]} & 0.01 [0.731] & \textcolor[rgb]{ 1,  0,  0}{-0.19 [0.006]} & \textcolor[rgb]{ 1,  0,  0}{0.06 [0.004]} & 0.05 [0.649] & \textcolor[rgb]{ 1,  0,  0}{-0.62 [0.001]} & \textcolor[rgb]{ 1,  0,  0}{0.05 [0.049]} & \textcolor[rgb]{ 1,  0,  0}{0.25 [0.033]} & 0.14 [0.334]\\
V216 & 0.02 [0.539] & \textcolor[rgb]{ 1,  0,  0}{0.19 [0.001]} & 0.01 [0.953]& \textcolor[rgb]{ 1,  0,  0}{-0.09 [0.015]} & \textcolor[rgb]{ 1,  0,  0}{0.03 [0.000]} & 0.02 [0.811]& \textcolor[rgb]{ 1,  0,  0}{-1.24 [0.000]} & -0.02 [0.925] & 0.00 [0.591] & 0.04 [0.789] \\
V267 & 0.09 [0.091] & \textcolor[rgb]{ 1,  0,  0}{0.06 [0.031]} & 0.00 [0.947] & \textcolor[rgb]{ 1,  0,  0}{-0.26 [0.005]} & \textcolor[rgb]{ 1,  0,  0}{0.04 [0.005]} & 0.05 [0.328] & \textcolor[rgb]{ 1,  0,  0}{-0.24 [0.042]} & 0.03 [0.320] & \textcolor[rgb]{ 1,  0,  0}{0.15 [0.016]} & 0.09 [0.672] \\
V307 & 0.01 [0.804] & \textcolor[rgb]{ 1,  0,  0}{0.18 [0.000]} & 0.01 [0.996]& \textcolor[rgb]{ 1,  0,  0}{-0.07 [0.001]} & \textcolor[rgb]{ 1,  0,  0}{0.04 [0.001]} & 0.02 [0.559] & \textcolor[rgb]{ 1,  0,  0}{-1.31 [0.004]} & -0.01 [0.912] & 0.00 [0.788] & 0.04 [0.389]\\
  \bottomrule
  \end{tabular}\\

Note: 1) Values in square brackets are corresponding p-values. \\
\phantom{Note: }2) Responses in blue are the responses that a larger value means better health status.\\
\phantom{Note: }3) Values in {\color{red}red} are significant estimates ($p<0.05$).\\
\phantom{Note: }4) V27: Shimmer\textsubscript{PQ3}, V51: GQ\textsubscript{std\_cycle\_closed}, V52: GNE\textsubscript{mean}, V73: MFCC\textsubscript{mean\_1st\_coef}, V176: Log Entropy\textsubscript{det\_1st\_coef}, V216: Log Entropy\textsubscript{app\_1st\_coef}, V267: Log Entropy\textsubscript{det\_LT\_1st\_coef}, V307: Log Entropy\textsubscript{app\_LT\_1st\_coef}. 
\end{table}
\vfill
\raisebox{-0pt}{\makebox[\linewidth]{\thepage}}

\clearpage
\thispagestyle{empty}
\begin{table}[!h]
\caption{Estimated loading matrix and coefficients of MRSS model (Motor Model)}\label{wbtab:estimated}
\hspace{-0.10\textwidth}
\begin{tabular}{cccccccccc}
    \toprule
    & \multicolumn{4}{c}{loading $\lambda$} & \multicolumn{5}{c}{coefficients $\beta$} \\
    \cmidrule(r){2-5}\cmidrule(l){6-10}
  Response & $b^{(1)}$ & $b^{(2)}$ & $v^{(1)}$ & $v^{(2)}$ & age & gender & smoke & PD history & brain stimu \\
  \midrule
  A\_walking &   & \textcolor[rgb]{ 1,  0,  0}{0.59 [0.049]} & \textcolor[rgb]{ 1,  0,  0}{5.47 [0.000]} & \textcolor[rgb]{ 1,  0,  0}{-1.03 [0.010]} & \textcolor[rgb]{ 1,  0,  0}{1.02 [0.000]} & -0.16 [0.531] & -0.65 [0.102] & -0.49 [0.067] & -2.27 [0.059] \\
  A\_tapping &   & \textcolor[rgb]{ 1,  0,  0}{0.64 [0.034]} & \textcolor[rgb]{ 1,  0,  0}{6.06 [0.000]} & -0.97 [0.226] & \textcolor[rgb]{ 1,  0,  0}{0.76 [0.002]} & 0.16 [0.509] & -0.55 [0.168] & -0.34 [0.170] & -2.58 [0.065] \\
  \textcolor[rgb]{ .267,  .447,  .769}{median $C_1$} & -0.02 [0.948] & \textcolor[rgb]{ 1,  0,  0}{0.77 [0.011]} & \textcolor[rgb]{ 1,  0,  0}{-1.04 [0.000]} & -0.02 [0.935] & \textcolor[rgb]{ 1,  0,  0}{-0.16 [0.006]} & 0.00 [0.989] & \textcolor[rgb]{ 1,  0,  0}{-0.86 [0.001]} & -0.17 [0.502] & -0.55 [0.069] \\
  sd $C_2$ & 0.02 [0.965] & \textcolor[rgb]{ 1,  0,  0}{-0.72 [0.041]} & \textcolor[rgb]{ 1,  0,  0}{1.02 [0.000]} & -0.03 [0.916] & \textcolor[rgb]{ 1,  0,  0}{0.23 [0.019]} & 0.03 [0.891] & 0.16 [0.513] & 0.12 [0.621] & -0.16 [0.510] \\
  \textcolor[rgb]{ .267,  .447,  .769}{dfa Y} & 0.04 [0.858] & \textcolor[rgb]{ 1,  0,  0}{0.98 [0.005]} & \textcolor[rgb]{ 1,  0,  0}{-0.33 [0.030]} & 0.00 [0.995] & \textcolor[rgb]{ 1,  0,  0}{-0.16 [0.039]} & 0.05 [0.827] & 0.48 [0.056] & 0.02 [0.948] & -0.16 [0.509] \\
  range Tap Inter & \textcolor[rgb]{ 1,  0,  0}{0.58 [0.020]} & -0.27 [0.376] & \textcolor[rgb]{ 1,  0,  0}{0.43 [0.033]} & 0.32 [0.205] & \textcolor[rgb]{ 1,  0,  0}{0.15 [0.027]} & -0.06 [0.825] & -0.17 [0.486] & 0.02 [0.950] & 0.16 [0.510] \\
  \textcolor[rgb]{ .267,  .447,  .769}{number Taps} & \textcolor[rgb]{ 1,  0,  0}{-0.42 [0.027]} & 0.36 [0.231] & \textcolor[rgb]{ 1,  0,  0}{-0.39 [0.042]} & \textcolor[rgb]{ 1,  0,  0}{-0.47 [0.018]} & \textcolor[rgb]{ 1,  0,  0}{-0.47 [0.000]} & -0.05 [0.841] & 0.13 [0.594] & -0.16 [0.510] & 0.02 [0.947] \\
  sd Tap Inter & \textcolor[rgb]{ 1,  0,  0}{0.64 [0.034]} & -0.23 [0.450] & \textcolor[rgb]{ 1,  0,  0}{0.34 [0.025]} & 0.37 [0.080] & \textcolor[rgb]{ 1,  0,  0}{0.47 [0.000]} & 0.08 [0.737] & -0.09 [0.728] & -0.02 [0.946] & -0.02 [0.948] \\  
  \bottomrule
\end{tabular}\\

Note: 1) Values in square brackets are corresponding p-values. \\
\phantom{Note: }2) Responses in blue are the responses that a larger value means better health status.\\
\phantom{Note: }3) Values in {\color{red}red} are significant estimates ($p<0.05$).
\end{table}
\vfill
\raisebox{-0pt}{\makebox[\linewidth]{\thepage}}

\clearpage
\thispagestyle{empty}
\begin{table}[!h]
  \centering
\caption{Estimated coefficients of GLMM (Motor Model)}\label{wbtab:estglmm}
\begin{tabular}{ccccccc}
  \toprule
Response & (Intercept) & \makecell[c]{A\_(walking \\or tapping)} & PD & age & gender & medication  \\
\midrule
A\_walking & \textcolor[rgb]{ 1,  0,  0}{-7.95 [0.000]} &   & \textcolor[rgb]{ 1,  0,  0}{3.42 [0.002]} & 0.19 [0.121] & 0.27 [0.207] & \textcolor[rgb]{ 1,  0,  0}{2.55 [0.000]} \\
A\_tapping & \textcolor[rgb]{ 1,  0,  0}{-8.73 [0.000]} &   & \textcolor[rgb]{ 1,  0,  0}{4.44 [0.000]} & 0.18 [0.107] & -0.05 [0.748] & \textcolor[rgb]{ 1,  0,  0}{2.76 [0.000]} \\
\textcolor[rgb]{ .267,  .447,  .769}{median $C_1$} & 0.07 [0.765] & -0.01 [0.617] & \textcolor[rgb]{ 1,  0,  0}{-0.96 [0.007]} & 0.06 [0.363] & 0.23 [0.054] & \textcolor[rgb]{ 1,  0,  0}{0.49 [0.048]} \\
sd $C_2$ & \textcolor[rgb]{ 1,  0,  0}{-0.73 [0.000]} & \textcolor[rgb]{ 1,  0,  0}{0.03 [0.002]} & 0.46 [0.126] & -0.04 [0.395] & \textcolor[rgb]{ 1,  0,  0}{0.86 [0.000]} & -0.10 [0.621] \\
\textcolor[rgb]{ .267,  .447,  .769}{dfa Y} & \textcolor[rgb]{ 1,  0,  0}{0.27 [0.042]} & \textcolor[rgb]{ 1,  0,  0}{-0.03 [0.039]} & -0.37 [0.071] & -0.01 [0.671] & -0.10 [0.152] & 0.20 [0.151] \\
range Tap Inter & \textcolor[rgb]{ 1,  0,  0}{-0.46 [0.000]} & \textcolor[rgb]{ 1,  0,  0}{-0.03 [0.002]} & \textcolor[rgb]{ 1,  0,  0}{0.82 [0.000]} & \textcolor[rgb]{ 1,  0,  0}{0.09 [0.024]} & -0.11 [0.106] & -0.25 [0.057] \\
\textcolor[rgb]{ .267,  .447,  .769}{number Taps} & \textcolor[rgb]{ 1,  0,  0}{0.31 [0.037]} & \textcolor[rgb]{ 1,  0,  0}{0.05 [0.000]} & \textcolor[rgb]{ 1,  0,  0}{-0.72 [0.003]} & -0.04 [0.506] & 0.15 [0.112] & 0.23 [0.203] \\
sd Tap Inter & \textcolor[rgb]{ 1,  0,  0}{-0.33 [0.004]} & \textcolor[rgb]{ 1,  0,  0}{-0.03 [0.001]} & \textcolor[rgb]{ 1,  0,  0}{0.55 [0.003]} & 0.04 [0.375] & -0.06 [0.359] & -0.24 [0.078] \\
\bottomrule
\end{tabular}\\[10pt]

\begin{tabular}{ccccc}
  \toprule
  Response & smoke & PD history & brain stimu & morning \\
  \midrule
  A\_walking & 0.26 [0.153] & 0.30 [0.282] & -0.54 [0.212] & 0.36 [0.564] \\
  A\_tapping & 0.23 [0.106] & 0.16 [0.401] & -0.49 [0.090] & 0.25 [0.649] \\
  \textcolor[rgb]{ .267,  .447,  .769}{median $C_1$} & -0.09 [0.368] & 0.18 [0.202] & 0.07 [0.772] & -0.16 [0.326] \\
  sd $C_2$ & -0.02 [0.780] & -0.14 [0.237] & -0.35 [0.082] & 0.04 [0.746] \\
  \textcolor[rgb]{ .267,  .447,  .769}{dfa Y} & 0.00 [0.949] & 0.00 [0.955] & 0.00 [0.980] & 0.07 [0.898] \\
  range Tap Inter & 0.00 [0.976] & 0.09 [0.277] & 0.08 [0.560] & 0.26 [0.356] \\
  \textcolor[rgb]{ .267,  .447,  .769}{number Taps} & 0.03 [0.731] & -0.10 [0.365] & -0.10 [0.557] & -0.36 [0.266] \\
  sd Tap Inter & 0.00 [0.951] & 0.11 [0.176] & 0.10 [0.444] & \textcolor[rgb]{ 1,  0,  0}{0.30 [0.035]} \\  
\bottomrule
\end{tabular}\\[5pt]

\raggedright
Note: 1) Values in square brackets are corresponding p-values. \\
\phantom{Note: }2) Responses in blue are the responses that a larger value means better health status.\\
\phantom{Note: }3) Values in {\color{red}red} are significant estimates ($p<0.05$).
\end{table}
\vfill
\raisebox{-0pt}{\makebox[\linewidth]{\thepage}}
\end{landscape}
\clearpage
\bibliographystyle{chicago}  \bibliography{ref}

\begin{thebibliography}{}

\bibitem[\protect\citeauthoryear{Alaa and van~der Schaar}{Alaa and van~der
  Schaar}{2019}]{alaa2019attentive}
Alaa, A. and M.~van~der Schaar (2019).
\newblock Attentive state-space modeling of disease progression.
\newblock {\em NeurIPS 2019\/}~{\em 32}, 1--11.

\bibitem[\protect\citeauthoryear{Baumeister and Montag}{Baumeister and
  Montag}{2019}]{baumeister2019digital}
Baumeister, H. and C.~Montag (2019).
\newblock {\em Digital Phenotyping and Mobile Sensing}.
\newblock Springer.

\bibitem[\protect\citeauthoryear{Beck and Tetruashvili}{Beck and
  Tetruashvili}{2013}]{beck2013convergence}
Beck, A. and L.~Tetruashvili (2013).
\newblock On the convergence of block coordinate descent type methods.
\newblock {\em SIAM journal on Optimization\/}~{\em 23\/}(4), 2037--2060.

\bibitem[\protect\citeauthoryear{Bengtsson and Cavanaugh}{Bengtsson and
  Cavanaugh}{2006}]{bengtsson2006improved}
Bengtsson, T. and J.~E. Cavanaugh (2006).
\newblock An improved akaike information criterion for state-space model
  selection.
\newblock {\em Computational Statistics \& Data Analysis\/}~{\em 50\/}(10),
  2635--2654.

\bibitem[\protect\citeauthoryear{Bhidayasiri and Truong}{Bhidayasiri and
  Truong}{2008}]{bhidayasiri2008motor}
Bhidayasiri, R. and D.~D. Truong (2008).
\newblock Motor complications in parkinson disease: clinical manifestations and
  management.
\newblock {\em Journal of the Neurological Sciences\/}~{\em 266\/}(1-2),
  204--215.

\bibitem[\protect\citeauthoryear{Boersma}{Boersma}{2006}]{boersma2006praat}
Boersma, P. (2006).
\newblock Praat: doing phonetics by computer.
\newblock {\em http://www.praat.org/\/}.

\bibitem[\protect\citeauthoryear{Bot, Suver, Neto, Kellen, Klein, Bare, Doerr,
  Pratap, Wilbanks, Dorsey, et~al.}{Bot et~al.}{2016}]{bot2016mpower}
Bot, B.~M., C.~Suver, E.~C. Neto, M.~Kellen, A.~Klein, C.~Bare, M.~Doerr,
  A.~Pratap, J.~Wilbanks, E.~R. Dorsey, et~al. (2016).
\newblock The mpower study, {P}arkinson disease mobile data collected using
  {R}esearch{K}it.
\newblock {\em Scientific data\/}~{\em 3\/}(1), 1--9.

\bibitem[\protect\citeauthoryear{Brookes}{Brookes}{2006}]{VOICEBOX}
Brookes, M. (2006).
\newblock Voicebox: Speech processing toolbox for matlab.
\newblock [Online; accessed 11-March-2021].

\bibitem[\protect\citeauthoryear{Camacho and Harris}{Camacho and
  Harris}{2008}]{camacho2008sawtooth}
Camacho, A. and J.~G. Harris (2008).
\newblock A sawtooth waveform inspired pitch estimator for speech and music.
\newblock {\em The Journal of the Acoustical Society of America\/}~{\em
  124\/}(3), 1638--1652.

\bibitem[\protect\citeauthoryear{Chan and Eisenstat}{Chan and
  Eisenstat}{2018}]{chan2018bayesian}
Chan, J.~C. and E.~Eisenstat (2018).
\newblock Bayesian model comparison for time-varying parameter vars with
  stochastic volatility.
\newblock {\em Journal of Applied Econometrics\/}~{\em 33\/}(4), 509--532.

\bibitem[\protect\citeauthoryear{Chan and Grant}{Chan and
  Grant}{2016}]{chan2016observed}
Chan, J.~C. and A.~L. Grant (2016).
\newblock On the observed-data deviance information criterion for volatility
  modeling.
\newblock {\em Journal of Financial Econometrics\/}~{\em 14\/}(4), 772--802.

\bibitem[\protect\citeauthoryear{Chatfield and Xing}{Chatfield and
  Xing}{2019}]{chatfield2019analysis}
Chatfield, C. and H.~Xing (2019).
\newblock {\em The analysis of time series: an introduction with R}.
\newblock CRC press.

\bibitem[\protect\citeauthoryear{De~Jong}{De~Jong}{1989}]{de1989smoothing}
De~Jong, P. (1989).
\newblock Smoothing and interpolation with the state-space model.
\newblock {\em Journal of the American Statistical Association\/}~{\em
  84\/}(408), 1085--1088.

\bibitem[\protect\citeauthoryear{De~Jong}{De~Jong}{1991}]{de1991diffuse}
De~Jong, P. (1991).
\newblock The diffuse kalman filter.
\newblock {\em The Annals of Statistics\/}, 1073--1083.

\bibitem[\protect\citeauthoryear{Dorsey, Glidden, Holloway, Birbeck, and
  Schwamm}{Dorsey et~al.}{2018}]{dorsey2018teleneurology}
Dorsey, E.~R., A.~M. Glidden, M.~R. Holloway, G.~L. Birbeck, and L.~H. Schwamm
  (2018).
\newblock Teleneurology and mobile technologies: the future of neurological
  care.
\newblock {\em Nature Reviews Neurology\/}~{\em 14\/}(5), 285.

\bibitem[\protect\citeauthoryear{Durbin and Koopman}{Durbin and
  Koopman}{1997}]{durbin1997monte}
Durbin, J. and S.~J. Koopman (1997).
\newblock Monte carlo maximum likelihood estimation for non-gaussian state
  space models.
\newblock {\em Biometrika\/}~{\em 84\/}(3), 669--684.

\bibitem[\protect\citeauthoryear{Durbin and Koopman}{Durbin and
  Koopman}{2000}]{durbin2000time}
Durbin, J. and S.~J. Koopman (2000).
\newblock Time series analysis of non-gaussian observations based on state
  space models from both classical and bayesian perspectives.
\newblock {\em Journal of the Royal Statistical Society: Series B (Statistical
  Methodology)\/}~{\em 62\/}(1), 3--56.

\bibitem[\protect\citeauthoryear{Durbin and Koopman}{Durbin and
  Koopman}{2012}]{durbin2012time}
Durbin, J. and S.~J. Koopman (2012).
\newblock {\em Time series analysis by state space methods}.
\newblock Oxford university press.

\bibitem[\protect\citeauthoryear{Gamerman, dos Santos, and Franco}{Gamerman
  et~al.}{2013}]{gamerman2013non}
Gamerman, D., T.~R. dos Santos, and G.~C. Franco (2013).
\newblock A non-gaussian family of state-space models with exact marginal
  likelihood.
\newblock {\em Journal of Time Series Analysis\/}~{\em 34\/}(6), 625--645.

\bibitem[\protect\citeauthoryear{Ghosh, Cheng, and Sun}{Ghosh
  et~al.}{2016}]{ghosh2016deep}
Ghosh, S., Y.~Cheng, and Z.~Sun (2016).
\newblock Deep state space models for computational phenotyping.
\newblock In {\em 2016 IEEE International Conference on Healthcare Informatics
  (ICHI)}, pp.\  399--402. IEEE.

\bibitem[\protect\citeauthoryear{Grunwald, Guttorp, and Raftery}{Grunwald
  et~al.}{1993}]{grunwald1993prediction}
Grunwald, G.~K., P.~Guttorp, and A.~E. Raftery (1993).
\newblock Prediction rules for exponential family state space models.
\newblock {\em Journal of the Royal Statistical Society: Series B
  (Methodological)\/}~{\em 55\/}(4), 937--943.

\bibitem[\protect\citeauthoryear{Harvey and Fernandes}{Harvey and
  Fernandes}{1989}]{harvey1989time}
Harvey, A.~C. and C.~Fernandes (1989).
\newblock Time series models for count or qualitative observations.
\newblock {\em Journal of Business \& Economic Statistics\/}~{\em 7\/}(4),
  407--417.

\bibitem[\protect\citeauthoryear{Ho, Bradshaw, and Iansek}{Ho
  et~al.}{2008}]{ho2008better}
Ho, A.~K., J.~L. Bradshaw, and R.~Iansek (2008).
\newblock For better or worse: The effect of levodopa on speech in
  {P}arkinson's disease.
\newblock {\em Movement disorders: official journal of the Movement Disorder
  Society\/}~{\em 23\/}(4), 574--580.

\bibitem[\protect\citeauthoryear{Hulme, Martin, Sperrin, Casson, Bucci, Lewis,
  and Peek}{Hulme et~al.}{2020}]{hulme2020adaptive}
Hulme, W.~J., G.~P. Martin, M.~Sperrin, A.~J. Casson, S.~Bucci, S.~Lewis, and
  N.~Peek (2020).
\newblock Adaptive symptom monitoring using hidden markov models--an
  application in ecological momentary assessment.
\newblock {\em IEEE Journal of Biomedical and Health Informatics\/}~{\em
  25\/}(5), 1770--1780.

\bibitem[\protect\citeauthoryear{Icaza and Jones}{Icaza and
  Jones}{1999}]{icaza1999state}
Icaza, G. and R.~Jones (1999).
\newblock A state-space em algorithm for longitudinal data.
\newblock {\em Journal of Time Series Analysis\/}~{\em 20\/}(5), 537--550.

\bibitem[\protect\citeauthoryear{Jain, Powers, Hawkins, and Brownstein}{Jain
  et~al.}{2015}]{jain2015digital}
Jain, S.~H., B.~W. Powers, J.~B. Hawkins, and J.~S. Brownstein (2015).
\newblock The digital phenotype.
\newblock {\em Nature Biotechnology\/}~{\em 33\/}(5), 462--463.

\bibitem[\protect\citeauthoryear{Jones}{Jones}{1993}]{jones1993longitudinal}
Jones, R.~H. (1993).
\newblock {\em Longitudinal data with serial correlation: a state-space
  approach}.
\newblock CRC Press.

\bibitem[\protect\citeauthoryear{Kantz and Schreiber}{Kantz and
  Schreiber}{2004}]{kantz2004nonlinear}
Kantz, H. and T.~Schreiber (2004).
\newblock {\em Nonlinear time series analysis}, Volume~7.
\newblock Cambridge university press.

\bibitem[\protect\citeauthoryear{Kitagawa}{Kitagawa}{1987}]{kitagawa1987non}
Kitagawa, G. (1987).
\newblock Non-gaussian state-space modeling of nonstationary time series.
\newblock {\em Journal of the American statistical association\/}~{\em
  82\/}(400), 1032--1041.

\bibitem[\protect\citeauthoryear{Klein}{Klein}{2003}]{klein2003state}
Klein, B.~M. (2003).
\newblock {\em State space models for exponential family data}.
\newblock Ph.\ D. thesis, Citeseer.

\bibitem[\protect\citeauthoryear{Lee, Turchioe, Creber, Biviano, Hickey, and
  Bakken}{Lee et~al.}{2021}]{lee2021phenotypes}
Lee, J., M.~R. Turchioe, R.~M. Creber, A.~Biviano, K.~Hickey, and S.~Bakken
  (2021).
\newblock Phenotypes of engagement with mobile health technology for heart
  rhythm monitoring.
\newblock {\em JAMIA open\/}~{\em 4\/}(2), ooab043.

\bibitem[\protect\citeauthoryear{Liang, Zheng, and Zeng}{Liang
  et~al.}{2019}]{liang2019survey}
Liang, Y., X.~Zheng, and D.~D. Zeng (2019).
\newblock A survey on big data-driven digital phenotyping of mental health.
\newblock {\em Information Fusion\/}~{\em 52}, 290--307.

\bibitem[\protect\citeauthoryear{Liu, Lu, Niu, and Wu}{Liu
  et~al.}{2011}]{liu2011mixed}
Liu, D., T.~Lu, X.-F. Niu, and H.~Wu (2011).
\newblock Mixed-effects state-space models for analysis of longitudinal dynamic
  systems.
\newblock {\em Biometrics\/}~{\em 67\/}(2), 476--485.

\bibitem[\protect\citeauthoryear{Marsden and Parkes}{Marsden and
  Parkes}{1976}]{marsden1976off}
Marsden, C.~D. and J.~Parkes (1976).
\newblock ``on-off'' effects in patients with parkinson's disease on chronic
  levodopa therapy.
\newblock {\em The Lancet\/}~{\em 307\/}(7954), 292--296.

\bibitem[\protect\citeauthoryear{Mermelstein}{Mermelstein}{1976}]{mermelstein1976distance}
Mermelstein, P. (1976).
\newblock Distance measures for speech recognition, psychological and
  instrumental.
\newblock {\em Pattern recognition and artificial intelligence\/}~{\em 116},
  374--388.

\bibitem[\protect\citeauthoryear{Michaelis, Gramss, and Strube}{Michaelis
  et~al.}{1997}]{michaelis1997glottal}
Michaelis, D., T.~Gramss, and H.~W. Strube (1997).
\newblock Glottal-to-noise excitation ratio--a new measure for describing
  pathological voices.
\newblock {\em Acta Acustica united with Acustica\/}~{\em 83\/}(4), 700--706.

\bibitem[\protect\citeauthoryear{Naylor, Kounoudes, Gudnason, and
  Brookes}{Naylor et~al.}{2006}]{naylor2006estimation}
Naylor, P.~A., A.~Kounoudes, J.~Gudnason, and M.~Brookes (2006).
\newblock Estimation of glottal closure instants in voiced speech using the
  dypsa algorithm.
\newblock {\em IEEE Transactions on Audio, Speech, and Language
  Processing\/}~{\em 15\/}(1), 34--43.

\bibitem[\protect\citeauthoryear{Neto, Perumal, Pratap, Bot, Mangravite, and
  Omberg}{Neto et~al.}{2017}]{neto2017analysis}
Neto, E.~C., T.~M. Perumal, A.~Pratap, B.~M. Bot, L.~Mangravite, and L.~Omberg
  (2017).
\newblock On the analysis of personalized medication response and
  classification of case vs control patients in mobile health studies: the
  mpower case study.

\bibitem[\protect\citeauthoryear{Seth, Barrett, and Barnett}{Seth
  et~al.}{2015}]{seth2015granger}
Seth, A.~K., A.~B. Barrett, and L.~Barnett (2015).
\newblock Granger causality analysis in neuroscience and neuroimaging.
\newblock {\em Journal of Neuroscience\/}~{\em 35\/}(8), 3293--3297.

\bibitem[\protect\citeauthoryear{Shephard and Pitt}{Shephard and
  Pitt}{1997}]{shephard1997likelihood}
Shephard, N. and M.~K. Pitt (1997).
\newblock Likelihood analysis of non-gaussian measurement time series.
\newblock {\em Biometrika\/}~{\em 84\/}(3), 653--667.

\bibitem[\protect\citeauthoryear{Sieberts, Schaff, Duda, Pataki, Sun, Snyder,
  Daneault, Parisi, Costante, Rubin, et~al.}{Sieberts
  et~al.}{2021}]{sieberts2021crowdsourcing}
Sieberts, S.~K., J.~Schaff, M.~Duda, B.~{\'A}. Pataki, M.~Sun, P.~Snyder, J.-F.
  Daneault, F.~Parisi, G.~Costante, U.~Rubin, et~al. (2021).
\newblock Crowdsourcing digital health measures to predict {P}arkinson’s
  disease severity: the {P}arkinson’s disease digital biomarker {DREAM}
  challenge.
\newblock {\em NPJ Digital Medicine\/}~{\em 4\/}(1), 1--12.

\bibitem[\protect\citeauthoryear{Skodda, Rinsche, and Schlegel}{Skodda
  et~al.}{2009}]{skodda2009progression}
Skodda, S., H.~Rinsche, and U.~Schlegel (2009).
\newblock Progression of dysprosody in {P}arkinson's disease over time—a
  longitudinal study.
\newblock {\em Movement disorders: official journal of the Movement Disorder
  Society\/}~{\em 24\/}(5), 716--722.

\bibitem[\protect\citeauthoryear{Snyder, Tummalacherla, Perumal, and
  Omberg}{Snyder et~al.}{2020}]{snyder2020mhealthtools}
Snyder, P., M.~Tummalacherla, T.~Perumal, and L.~Omberg (2020).
\newblock mhealthtools: A modular r package for extracting features from mobile
  and wearable sensor data.
\newblock {\em Journal of Open Source Software\/}~{\em 5\/}(47), 2106.

\bibitem[\protect\citeauthoryear{Tsanas}{Tsanas}{2010}]{tsanas2010new}
Tsanas, A. (2010).
\newblock New nonlinear markers and insights into speech signal degradation for
  effective tracking of {P}arkinson’s disease symptom severity.
\newblock {\em Age (years)\/}~{\em 64\/}(8.1), 63--6.

\bibitem[\protect\citeauthoryear{Tsanas}{Tsanas}{2012}]{tsanas2012accurate}
Tsanas, A. (2012).
\newblock {\em Accurate telemonitoring of {P}arkinson’s disease symptom
  severity using nonlinear speech signal processing and statistical machine
  learning}.
\newblock Ph.\ D. thesis, Oxford University, UK.

\bibitem[\protect\citeauthoryear{Tsanas, Little, McSharry, and Ramig}{Tsanas
  et~al.}{2009}]{tsanas2009accurate}
Tsanas, A., M.~Little, P.~McSharry, and L.~Ramig (2009).
\newblock Accurate telemonitoring of {P}arkinson’s disease progression by
  non-invasive speech tests.
\newblock {\em Nature Precedings\/}, 1--1.

\bibitem[\protect\citeauthoryear{Tsanas, Little, McSharry, and Ramig}{Tsanas
  et~al.}{2011}]{tsanas2011nonlinear}
Tsanas, A., M.~A. Little, P.~E. McSharry, and L.~O. Ramig (2011).
\newblock Nonlinear speech analysis algorithms mapped to a standard metric
  achieve clinically useful quantification of average {P}arkinson's disease
  symptom severity.
\newblock {\em Journal of the royal society interface\/}~{\em 8\/}(59),
  842--855.

\bibitem[\protect\citeauthoryear{Van~Ness, O’Leary, Byers, Fried, and
  Dubin}{Van~Ness et~al.}{2004}]{van2004fitting}
Van~Ness, P.~H., J.~O’Leary, A.~L. Byers, T.~R. Fried, and J.~Dubin (2004).
\newblock Fitting longitudinal mixed effect logistic regression models with the
  nlmixed procedure.
\newblock In {\em Proceedings of the 29th Annual SAS Users Group International
  Conference, Montreal, Canada}. Citeseer.

\bibitem[\protect\citeauthoryear{Velasco}{Velasco}{2020}]{velascomixed}
Velasco, L. L.~H. (2020).
\newblock {\em Mixed Effects State-Space Models for Longitudinal Data with
  Heavy Tails.}
\newblock Ph.\ D. thesis, Federal University of Rio de Janeiro.

\bibitem[\protect\citeauthoryear{Vidoni}{Vidoni}{1999}]{vidoni1999exponential}
Vidoni, P. (1999).
\newblock Exponential family state space models based on a conjugate latent
  process.
\newblock {\em Journal of the Royal Statistical Society: Series B (Statistical
  Methodology)\/}~{\em 61\/}(1), 213--221.

\bibitem[\protect\citeauthoryear{Wang, Zhang, Kr{\"o}se, and van Hoof}{Wang
  et~al.}{2021}]{wang2021optimizing}
Wang, S., C.~Zhang, B.~Kr{\"o}se, and H.~van Hoof (2021).
\newblock Optimizing adaptive notifications in mobile health interventions
  systems: Reinforcement learning from a data-driven behavioral simulator.
\newblock {\em Journal of Medical Systems\/}~{\em 45\/}(12), 1--8.

\bibitem[\protect\citeauthoryear{Willis, Schootman, Evanoff, Perlmutter, and
  Racette}{Willis et~al.}{2011}]{willis2011neurologist}
Willis, A., M.~Schootman, B.~Evanoff, J.~Perlmutter, and B.~Racette (2011).
\newblock Neurologist care in {P}arkinson disease: a utilization, outcomes, and
  survival study.
\newblock {\em Neurology\/}~{\em 77\/}(9), 851--857.

\bibitem[\protect\citeauthoryear{{World Health Organization}}{{World Health
  Organization}}{2004}]{world2004atlas}
{World Health Organization} (2004).
\newblock {\em Atlas: country resources for neurological disorders 2004:
  results of a collaborative study of the World Health Organization and the
  World Federation of Neurology}.
\newblock World Health Organization.

\bibitem[\protect\citeauthoryear{Wroge, {\"O}zkanca, Demiroglu, Si, Atkins, and
  Ghomi}{Wroge et~al.}{2018}]{wroge2018parkinson}
Wroge, T.~J., Y.~{\"O}zkanca, C.~Demiroglu, D.~Si, D.~C. Atkins, and R.~H.
  Ghomi (2018).
\newblock Parkinson’s disease diagnosis using machine learning and voice.
\newblock In {\em 2018 IEEE Signal Processing in Medicine and Biology Symposium
  (SPMB)}, pp.\  1--7. IEEE.

\bibitem[\protect\citeauthoryear{Zhou and Tang}{Zhou and
  Tang}{2014}]{zhou2014estimating}
Zhou, J. and A.~Tang (2014).
\newblock Estimating linear mixed-effects state space model based on
  disturbance smoothing.

\end{thebibliography}
\end{document}